\documentclass[letterpaper, 11pt]{article}

\usepackage{mathtools}
\usepackage{array}
\usepackage{enumitem}
\usepackage{amsmath,amsfonts,amssymb,bm}
\usepackage{titlesec}
\usepackage{pgfgantt,rotating}
\usepackage{float}
\usepackage{graphicx}
\usepackage{sidecap}
\usepackage{wrapfig}
\usepackage{bold-extra}
\usepackage{authblk}
\usepackage{appendix}
\usepackage{cite}
\newcommand{\etal}{{\it et al. }}

\usepackage[margin=1in]{geometry}

\newcommand{\beq}{\begin{equation}}
\newcommand{\eeq}{\end{equation}}
\newcommand{\bgqar}{\begin{eqnarray}}
\newcommand{\enqar}{\end{eqnarray}}
\newcommand{\bgqarn}{\begin{eqnarray*}}
\newcommand{\enqarn}{\end{eqnarray*}}
\newcommand{\bgary}{\begin{array}}
\newcommand{\enary}{\end{array}}

\graphicspath{{figures/}}

\title{Guided Wave–Based Structural Awareness Under Varying Operating States via Manifold Representations}

\author{Yiming Fan}
\author{Dimitris G Giovanis}
\author{Fotis Kopsaftopoulos\footnote{Corresponding author.}}

\affil{\small Intelligent Structural Systems Laboratory (ISSL) \\ Department of Mechanical, Aerospace and Nuclear Engineering \\ Rensselaer Polytechnic Institute, Troy, NY, USA \\ Email: \{fany5,kopsaf\}@rpi.edu }

\affil{\small  Department of Civil and Systems Engineering \\ Johns Hopkins University \\ Email: \{dgiovan1\}@jhu.edu }

\date{\today}

\begin{document}

\maketitle

%-----------------------------------------------------------------------------------------------------
% Abstract
%-----------------------------------------------------------------------------------------------------

\begin{abstract}
Guided wave–based structural health monitoring (SHM) remains a powerful strategy for identifying early-stage defects and safeguarding vital aerospace structures. Yet, its practical use is often hindered by the enormous, high-dimensional data streams produced by sensor arrays operating at megahertz sampling rates, coupled with the added complexity of shifts in environmental and operational conditions (EOCs). Studies have explored various data-compression approaches that retain critical diagnostic details in a lower-dimensional latent space. While conventional techniques can streamline dimensionality to some extent, they do not always capture the nonlinear interactions typical of guided waves. Manifold learning, as illustrated by Diffusion Maps, tackles these nonlinearities by deriving low-dimensional embeddings directly from wave signals, minimizing the need for manual feature extraction. In parallel, developments in deep learning—particularly autoencoders—provide an encoder-decoder model for both data compression and reconstruction. Convolutional autoencoders (CAEs) and variational autoencoders (VAEs) have been particularly effective for guided wave applications. However, current methods can still struggle to maintain accurate state estimation under changing EOCs, and they are often limited to a single task. In response, the proposed framework adopts a two-fold strategy: it compresses high-dimensional signals into lower-dimensional representations and then leverages those representations to both estimate structural states and reconstruct the original data, even as conditions vary. Applied to two real-world use-cases, this integrated method has proven its ability to preserve and retrieve key damage signatures under noise, shifting operational parameters, and other complicating factors.
\end{abstract}

%-----------------------------------------------------------------------------------------------------
%                              Important conventions and symbols -- Acronyms
%-----------------------------------------------------------------------------------------------------

\clearpage

\section{Introduction} \label{sec:intro}
Guided waves are a special category of ultrasonic waves which can propagate through structural surfaces or thin-walled structures. Their unique characteristics, particularly their dispersion, reflect both material composition and structural geometrics. Given their capability to traverse long distance with minimal energy attenuation and sensitivity to multiple types of defects, i.e., cracks, corrosion or delaminations, etc, guided waves, have proven invaluable for monitoring systems across such as aerosapce, civil and mechanical sectors \cite{farrar2007introduction,diamanti2010structural}. A major benefit of guided waves applications in structural health monitoring (SHM) is their ability for early stage detection of flaws by sending and recording ultrasonic signals at chosen locaitons, using which practitioners can find minor anomalies while still manageable. Maintaining structural integrity is critical for both economics and safety \cite{Amer-Kopsaftopoulos19a,Amer-Kopsaftopoulos20}. In aerospace, for instance, an unseen crack in a fuselage panel can escalate into a critical failure mid-flight, leading to severe accident. Guided wave-based SHM can be implemented autonomously, offering operators a chance to address defects well before they lead to unplanned downtime or catastrophic breakdown \cite{2016-9,2017-5}.

Despite the advantages, guided wave approaches typically require large volumes of sensor data, which brings its own challenges \cite{Janapati-etal16}. Many inspection systems place numerous sensors througout a structure to form sensor arrays while capturing time-series data at magahertz sampling rate which leads to substantial raw data streams to be stored and processed. With dozens or hundreds of sensors installed, the size and dimensionality of the data may quickly become overwhelming, making it difficult in managing the datasets in real time. Moreover, environmental and operating conditions (EOCs), ranging from temperature to humidity, also affect the signals’ appearance. It then becomes critical to isolate the genuine damage indicators from the routine variations \cite{abbassi2023evaluation, eybpoosh2017energy}. 

These obstacles underscore the urgency of reducing data dimensionality while preserving key information for structural health diagnostics. To tackle the issues, researchers have employed data compression procedures to project the raw data onto a “latent space” \cite{vega2022variational}. This approach is also known as manifold learning. 
The representations in such a space preserve the features of dataset with fewer dimensions by naturally accentuate key damage-related patterns while suppressing irrelevant or noisy content. For guided waves data, with their overlapping modes and reflections, the compressed representation can reveal relationships that can otherwise be concealed. 
Classical approaches such as Principal Component Analysis (PCA) has been employed to reduce high-dimensional data in a variety of signal processing domains by identifying axes of maximum variance and keeping only the major components thereby simplifying the data \cite{mackiewicz1993principal, el2021adaptive}. This approach has shown success in providing quick dimensionality reduction. Nevertheless, as a linear data compression technique, PCA may not fit with complex nonlinear interactions in guided wave phenomena. Other linear data compression approaches such as Independent Component Analysis (ICA) and Partial Least Squares (PLS) also face similar issues \cite{ghavamian2018detection, mitra2016guided}. Under this concern, the SHM community increasingly recognizes that more advanced nonlinear methods are necessary, especially when dealing with damage signatures that require nonlinear transformations for better detection. Given the complex and nonlinear nature of wave propagation, researchers have turned to manifold learning strategies, such as Diffusion Maps (DMaps), Isomap and Locally Linear Embedding (LLE) \cite{coifman2006diffusion, balasubramanian2002isomap, roweis2000nonlinear}. These algorithms preserve the underlying data geometry, which is crucial for capturing wave interactions in complex structural configurations. Of particular interest, DMaps construct a weighted graph composed by data points, upon which a Markov chain and eigenvectors that reflect the primary modes of variation can be derived. Huang \etal proposes a hybrid deep learning framework for defect detection in carbon-fiber-reinforced polymer (CFRP) plates, combining diffusion models for data augmentation and DenseNet for defect classification \cite{huang2023hybrid}. Chinde \etal presents a spectral diffusion map approach for SHM of wind turbine blades using time-series data from distributed sensors \cite{chinde2015spectral}.

Simultaneoulsy, breakthroughs in deep learning have paved the way for newer dimensionality reduction strategies \cite{nerlikar2024physics, ewald2019deepshm}. Autoencoders (AEs), for example, have gained attention due to its special structure and characteristics. An autoencoder has two major components which are, an encoder that maps input data to a latent space, and a decoder that reconstructs the original data from the latent representation. By minimizing the discrepancy which is often measured by mean squared error between the inputs and their reconstructions, the network learns a compact encoding that highlights the most informative features. By condensing data into a bottleneck layer and then striving to expand it back to the original shape, data compression procedure is naturally adopted by AEs. In SHM scenarios, an AE that is exposed solely to signals drawn from the “healthy” structural state can subsequently flag damage by exhibiting significantly higher reconstruction error once anomalies enter the input stream. For example, Lee \etal presents an automated technique for detecting and classifying fatigue damage in carbon fiber reinforced polymer (CFRP) composite structures using ultrasonic Lamb waves and a deep autoencoder (DAE) \cite{lee2022automated}. The proposed method accurately detected and classified fatigue damage without manual feature extraction, showing high sensitivity and reliability during progressive fatigue cycles. Yang \etal proposes an unsupervised damage detection method using an optimal autoencoder for long-term structural health monitoring under varying environmental conditions \cite{yang2023unsupervised}. The approach enables robust, long-term and unsupervised damage detection in dynamic environments without requiring pristine training data, making it practical for real-world SHM applications.

Over time, several variations of the autoencoder framework have emerged. For example, Convolutional autoencoders (CAEs) introduce convolutional operations in both encoder and decoder components by imploying convolutional layers to detect local patterns which is originally popularized for image processing. Due to the capability to capture both spatial and temporal variations in the waveform, CAE filters can naturally capture changes in wave modes, reflections or potential defects. Although CAEs often need larger datasets and more hyperparameter tuning (e.g., selecting filter sizes or layer depths), they have already shown promise in detecting subtle anomalies in real and simulated structural environments. For example, Rautela \etal proposes an unsupervised CAE-based feature learning approach for delamination detection in composite panels using wavelet-enhanced guided wave representations, addressing the limitations of supervised learning due to the difficulty in collecting labeled damage datasets \cite{rautela2022delamination}. Shang \etal presents a deep convolutional denoising autoencoder (DCDAE) for detecting bridge damage using vibration data. The model has been proved to be robustness even under high noise and temperature variations \cite{shang2021vibration}.

Another popular type of autoencoder that worths mentioning is variational autoencoders (VAEs) which introduces a probabilistic twist to the autoencoder. Instead of mapping inputs to a fixed set of latent variables, VAEs learn distributions over these variables, typically assuming a Gaussian form as prior. During training, the encoder outputs both a mean and a variance for the latent representation, and a sampling step is incorporated to learn the distribution. This design enables VAEs to capture uncertainty about the data-generating process, which can be especially beneficial in SHM scenarios where variability stems from environmental changes or sensor noise. VAEs also allow for generating synthetic signals by sampling from the latent distributions, which can be useful if one wants to simulate damage conditions or enrich a sparse dataset. Hwang \etal combines VAE and Support Vector Rgression (SVR) to offer a robust approach for real-time monitoring, detection, and classification of fatigue damage in composite structures under complex loading conditions \cite{hwang2025fatigue}. Kulkarni \etal introduces a Physics-Informed Variational Autoencoder (PI-VAE) to expand sparse measurements into full-field representations while detecting damage in structures under thermal excitation \cite{kulkarni2024full}. Junges \etal applies VAE to mitigate the impact of temperature variations on ultrasonic guided wave-based SHM \cite{junges2024mitigating}.

However, current methods can still struggle to maintain accurate state estimation under changing EOCs, and they are often limited to a single task. In response, the proposed framework adopts a two-fold strategy: it compresses high-dimensional signals into lower-dimensional representations and then leverages those representations to both estimate structural states and reconstruct the original data, even as conditions vary.
When structure condition itself and multiple external parameters fluctuate at once, it becomes difficult for conventional methods to accurately estimate the system state. To tackle this issue, a versatile framework that integrates different compression and expansion techniques is proposed, allowing for both state forecasting and data reconstruction at specific states even as conditions evolve. In the setup, dimensionality reduction can be performed with diffusion maps or encoders, while Laplacian pyramids or decoders enable data expansion. The framework was validated on two distinct scenarios featuring varied noise intensities: one examining a static system with changing structural damage levels and loads, and another involving a wind tunnel test where signals were influenced by angle-of-attack (AoA) and airspeed variations. Data augmentation was further employed to enhance the performance of the VAE-based module. Results indicate that, under static settings, the proposed approach precisely reconstructs signals and identifies states, and it continues to deliver reliable outcomes even under challenging conditions.

%-----------------------------------------------------------------------------------------------------
% PROBLEM STATEMENT                                         
%-----------------------------------------------------------------------------------------------------
\section{Problem Statement} \label{sec:problem}

The \emph{goal} of this work is to investigate the latent space representations from both deterministic and probabilistic models and how they affect the model performance on solving both forward and inverse SHM tasks. The near-perfect prediction accuracy from the first test case leads to a more challenging one where a wing is tested in a wind tunnel. Rather than collected in a static condition without significant disturbance, this time the data is received with high levels of noise due to airflow and structural complexity. Additional dimensionality reduction and generative methods using VAEs are employed to address the amplified uncertainty. The results reveal that the compression-expansion hybrid approaches can maintain diagnostic accuracy even when ambient conditions vary considerably, demonstrating the importance of adaptable methods for real-world SHM deployments.

The main contributions of this study are as follows: 
\begin{itemize}
\item Implementation and comparison of various nonlinear data compression techniques, including both deterministic and probabilistic models.
\item Data augmentation through sampling methods based on distributions learned from probabilistic machine learning models.
\item Application to challenging real-world test cases involving complex structures under noisy conditions.
\item Comprehensive evaluation of different models in terms of data compression and expansion performance.
\end{itemize}

\section{Data Representation and Explanation}

In this study, model identification is based on noise-corrupted output data records corresponding to a sample of the admissible structure and operating conditions. Each data record corresponds to a specific value of the state vector $\bm{k}$, which, without loss of generality, is assumed to be two-dimensional. The dataset consists of $N$ such records, each represented as a vector in $\mathbb{R}^m$. A sample of $M_1$ values is used for the first measurable variable $k^1$ (first element of vector $\bm{k}$), while a sample of $M_2$ values is used for the second externally measurable variable $k^2$ (second element of vector $\bm{k}$). 

Each experiment is characterized
by a specific element of $\bm{k}$, say $\bm{k} = [k_i^1,k_j^2]$. This vector is, for simplicity of notation, also designated as the duplet $k_{i,j}=(k_i^1,k_j^2)$ (the first term designating the value of $k^1$ and the second that of $k^2$). As a result, a total of $\sum_{i=1}^{M_1} \sum_{j=1}^{M_2} n_{i,j}$ experiments ($n_{i,j}$ is the number of trials repeated at state $k_{i,j}$ to decrease noise effects) are performed, with the complete
series covering the required range of each scalar parameter, say $[k_{min}^1, k_{max}^1]$ and $[k_{min}^2, k_{max}^2]$, via the
discretizations $k^1=k_1^1,k_2^1,...,k_{M_1}^1$ and $k^2=k_1^2,k_2^2,...,k_{M_2}^2$.

Data records from different structure and operating points corresponding to various values of the operating parameter vector can be written as:
\begin{equation}
 \bm{Y}\triangleq \{y_{\bm{k}}[t]|\bm{k}\triangleq[k^1,k^2]^T, t=1,...,m, k^1\in{k_1^1,k_2^1,...,k_{M_1}^1}, k^2\in{k_1^2,k_2^2,...,k_{M_2}^2}\}
 \label{eq:ae_eqn1}
\end{equation}

In this expression, $t$ designates normalized discrete time (the corresponding analog time being $t\cdot T$ with $T$ standing for the sampling period), $y_{\bm{k}}[t]$ the noise-corrupted output signals corresponding to $\bm{k}$. Each data record can then be expressed as $\bm{y}_{\bm{k}}\in\mathbb{R}^{m}$.
% $N$ stands for the signal length (in samples) corresponding to each experiment (each $\bm{k}$).
%-----------------------------------------------------------------------------------------------------

\section{Background Knowledge}

%%%%%%%%%%%%%%%%%%%%%%%%%%%%%%%

\subsection{Diffusion Maps}
\label{S:dmaps}
\noindent
Originally introduced by Coifman and Lafon \cite{coifman2006diffusion}, the diffusion maps (DMaps) method is a powerful approach for dimension reduction of a finite dataset 
\begin{equation}
\mathbf{y} = \{\mathbf{y}_1, \ldots, \mathbf{y}_N\} \subset \mathbb{R}^m,
\end{equation}
where $m$ denotes the ambient dimension of each sample (e.g., the number of time steps in a signal). Under the assumption that these samples lie on a low-dimensional manifold $\mathcal{M} \subset \mathbb{R}^m$, DMaps provides a set of intrinsic coordinates that capture the geometry of $\mathcal{M}$. The first step of the algorithm involves constructing a weighted graph on the data by defining a Markov transition matrix. A positive semi-definite kernel (often chosen to be Gaussian),
\begin{equation}
    \mathbf{K}(\mathbf{y}_i, \mathbf{y}_j) 
    = \exp \!\Bigl(-\frac{\|\mathbf{y}_i - \mathbf{y}_j\|_2^2}{2\epsilon}\Bigr),
\end{equation}
is used, where $\epsilon$ is a bandwidth parameter. Small pairwise distances $\|\mathbf{y}_i - \mathbf{y}_j\|_2$ correspond to higher similarity. As $\epsilon \to 0$ and $N \to \infty$, this kernel approximates the Neumann heat kernel $e^{-\Delta t}$ on $\mathcal{M}$, with $\Delta$ denoting the Laplace--Beltrami operator; its eigenfunctions reveal intrinsic manifold coordinates. Because density variations in the data can bias the operator, an initial normalization step is often performed:
\begin{equation}
    \tilde{\mathbf{K}} 
    = \mathbf{P}^{-\alpha} \,\mathbf{K}\,\mathbf{P}^{\alpha},
\end{equation}
where
\begin{equation}
    \mathbf{P}_{ii} 
    = \sum_{j=1}^{N} \mathbf{K}(\mathbf{y}_i, \mathbf{y}_j),
\end{equation}
and $\alpha$ modulates how strongly the sampling density influences the geometry. Setting $\alpha=1$ ignores density effects to approximate the Laplace--Beltrami operator directly; choosing $\alpha=0$ maximizes the density influence (an accurate geometric approximation then requires uniform sampling). A second normalization step produces the row-stochastic matrix
\begin{equation}
    \mathbf{L}(\mathbf{y}_i, \mathbf{y}_j)
    = \frac{\tilde{\mathbf{K}}(\mathbf{y}_i, \mathbf{y}_j)}
           {\sum_{j=1}^{N} \tilde{\mathbf{K}}(\mathbf{y}_i, \mathbf{y}_j)},
\end{equation}
which admits the eigendecomposition
\begin{equation}
\mathbf{L}\,\mathbf{v}_i 
= \lambda_i \,\mathbf{v}_i.
\end{equation}
By choosing the first $d<m$ eigenvectors in a parsimonious manner \cite{dsilva2018parsimonious}, a $d$-dimensional reduction is achieved via
\begin{equation}
    \mathbf{y}_i \;\mapsto\; 
    \bigl\{\lambda_1^t\,\mathbf{v}_1(\mathbf{y}_i),\,\ldots,\,
           \lambda_d^t\,\mathbf{v}_d(\mathbf{y}_i)\bigr\},
    \label{eq:diff-coord}
\end{equation}
where $t$ can be viewed as both a time parameter for the Markov chain and a scale parameter on the graph. A common choice is $t = 1$. The effectiveness of DMaps in capturing the manifold's structure relies on an appropriate kernel that measures similarity between points. Although a Gaussian kernel based on Euclidean distances is standard, certain applications benefit from custom distance metrics.

To recover points $\mathbf{y}_i$ in the original space from their reduced coordinates $\boldsymbol{\varphi}_i$, a multi-scale technique called Laplacian pyramids \cite{rabin2012heterogeneous} can be employed. Suppose a set of latent variables $\{\boldsymbol{\varphi}_i\}_{i=1}^{N}$ and corresponding outputs $\{\mathbf{y}_i\}_{i=1}^{N}$ are known. The objective is to approximate the mapping
\begin{equation}
    f^{-1}:\boldsymbol{\varphi}\;\mapsto\;\mathbf{y}.
\end{equation}
At each scale $l$, a Gaussian kernel in the latent space is defined as
\begin{equation}
W_l(\boldsymbol{\varphi}_i, \boldsymbol{\varphi}_j)
= \exp\!\Bigl(-\frac{\|\boldsymbol{\varphi}_i - \boldsymbol{\varphi}_j\|^2}{\sigma_l}\Bigr),
\end{equation}
where $\sigma_l$ indicates the scale. A smoothing operator then follows:
\begin{equation}
K_l(\boldsymbol{\varphi}_i, \boldsymbol{\varphi}_j) 
= \frac{W_l(\boldsymbol{\varphi}_i, \boldsymbol{\varphi}_j)}
       {\sum_{j} W_l(\boldsymbol{\varphi}_i, \boldsymbol{\varphi}_j)},
\end{equation}
which provides an approximation
\begin{equation}
s_l(\boldsymbol{\varphi}_k)
= \sum_{i=1}^{N} K_l(\boldsymbol{\varphi}_i, \boldsymbol{\varphi}_k)\,\mathbf{y}_i.
\end{equation}
The difference
\begin{equation}
d_l(\boldsymbol{\varphi}_k)
= \mathbf{y}_k - s_{l-1}(\boldsymbol{\varphi}_k)
\end{equation}
is fed back into the next level of refinement, and iterations continue until the desired reconstruction accuracy is achieved. This approach yields a practical method for inverting the diffusion map and retrieving high-dimensional samples from their low-dimensional parameterizations.

\subsection{Convolutional Autoencoder}
\label{S:cae}
\noindent
Convolutional Autoencoder (CAE) is a type of deep learning model that extends traditional autoencoders by incorporating convolutional layers instead of fully connected layers. Unlike conventional autoencoders with only fully connected (FC) layers which flatten input data, its architecture is particularly effective for temporally and spatially structured data, such as time series and images. CAEs maintain spatial and temporal structure through convolutional and pooling operations, making them more efficient in capturing hierarchical features. A CAE consists of two main components: 
\begin{itemize}
    \item \textbf{Encoder:} Extracts hierarchical spatial or temporal features from the input using convolutional and pooling layers. The encoder compresses the input data $\mathbf{y}$ into a low-dimensional latent representation $\mathbf{z}\in\mathbb{R}^{d}$, where $d$ is the pre-defined latent space width, the latent vector can then be regarded as a compressed yet informative representation of the original data $\mathbf{y}$.
    \item \textbf{Decoder:} Reconstructs the original input from the latent space using upsampling and transposed convolution layers.
\end{itemize}

Given an input dataset \( Y = \{ \mathbf{y}_i \}_{i=1}^{N} \), where each sample $\mathbf{y} \in \mathbb{R}^{m}$ is a time series, CAEs encode each sample into a compressed latent space representation \( Z = \{ \mathbf{z}_i \}_{i=1}^{N} \) through the encoder and reconstruct an approximate version \( \hat{Y} \) through the decoder. The encoding process can be expressed as:

\begin{equation}
    \mathbf{z} = f_{\theta}(\mathbf{y}) = \sigma(\mathbf{W} * \mathbf{y} + \mathbf{b})
\end{equation}
where \( f_{\theta} \) is the encoder which consists of convolutional layers followed by activation functions and pooling layers; $\mathbf{W}$ and $\mathbf{b}$ are the learnable convolution kernel and bias; $\sigma$ is a non-linear activation function. The decoding process can be expressed as:

\begin{equation}
    \hat{\mathbf{y}} = g_{\phi}(\mathbf{z}) = \sigma(\mathbf{W'} * \mathbf{z} + \mathbf{b'})
\end{equation}
where \( g_{\phi} \) is the decoder consists of transposed convolution (deconvolution) and upsampling layers, $\mathbf{W'} $ and $\mathbf{b'}$ are the parameters of the decoder.

CAEs are typically trained to minimize the reconstruction error between the original input and the reconstructed output. The loss function applied in this study is Mean Squared Error (MSE):

\[
\mathcal{L}_{\text{CAE}} = \frac{1}{N} \sum_{i=1}^{N} \| \mathbf{y}_i - g_{\phi}(f_{\theta}(\mathbf{y}_i)) \|^2
\]

% where \( f_{\theta} \) is the encoder, \( g_{\phi} \) is the decoder, \( \mathbf{y}_i \) is the original input, \( \hat{\mathbf{y}}_i = g_{\phi}(f_{\theta}(\mathbf{y}_i)) \) is the reconstructed output.

\subsection{Variational Autoencoder}
\label{S:vae}
\noindent

%%%%%%%% new version starts
A Variational Autoencoder (VAE) is another variant of autoencoder that builds upon traditional autoencoders by introducing a probabilistic approach to latent space representation. Unlike standard autoencoders, which map input data to a fixed latent representation, VAEs learn a probability distribution over the latent space, enabling controlled sampling and generative capabilities. Given the distribution, new samples can be drawn continuously from the latent space, enabling diverse and realistic outputs, allowing for dataset expansion, and potentially enhancing model performance. A VAE consists of two primary components of the neural network: inferential network (encoder) and generative network (decoder). The former part maps the input data \( \mathbf{y} \) into a latent probability distribution instead of a fixed vector. This process approximates the posterior distribution \( q_{\phi}(\mathbf{z}|\mathbf{y}) \) using a multivariate Gaussian with learned mean \( \mu \) and variance \( \sigma^2 \). The decoder then reconstructs the original data \( \hat{\mathbf{y}} \) from the latent representation \( \mathbf{z} \), modeling the conditional probability distribution \( p_{\theta}(\hat{\mathbf{y}}|\mathbf{z}) \).

The dataset \( Y = \{\mathbf{y}_i\}_{i=1}^{N} \) consists of independently sampled data points. The latent variable \( Z \) is defined in a lower-dimensional space, allowing the generative model to approximate the original dataset by producing \( \hat{Y} \). The encoder learns an approximate posterior distribution over the latent variable \( z \) as follows:

\[
q_{\phi}(\mathbf{z}|\mathbf{y}) = \mathcal{N}(\mathbf{z}; \mu(\mathbf{y}), \sigma^2(\mathbf{y}) \mathbf{I})
\]

Instead of directly sampling \( \mathbf{z} \), VAE introduces an auxiliary variable \( \epsilon \) sampled from a standard normal distribution \( \mathcal{N}(0,1) \), using the reparameterization trick:

\[
\mathbf{z} = \mu + \sigma \odot \epsilon, \quad \epsilon \sim \mathcal{N}(\mathbf{0}, \mathbf{I})
\]

This transformation allows backpropagation through the stochastic layer. Within the generative process, The decoder reconstructs output \( \mathbf{y}^* \) given \( \mathbf{z} \), as $p_{\theta}(\mathbf{y}^*|\mathbf{z})$. The likelihood function of \( \mathbf{y}^* \) herein is typically modeled as a Gaussian distribution. The training objective of a VAE is to maximize the variational lower bound (ELBO), which is equivalent to minimizing the sum of reconstruction error and Kullback-Leibler (KL) divergence between the approximate posterior \( q_{\phi}(\mathbf{z}|\mathbf{y}) \) and the true posterior \( p(\mathbf{z}|\mathbf{y}) \) and can be expressed as:

\[
\mathcal{L}_{\text{VAE}} = \mathbb{E}_{q_{\phi}(\mathbf{z}|\mathbf{y})} [\log p_{\theta}(\mathbf{y}|\mathbf{z})] - D_{\text{KL}}(q_{\phi}(\mathbf{z}|\mathbf{y}) \parallel p(\mathbf{z}))
\]
where the first term represents the reconstruction loss, ensuring that the generated data \( \mathbf{y}^* \) closely resembles the original input \( \mathbf{y} \) and the second term is the KL divergence, which regularizes the latent space to be close to a predefined prior distribution \( p(\mathbf{z}) \), assumed in this work to be a standard normal distribution \( \mathcal{N}(\mathbf{0}, \mathbf{I}) \)).

% \subsection{State Prediction using K-Nearest Neighbors (KNN)}
% After extracting features using VAE, classification is performed via KNN.
% \begin{equation}
%     \hat{s} = \arg\max_{s} \sum_{i=1}^{k} \mathbb{1} (s_i = s)
% \end{equation}
% where $s_i$ represents the state of the $i$-th neighbor.

%-----------------------------------------------------------------------------------------------------
% Methodology and flowchart
%-----------------------------------------------------------------------------------------------------

\section{Methodology}

\begin{figure}[!t]
\centering
\includegraphics[scale=0.7]{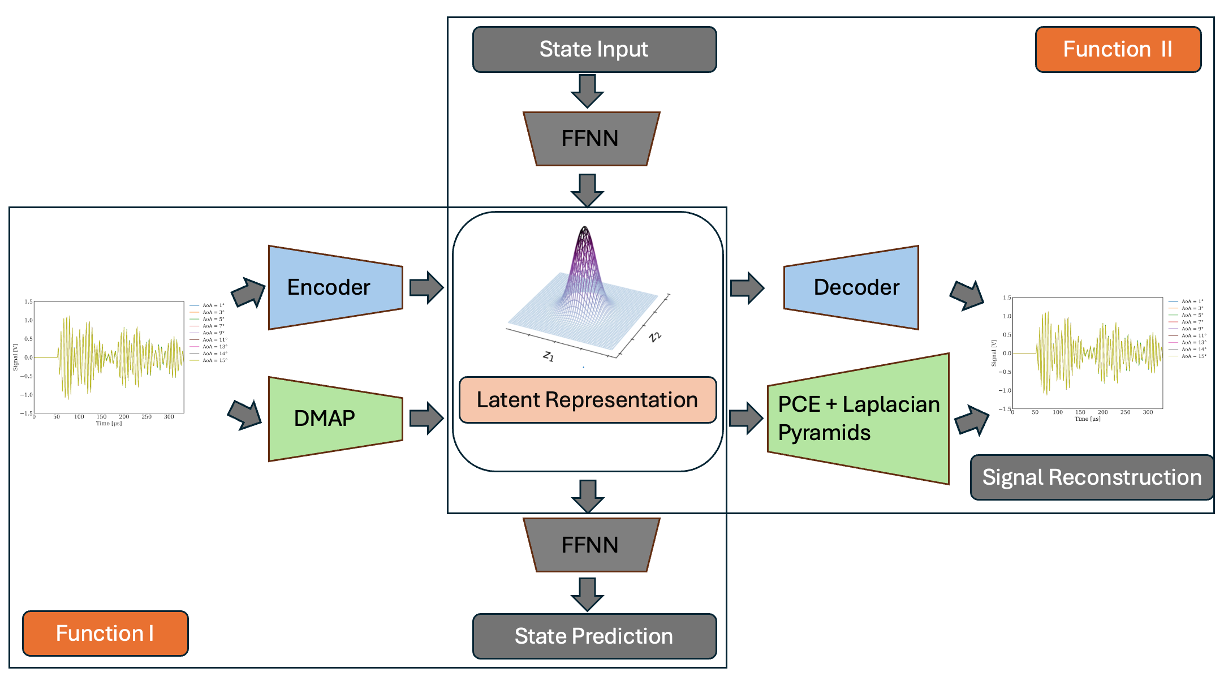}
\caption{The flowchart that demonstrates the procedures of state prediction and signal reconstruction achieved by two proposed approaches.}.
\label{fig:flowchart} 
\vspace{10pt}
\end{figure}

\noindent
In their earlier works \cite{fan2024data,fan2025unified}, the authors proposed a dual-function technique encompassing both state prediction and signal reconstruction via two distinct models, as depicted in Figure~\ref{fig:flowchart}. Within this framework, the data compression model generate latent features from high-dimensional data. Although these latent coordinates may arise from different mathematical underpinnings, they demonstrate comparable sensitivity to damage levels and various operating states.

Capitalizing on this property, latent variables derived by various methods are then mapped to state vectors through a feedforward neural network (FFNN). This arrangement enables data collected under diverse conditions to be compressed into a lower-dimensional domain and reliably classified, thereby reducing computational overhead. Conversely, for a given state, another FFNN can generate the latent coordinates, which are subsequently expanded to reconstruct the original signals via either Laplacian pyramids or a decoder. Both implementations of these functions are investigated and compared against two distinct datasets, showcasing their generalized performance in differing environments.

An important point is that the proposed framework can integrate a wide variety of compression and expansion approaches, as long as it is possible to form a reliable latent space representation. Since the input signals in this research include stochastic time series whose dynamics are often challenging for purely linear models, three nonlinear data compression techniques were employed: diffusion map, CAE and VAE. Their respective performances were then evaluated across multiple scenarios characterized by varying degrees of complexity, illustrating how each method contributes to the overall efficacy of the approach.

\subsection{Latent Space Representation}
%\noindent \textbf{1. Convolutional Autoencoder}  \vspace{12pt} 

In contrast to conventional feature extraction techniques, artificial neural networks (ANNs) can automatically extract nonlinear features, eliminating the need for tedious computations related to state-sensitive features. Within this framework, the convolutional autoencoder (CAE) is selected as the feature extraction tool, as it allows for extensive compression of inputs through the encoder and achieves low reconstruction error. While this study focuses on damage level and external load as the primary variables, it is straightforward to incorporate additional factors into this framework, following the same construction approach, thereby enhancing its adaptability to various working conditions and scenarios. 

% Denote the signal data as , where $N$ is the signal length.
As data $\bm{y}_{\bm{k}}\in\mathbb{R}^{N}$ pass through the encoder portion, it will be compressed by the convolutional layers as well as by the following downsampling and dense layers. Maximum compression occurs in the bottleneck layer, which also serves as the connector of the encoder and the decoder. The latent space vector, also called feature space, is derived from this layer. Denoting as $\bm{z}_{\bm{k}}\in\mathbb{R}^{D}$, where $D$ is the predefined latent space width, the latent vector can then be regarded as a compressed yet informative representation of the original data $\bm{y}_{\bm{k}}$. To express the latent space variable under different states clearly, 
$\bm{z}_{m}(k^1,k^2)$
will be used later in this paper, where $m$ indicates the $m$-th latent space variable.

% The state vector is denoted as $\mathbf{s}$. Since there are two state factors in this work, $\mathbf{s}$ can also be expressed as 
% $\mathbf{s} = [s_1,s_2]$
% , where 
% $s_i\in\mathbb{N}$ is the state index of $i$-th state factor. Note indices are applied here to represent the level of each state factor for convenient application, to be specific, integers 0-4 herein represent for the five damage levels and five loads that range from 0 to 20kN with a 5kN increment, e.g., $\mathbf{s} =[2,3]$ means the state at damage level 2, 15kN. To express the latent space variable under different states clearly, 
% $\mathbf{z}_{m,j}(s_1,s_2)$
% will be used later in this paper, where $m$ means the $m$-th model and $j$ means the $j$-th latent variable.

\subsection{State Estimation}
%\vspace{12pt}
%\noindent \textbf{2.	State Estimation Given Signals}  \vspace{12pt} 

As shown in Figure \ref{fig:flowchart}, the test phase contains two parts. The trained NNs can be arranged in specific sequences to achieve different objectives. For the purpose of state estimation, the signals are initially projected through the trained encoder, resulting in a compressed vector. This compressed data is then input into the first trained FFNN to generate the corresponding state vector estimates. This process can be expressed mathematically as follows:

\begin{equation}
 \widehat{\bm{k}}=\varphi_{1}(\bm{z})= \varphi_{1}(\alpha_{yz}(\bm{W}_{yz}\bm{y}+\bm{b}_{yz}))
 \label{eq:ae_eqn2}
\end{equation}
where $\varphi_{1}$ is the 1st FFNN function in the state estimation branch, $\alpha_{yz}$ is the nonlinear function for the encoder, $\bm{W}_{yz}$ and $\bm{b}_{yz}$ are weights and bias for encoder, respectively.

\subsection{Signal Reconstruction}
%\vspace{12pt}
%\noindent \textbf{3.	Signal Reconstruction Given States}  \vspace{12pt} 

For signal reconstruction propose, the inputs are the states with or without the paths information. Whether the paths information is needed depends on the model types that will be discussed later. The second FFNN aims to find the corresponding latent vectors, which will be further streched by the decoder to obtain the reconstructed signal. This process can be expressed as:
\begin{equation}
 \widehat{\bm{y}}=\alpha_{zy}(\bm{W}_{zy}\bm{z}+\bm{b}_{zy})=\alpha_{zy}(\bm{W}_{zy}{\varphi_{2}(\bm{k})}+\bm{b}_{zy})
 \label{eq:ae_eqn3}
\end{equation}

where $\varphi_{2}$ is the 2nd FFNN function in the signal reconstruction branch, $\alpha_{zy}$ is the nonlinear function for the decoder, $\bm{W}_{zy}$ and $\bm{b}_{zy}$ are weights and bias for decoder, respectively.

To assess the signal reconstruction performance, RSS/SSS($\%$) is used as the criteria to quantify how well the reconstructed signals match with the original ones. Given the amplitude of original signal and the reconstructed signal at time $t$ as $y[t]$ and $\widehat{y}[t]$ respectively, the formulation of RSS/SSS($\%$) is:

\begin{equation}
 RSS/SSS(\%) =
    \frac{\sum_{t=1}^{N}(y[t]-\widehat{y}[t])^2}{\sum_{t=1}^{N}{y[t]}^2}
 \label{eq:rss}
\end{equation}

\section{Experimental Setup}
\noindent
To assess the robustness and generalization of the proposed framework under varying real-world conditions, two distinct test cases featuring different state factors were conducted. The first test was carried out under static conditions, meaning both the internal structural changes and external environment remained constant throughout the signal acquisition phase. Consequently, this nearly noise-free setting ensured minimal discrepancies among signals taken under the same state, enabling the model to demonstrate highly reliable state prediction performance with very low error.

Building on the success observed in the static scenario, a second test was performed in a wind tunnel. This environment, characterized by airflow around the wing, introduced more noise into the acquired signals, thereby posing a greater challenge for accurate state prediction. Despite these complications, the framework’s predictive capabilities were put to the test, illustrating its potential to handle real-world complexities arising from fluctuating operational conditions.

\subsection{Test Case I}

% sample signal
\begin{figure}[t!]
    % \centering
    \begin{picture}(500,230)
    \put(50,10){\includegraphics[width=0.28\textwidth]{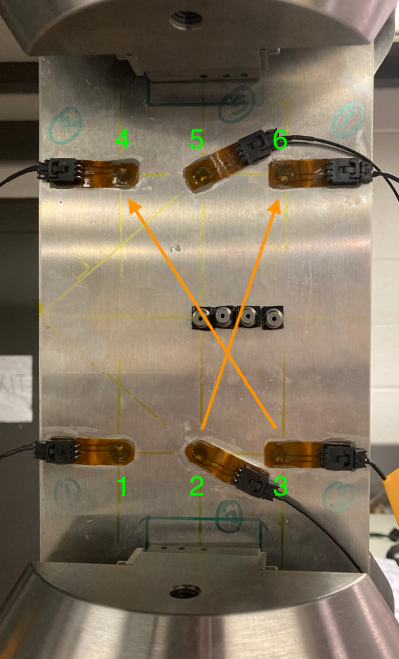}}
    \put(214,120){\includegraphics[width=0.38\textwidth]{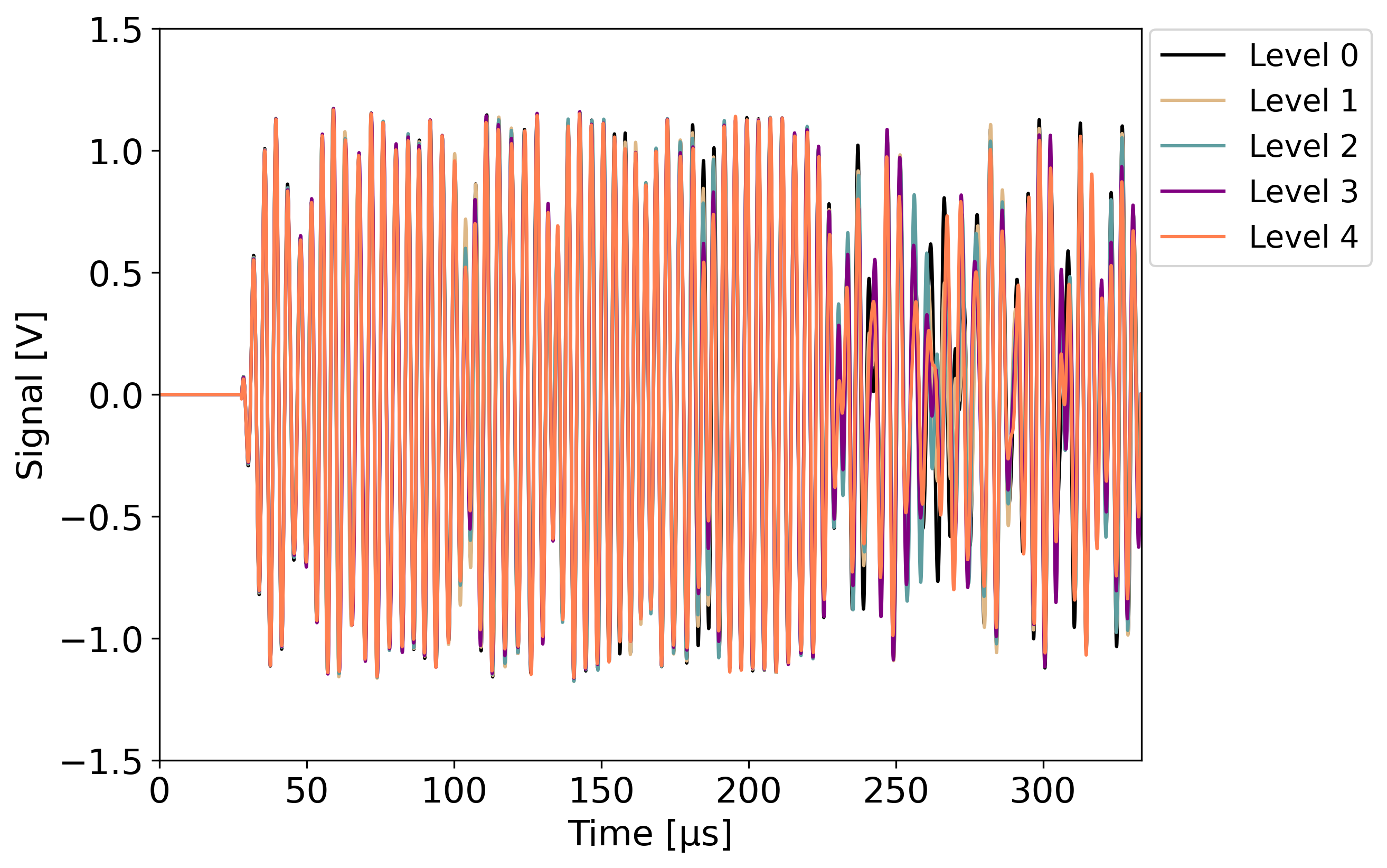}}
    \put(214,0){\includegraphics[width=0.38\textwidth]{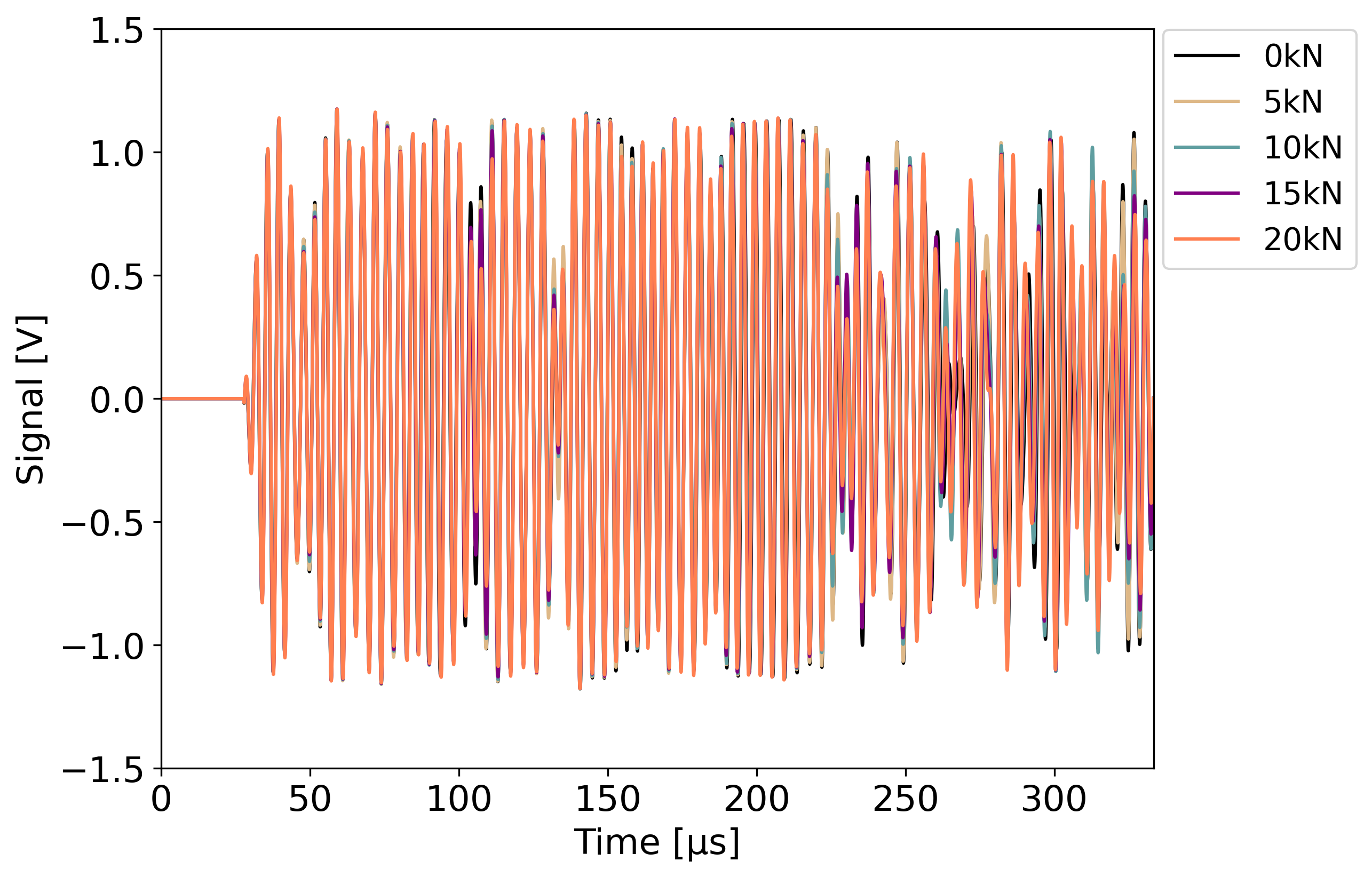}}
    
   %  \put(200,280){\color{black} \large {\fontfamily{phv}\selectfont \textbf{a}}}
   %  \put(410,280){\large {\fontfamily{phv}\selectfont \textbf{d}}}
   % \put(200,140){\large {\fontfamily{phv}\selectfont \textbf{b}}} 
   \put(380,140){\large {\fontfamily{phv}\selectfont \textbf{(b)}}}
   \put(200,0){\large {\fontfamily{phv}\selectfont \textbf{(a)}}} 
   \put(370,0){\large {\fontfamily{phv}\selectfont \textbf{(c)}}} 
    \end{picture} 
    \vspace{10pt}
    \caption{Experimental setup and sample signals collected in test case I. Panel (a): the Al plate in this test; panel (b): full signals under 5kN at all damage levels from path 3-4; Panel c: full signals at damage level 2 under all loads from path 2-6.}
\label{fig:signals_samples} 
% \vspace{-12pt}
\end{figure}
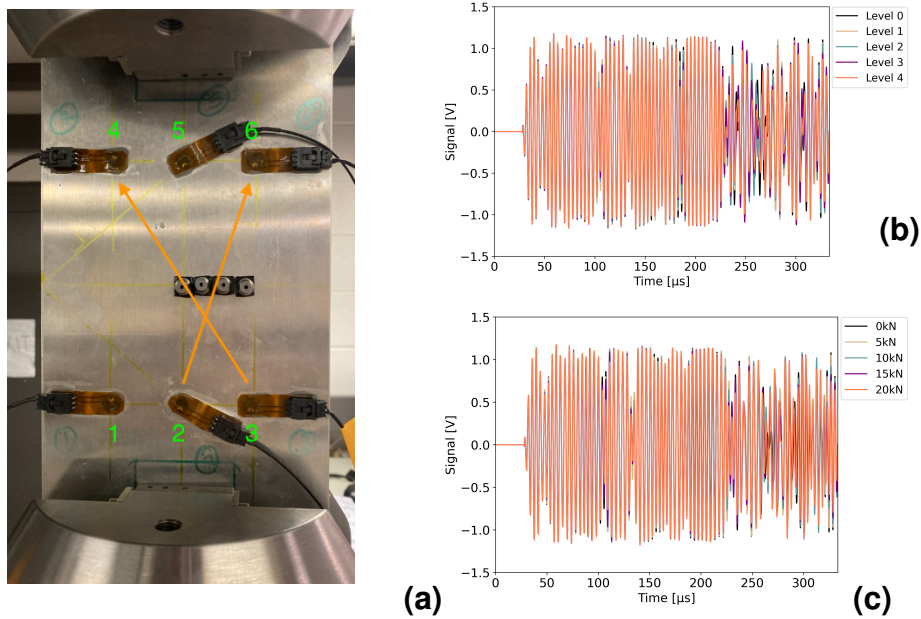

A Lamb wave-based active sensing scheme was adopted to validate the proposed framework using a 152.4\,mm\,$\times$\,304.8\,mm (6\,in\,$\times$\,12\,in) 6061 aluminum coupon of 0.093\,in thickness. Data gathering relied on a ScanGenie~III system (Acellent Technologies, Inc.). As indicated in Figure~\ref{fig:signals_samples}(a), six PZT-5A sensors were placed at predefined positions on the coupon. Based on their spatial assignments, sensors 1--3 alternately functioned as actuators by generating 5-peak tone burst sine wave inputs, while sensors 4--6 received the resulting signals.

A total of 25 unique conditions were examined, each merging a particular damage level with an external load level. Up to four 3-gram weights were placed centrally on the coupon, forming five distinct damage states (including a no-damage baseline). Meanwhile, five tensile loads of 0, 5, 10, 15, and 20\,kN were applied (Instron, Inc.) to represent different environmental/operational constraints. For each actuator-receiver path, 20 waveforms were acquired, except under the 20\,kN load, where only two waveforms were captured to explore the framework’s behavior with limited data.

Across nine sensor paths, 3{,}690 signals were recorded in total. Each time-series extended over 333.33\,$\mu$s at a sampling rate of 24\,MHz, yielding 8{,}000 data samples per waveform. For the 0--15\,kN load cases, eight of the 20 recordings per path were set aside for training, and the remaining 12 for testing. In contrast, under the 20\,kN load, one recording was assigned to training and the other to testing, thereby highlighting the model’s performance in data-scarce scenarios.

\subsection{Test Case II}

\begin{figure}[t!]
    \centering
    \begin{picture}(200,300)
    \put(-120,150){ \includegraphics[width=0.42\columnwidth]{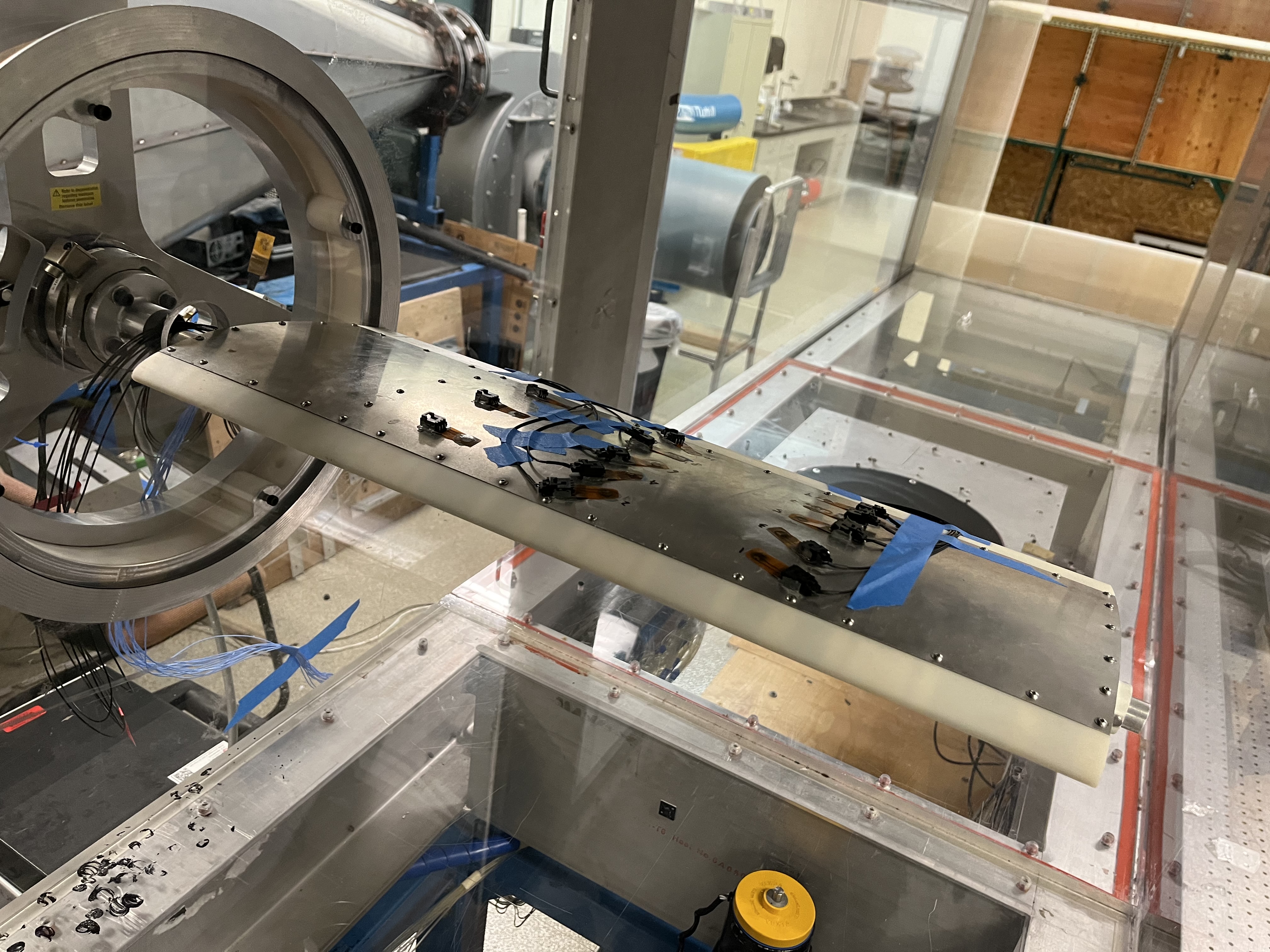}}
    \put(110,150){\includegraphics[width=0.42\columnwidth]{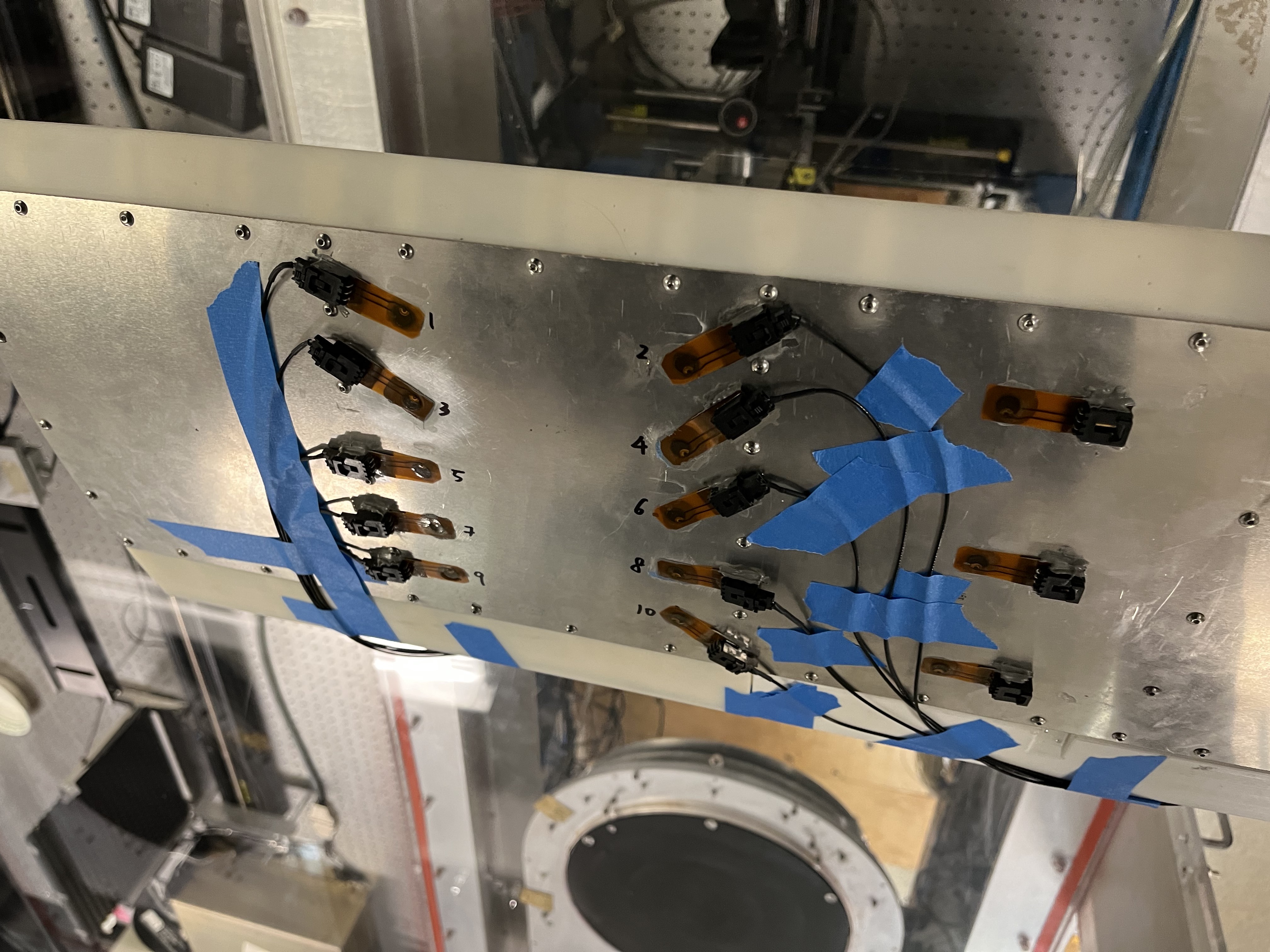}}
    \put(-130,0){ \includegraphics[width=0.47\columnwidth]{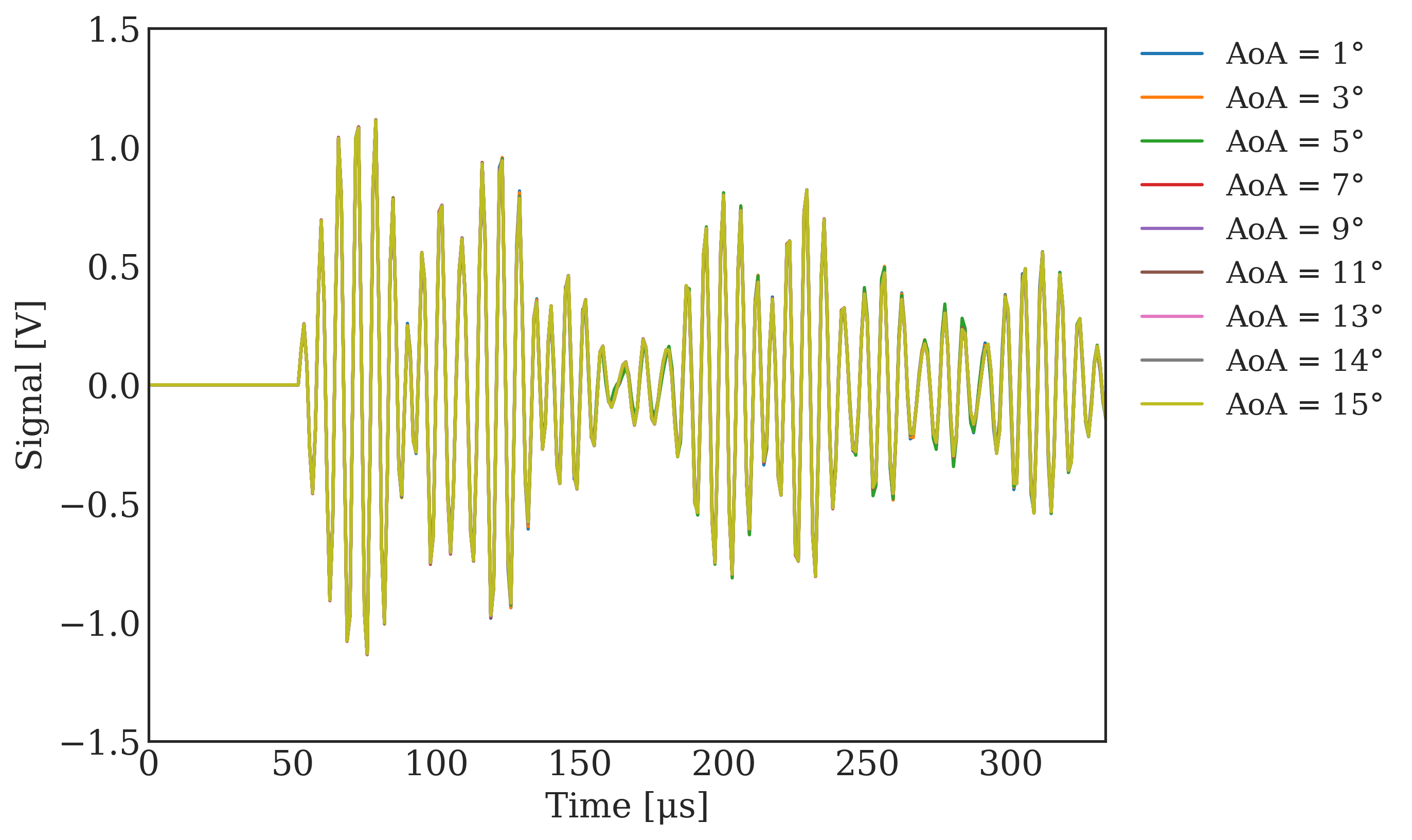}}
    \put(100,0){\includegraphics[width=0.49\columnwidth]{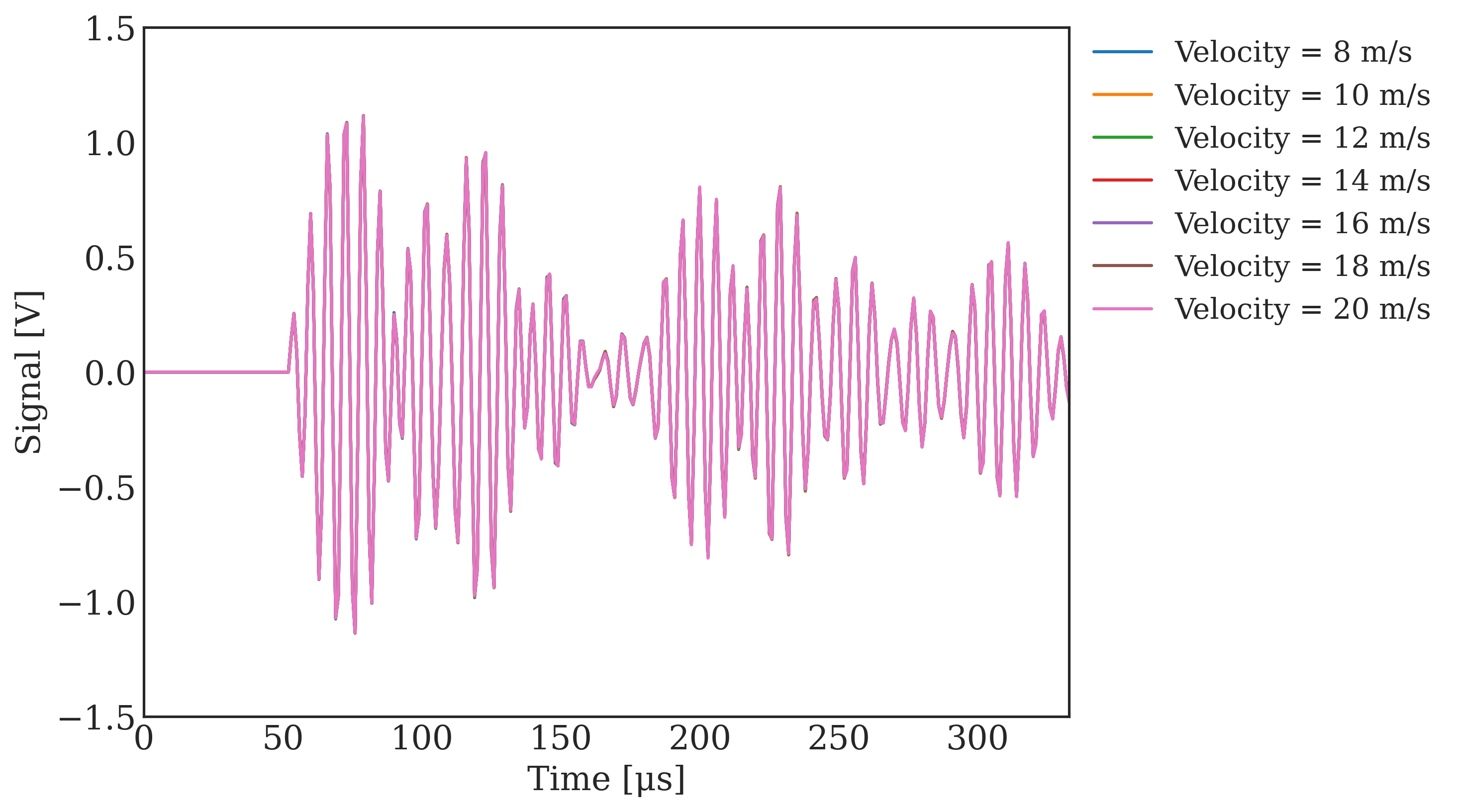}}
    % \put(-140,150){ \includegraphics[width=0.42\columnwidth]{figures/test2/setup/top.jpeg}}
    
    \put(84,156)
    {\large \textbf{(a)}}
    \put(310,156)
    {\large \textbf{(b)}}
    \put(80,16)
    {\large \textbf{(c)}}
    \put(310,16)
    {\large \textbf{(d)}}
    %\put(30,-40){\large \textbf{(c)}}    
    \end{picture}
    % \vspace{-0.5cm}
    
    \caption{Experimental setup and sample signals collected from test case II. (a)-(b): Setup with PZT array layout; (c): signals under different AoAs from sensor path 2-1; (d): signals under different airspeeds from sensor path 2-1.} 
    %; (c): Spectrogram of signal collected at 7°, 16 m/s from path 4-1; (d): Spectrogram of signal collected at 3°, 20 m/s from path 6-7
\label{fig:case2setup} %\vspace{-12pt}
\end{figure}
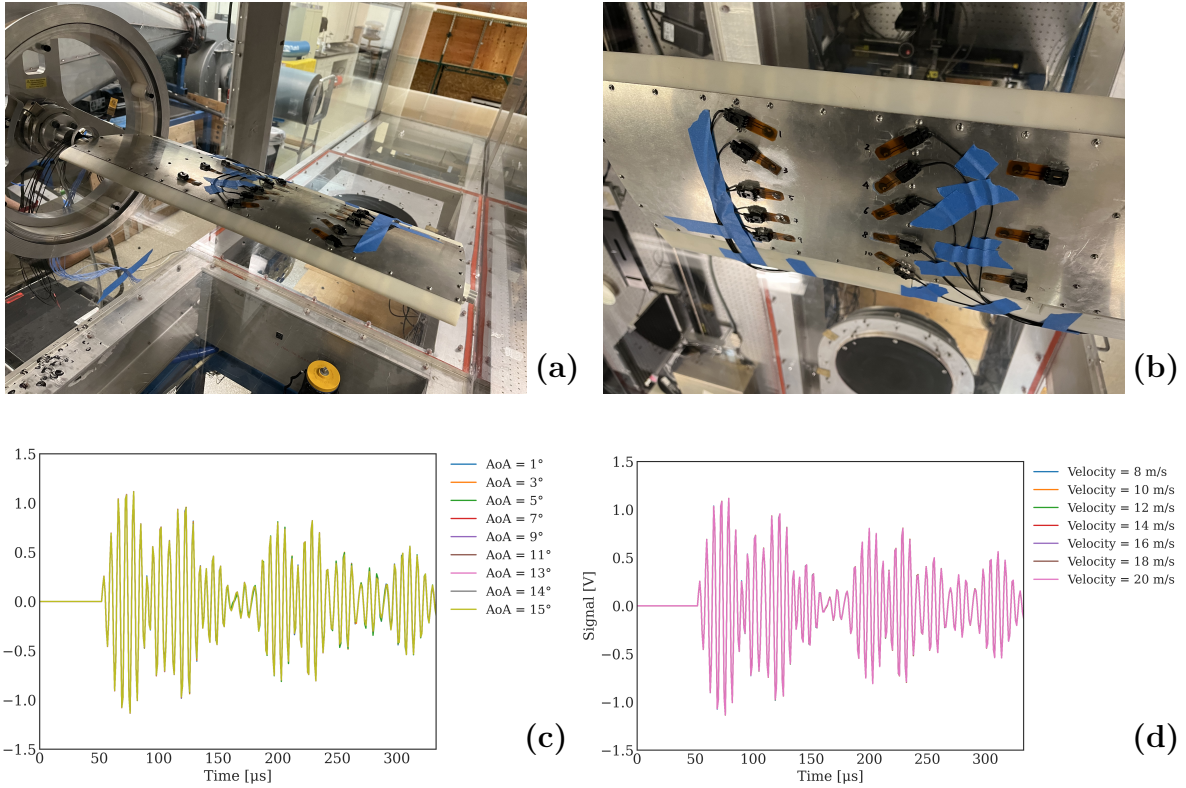 

The second experiment investigated a ``fly-by-feel'' wing, constructed in accordance with the NACA 4412 airfoil and boasting a 600\,mm wingspan along with a 250\,mm chord. Its principal load-bearing framework and outer skin were fashioned from 6061~aluminum, whereas the leading/trailing edges, control surfaces, and internal rib sections were additively manufactured (SLA). Before testing in a wind tunnel, numerical simulations ensured the wing’s structural integrity.

To record its dynamic responses, ten PZT sensors were positioned on the upper surface of the wing. During the wind tunnel trials, wind velocity increased in 2\,m/s increments, from 8\,m/s up to 20\,m/s, while the angle of attack was varied at nine discrete settings (1$^\circ$, 3$^\circ$, 5$^\circ$, 7$^\circ$, 9$^\circ$, 11$^\circ$, 13$^\circ$, 14$^\circ$, and 15$^\circ$). These factors gave rise to 63 distinct test conditions, each of which was evaluated five times. Subsequently, equivalent signals from reciprocal sensor paths effectively doubled the data per state, giving ten trials. Six trials were used for model training, and four were held back for validation. The setup of the wind tunnel experiment and the illustrative signals gathered at different wind speeds and angles of attack are highlighted in Figure \ref{fig:case2setup}.

In advance of final selection, preliminary tests served to identify those sensor paths most sensitive to state variations by examining DI distributions. Of 25 sensor paths initially considered, three were ultimately chosen for a in-depth study.

%-----------------------------------------------------------------------------------------------------
%                                        Results
%-----------------------------------------------------------------------------------------------------
\section{Indicative Results}

This section presents the results from both test cases in sequence. For each case, the latent representations are first examined to provide an intuitive understanding of how these coordinates vary with the underlying states, followed by an evaluation of the model's performance. In Test Case I, it was found that three latent variables offer sufficient accuracy for both damage level and loading condition, and this configuration is adopted throughout this study. For the CAE-based approach, seven latent variables are used, consistent with the findings of previous research \cite{fan2024data}. Unlike Test Case I, where the data were collected under near-static conditions with minimal noise, Test Case II involves wind tunnel experiments in which the wing undergoes vibration, making it more challenging for the model to accurately capture the state information.

It is worth noting that once the latent space with 7 variables is generated, the latent spaces for other cases can be obtained by simply selecting the first few columns of the original latent space. It was observed that three latent variables provide sufficient accuracy with respect to both damage level and loading condition in test case I, and this configuration is adopted in this work. For the CAE approach, seven latent variables are selected, following the findings of previous research \cite{fan2024data}.

\subsection{Test Case I}

\subsubsection{Latent Space Representation}

\begin{figure}[t!]
    \centering
    \begin{picture}(200,280)
    \put(-140,140){ \includegraphics[width=0.4\columnwidth]{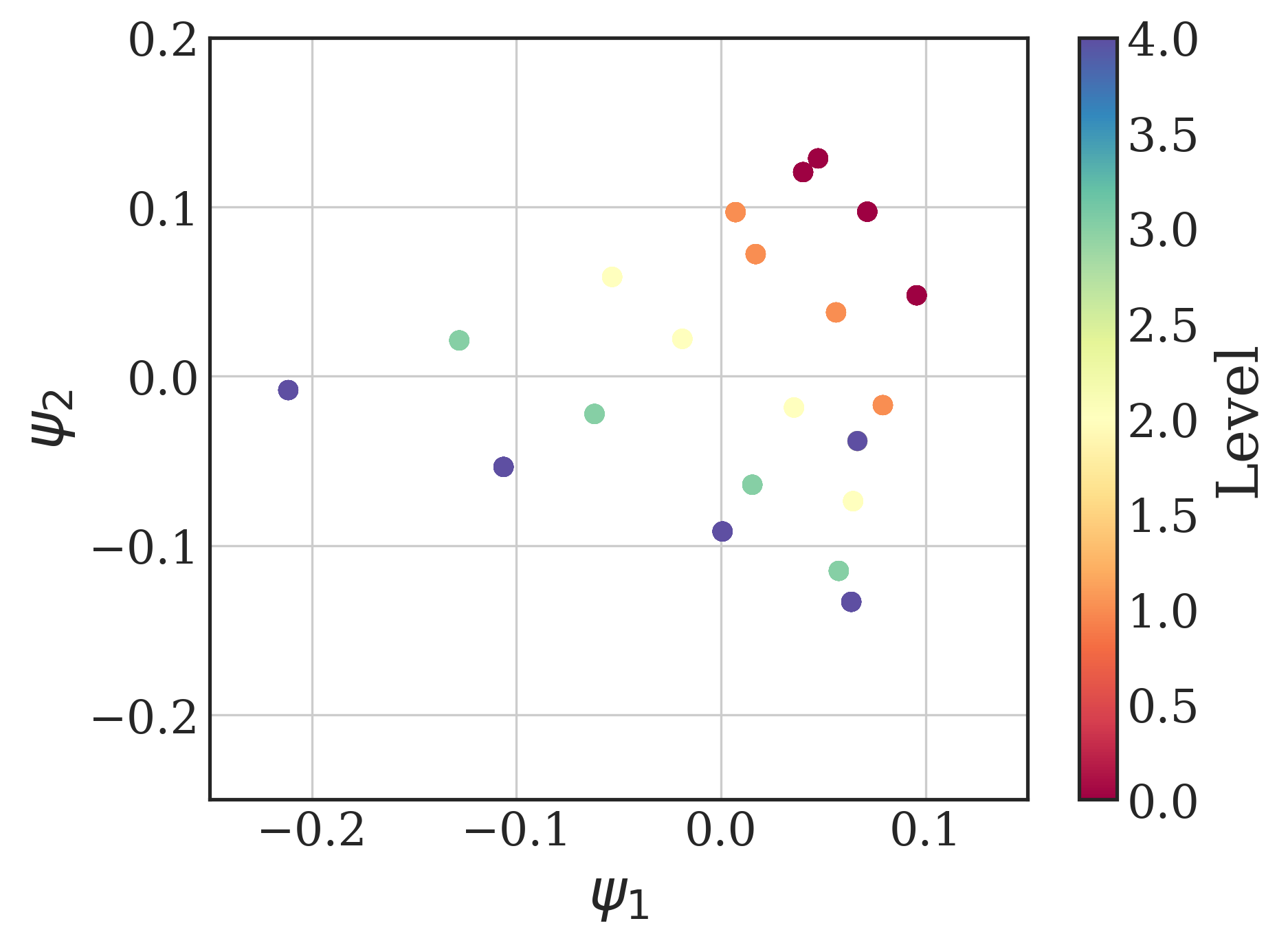}}
    \put(100,140){\includegraphics[width=0.4\columnwidth]{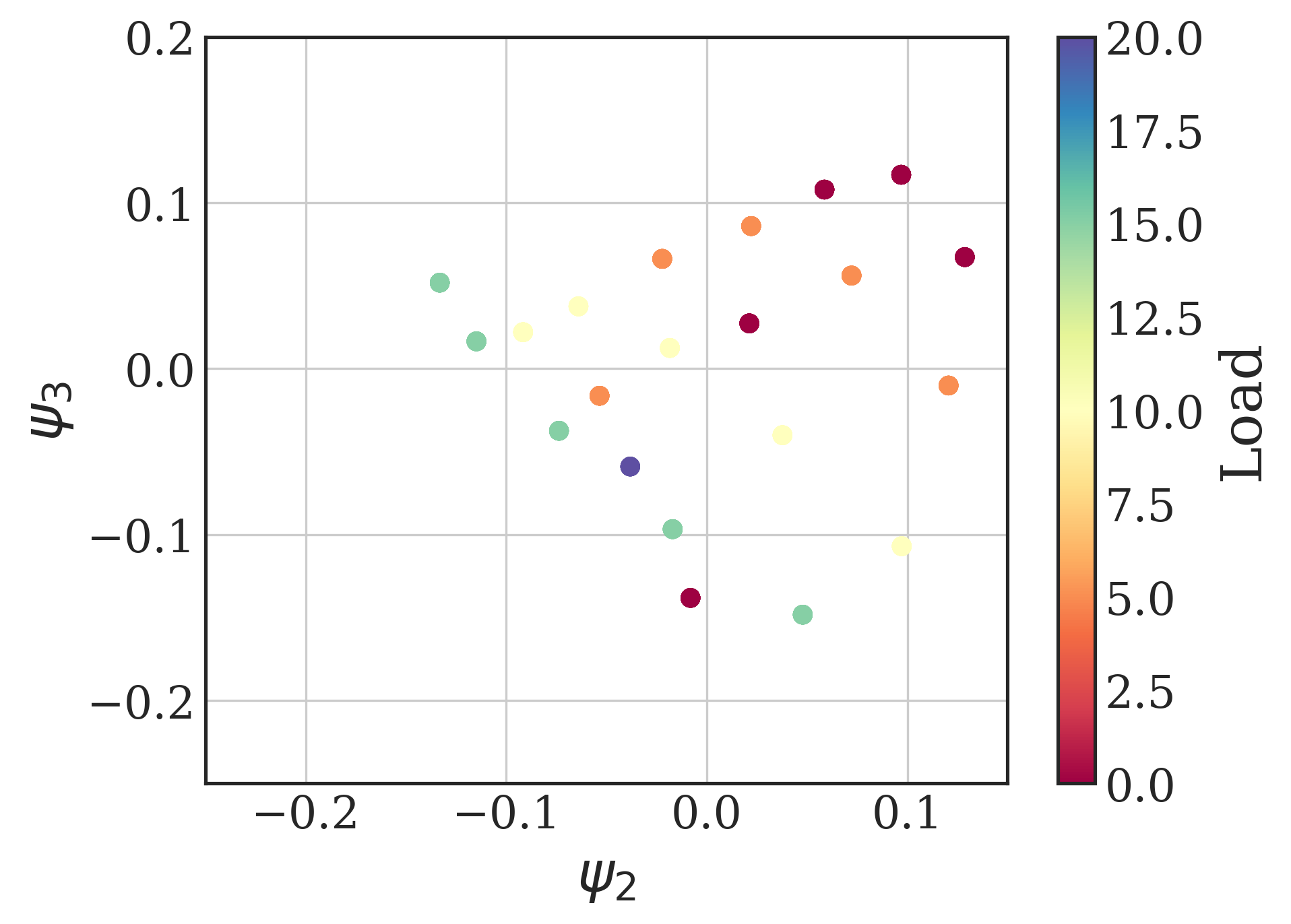}}
    \put(-140,0){ \includegraphics[width=0.4\columnwidth]{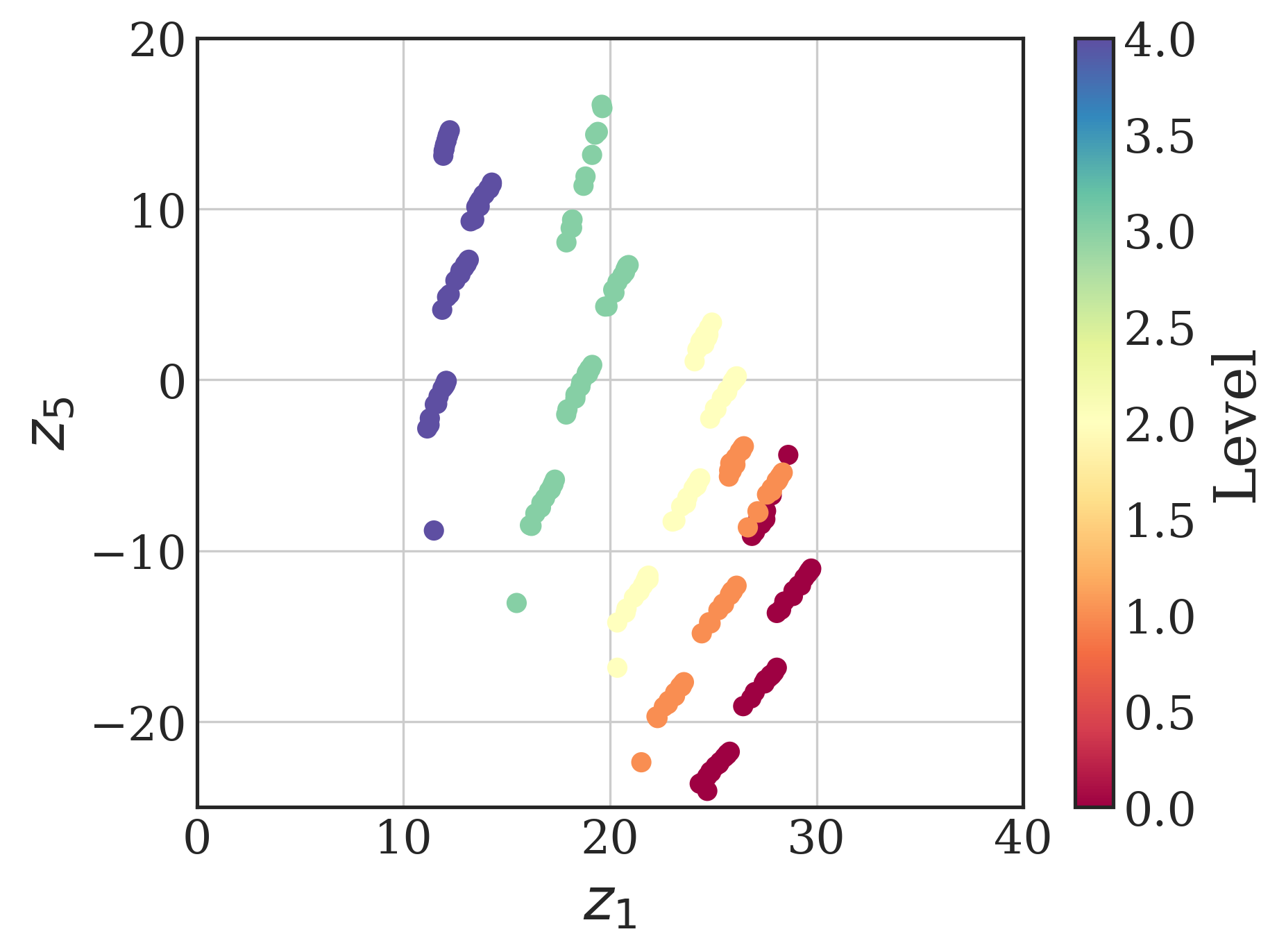}}
    \put(106,0){\includegraphics[width=0.39\columnwidth]{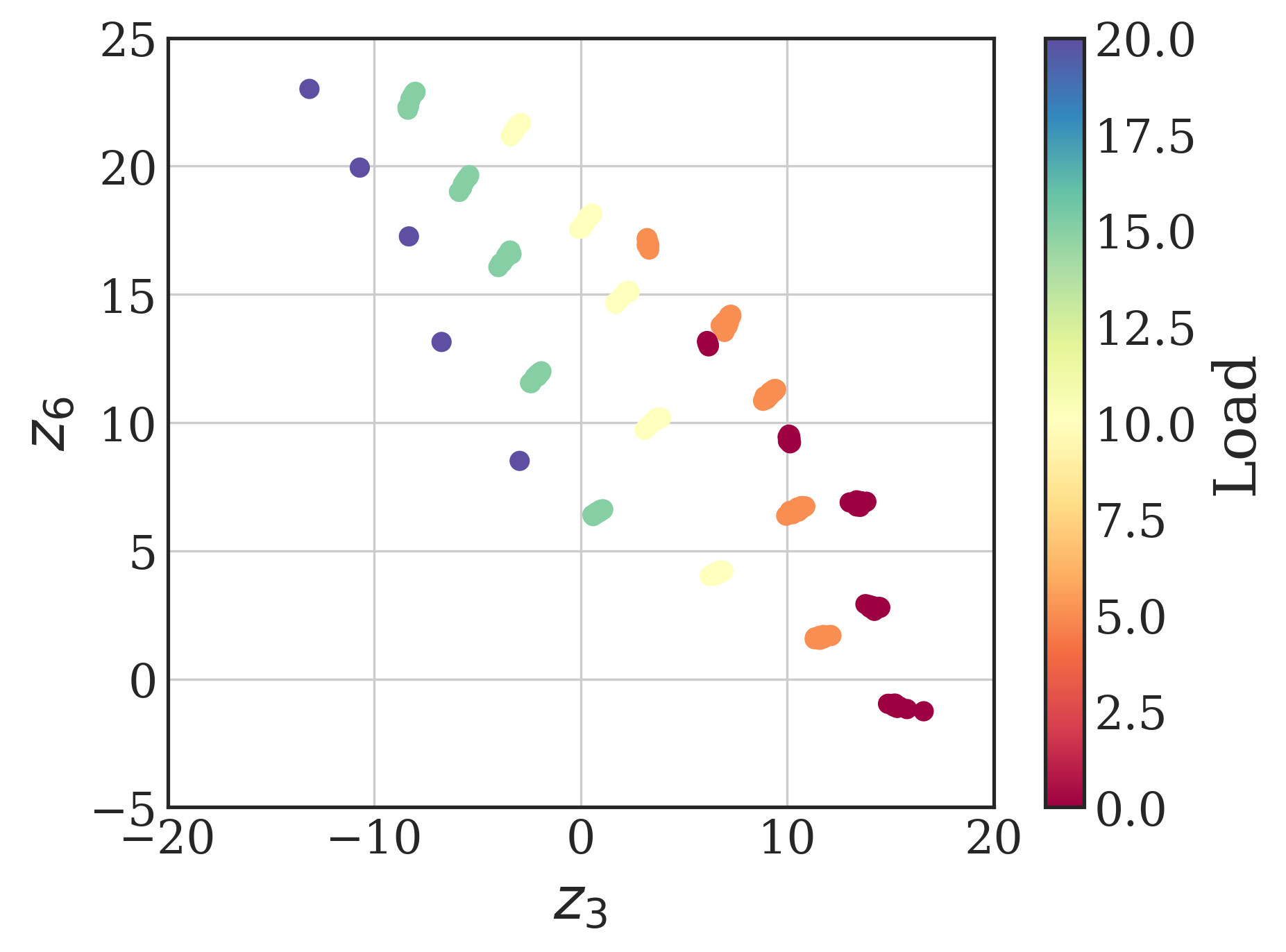}}
    
    %\put(40,-240){\includegraphics[width=0.7\columnwidth]{Figure/Candidacy_signal.png}}
    \put(50,156)
    {\large \textbf{(a)}}
    \put(290,156)
    {\large \textbf{(b)}}
    \put(50,16)
    {\large \textbf{(c)}}
    \put(290,16)
    {\large \textbf{(d)}}
    %\put(30,-40){\large \textbf{(c)}}    
    \end{picture}
    % \vspace{-0.5cm}
    
    \caption{Latent representation examples from test case I. (a)-(b):  Coordinates obtained from DMaps; (c)-(d): Latent variables from encoder.} 
\label{fig:lat1} %\vspace{-12pt}
\end{figure}
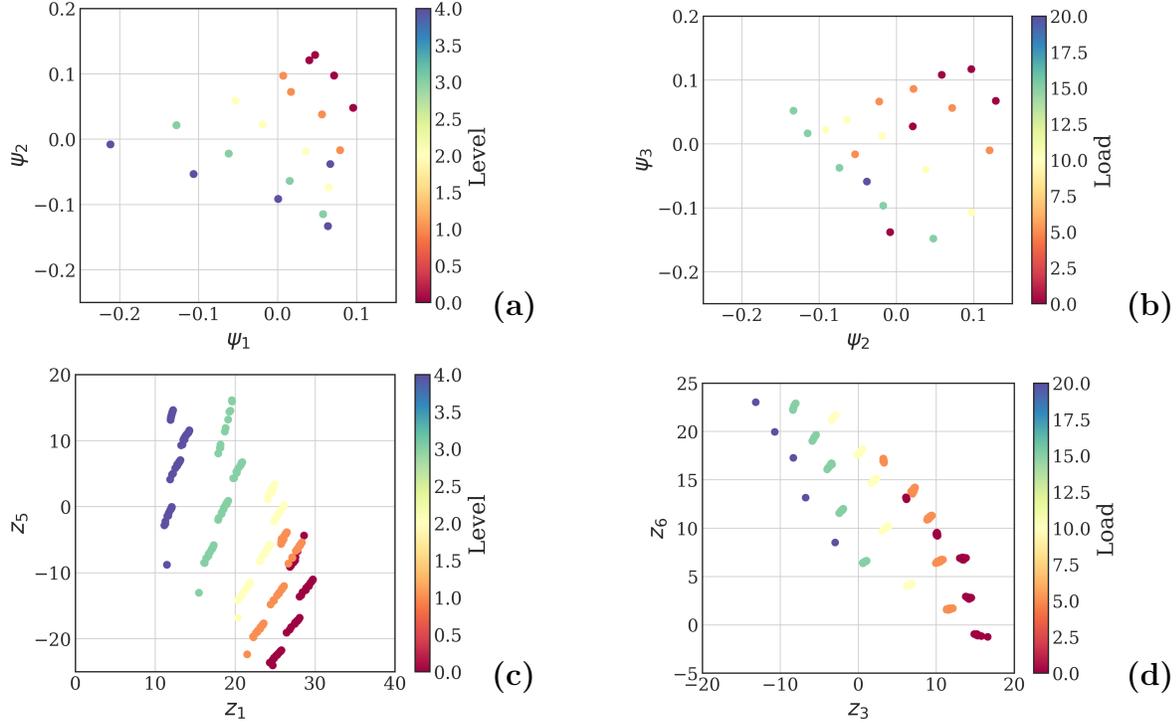 

Examples of latent space representations are shown in Figure \ref{fig:lat1}, where each plot includes all test points corresponding to the previously described states. Panels (a) and (b) display coordinates derived from diffusion maps, with panel (a) representing variations in damage levels and panel (b) reflecting changes under different external loads. In both cases, points associated with the same state tend to cluster within distinct regions. As the severity of damage increases, the clusters in panel (a) transition along a clear trajectory, changing from red to blue. A similar pattern is evident in panel (b) as the load increases. Panels (c) and (d) reveal even more distinct trends, suggesting that the CAE method may provide superior capability in differentiating data across states. The separation of latent coordinates and variables underscores their potential utility for state classification.

\subsubsection{State Prediction}

\begin{figure}[t!]
    \centering
    \begin{picture}(200,120)
    \put(-125,7){ \includegraphics[width=0.238\columnwidth]{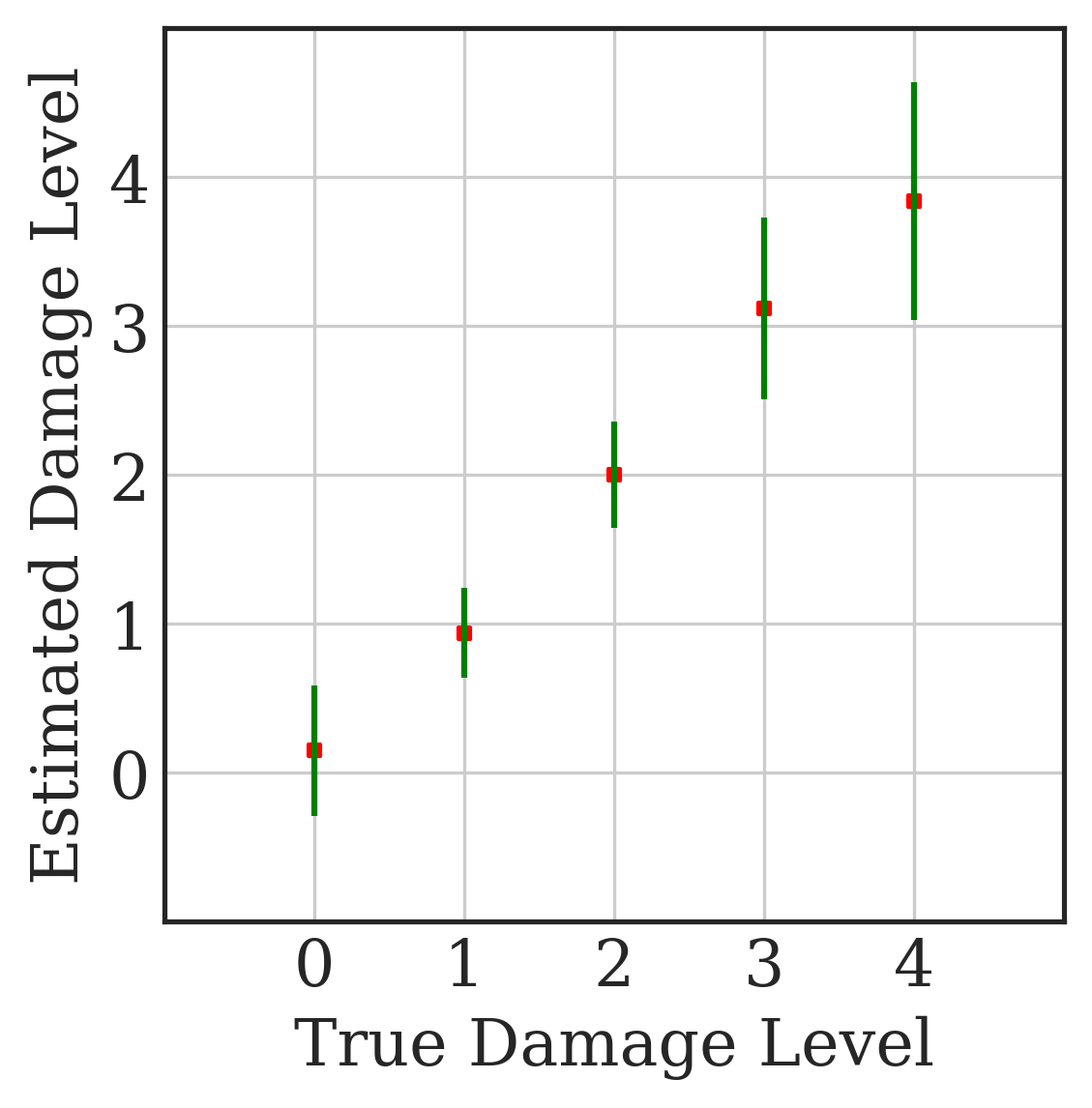}}
    \put(-10,10){\includegraphics[width=0.24\columnwidth]{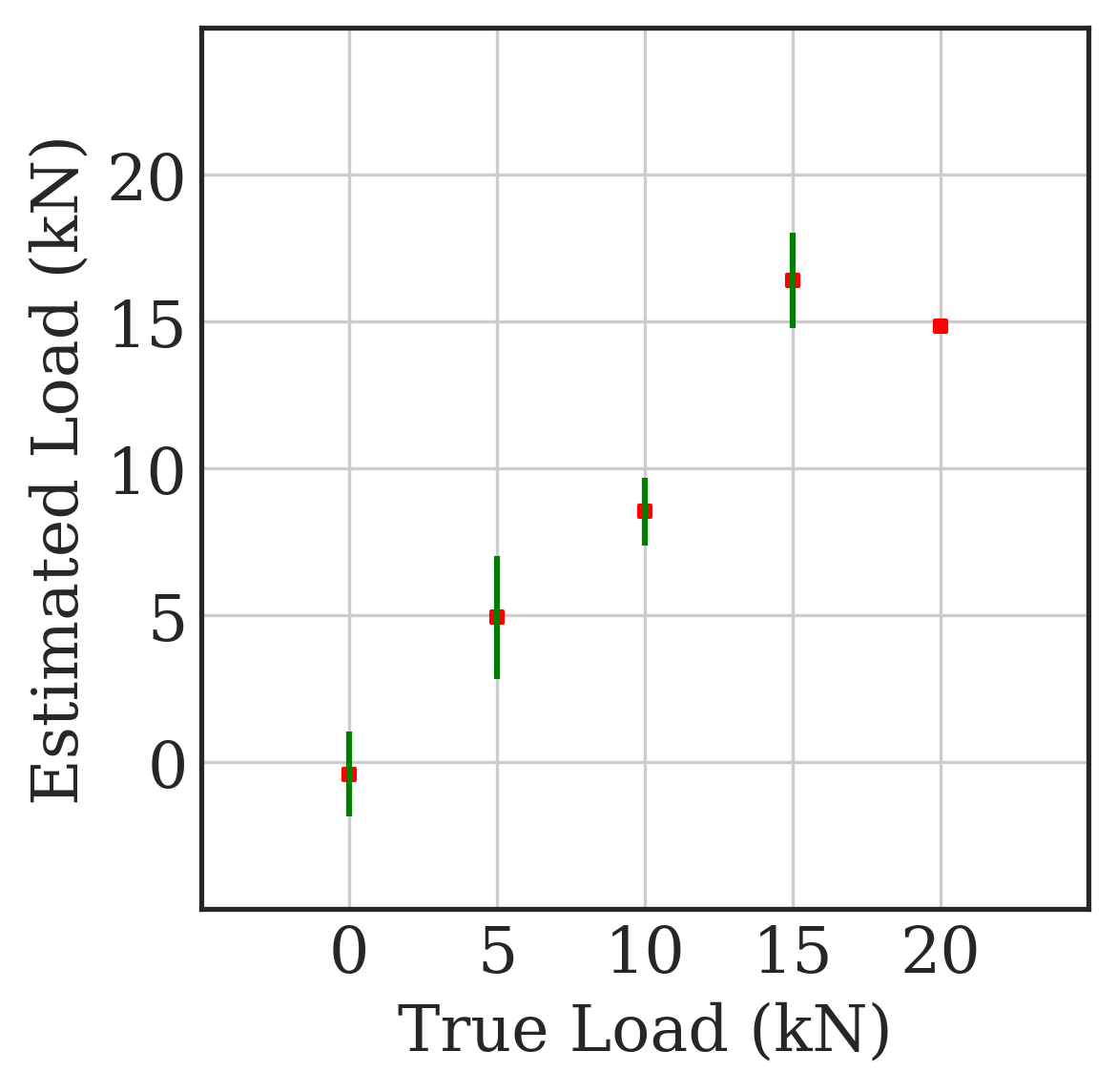}}
    \put(105,10){ \includegraphics[width=0.23\columnwidth]{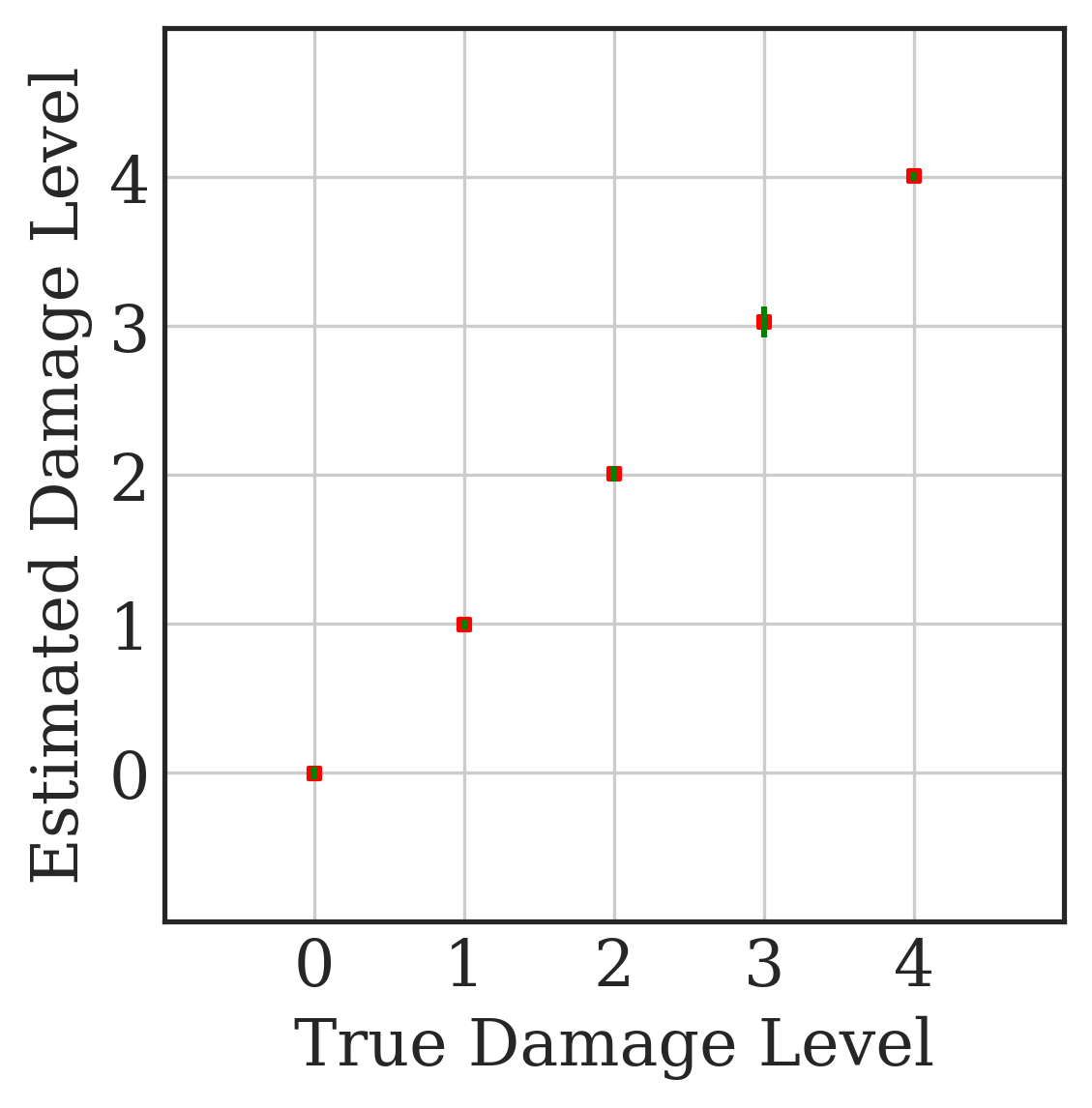}}
    \put(220,10){\includegraphics[width=0.24\columnwidth]{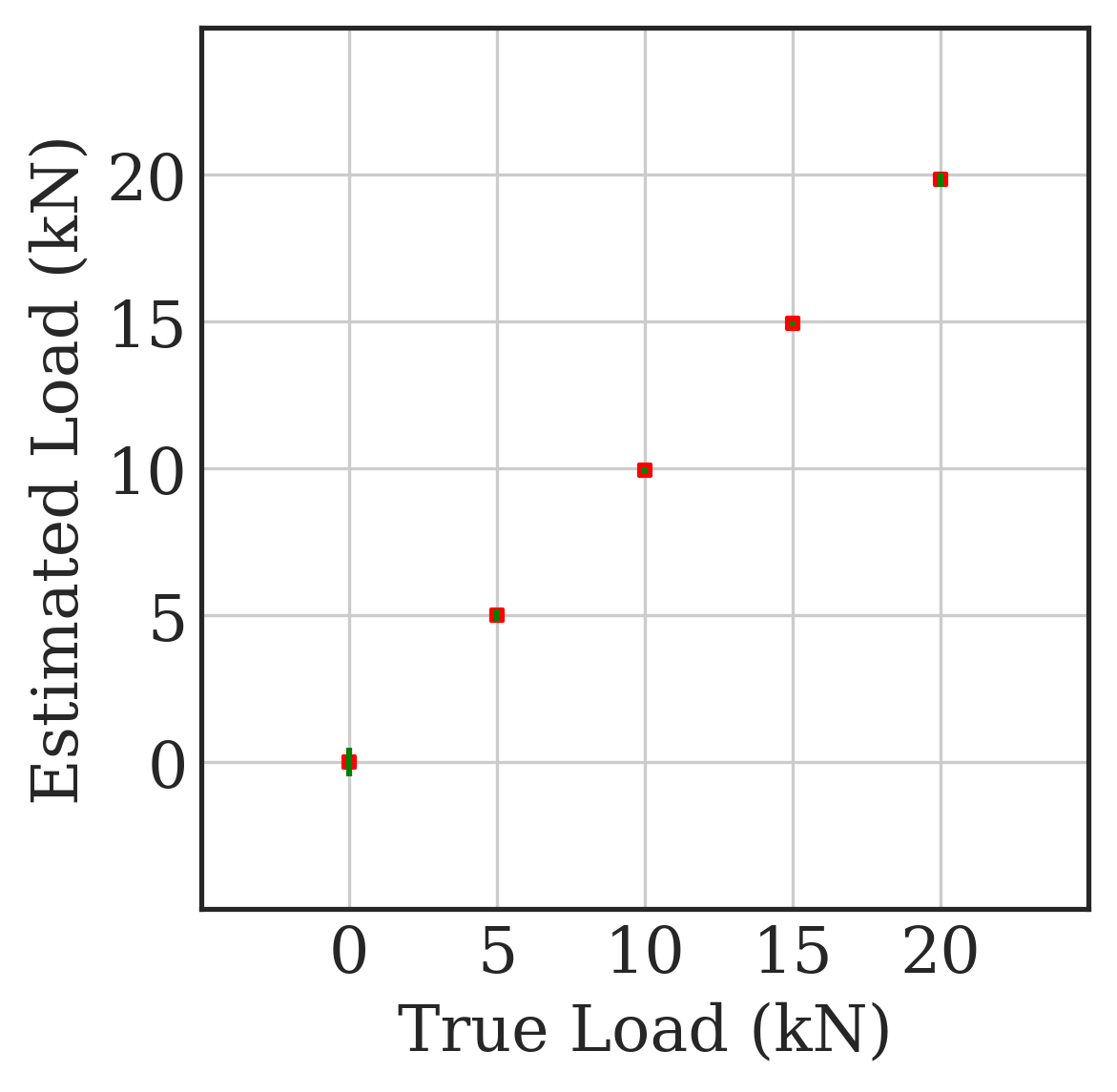}}
    
    %\put(40,-240){\includegraphics[width=0.7\columnwidth]{Figure/Candidacy_signal.png}}
    % \put(5,430){\large \textbf{(a)}}
    % \put(5,290){\large \textbf{(b)}}
    % \put(5,150){ \large \textbf{(c)}}
    % \put(5,10){\large \textbf{(d)}}  
    % \put(-110,194){\large \textbf{(a)}}
    % \put(105,194){\large \textbf{(b)}}
    \put(-120,15){ \large \textbf{(a)}}
    \put(-5,15){\large \textbf{(b)}}  
    \put(110,15){\large \textbf{(c)}}  
    \put(225,15){\large \textbf{(d)}}  
    \end{picture}
    % \vspace{-0.5cm}
    
    \caption{Error bar plots of state prediction results from test case I. (a)-(b): results using DMaps and FFNN; (c)-(d): results using encoder and FFNN.} 
\label{fig:state1} %\vspace{-12pt}
\end{figure}
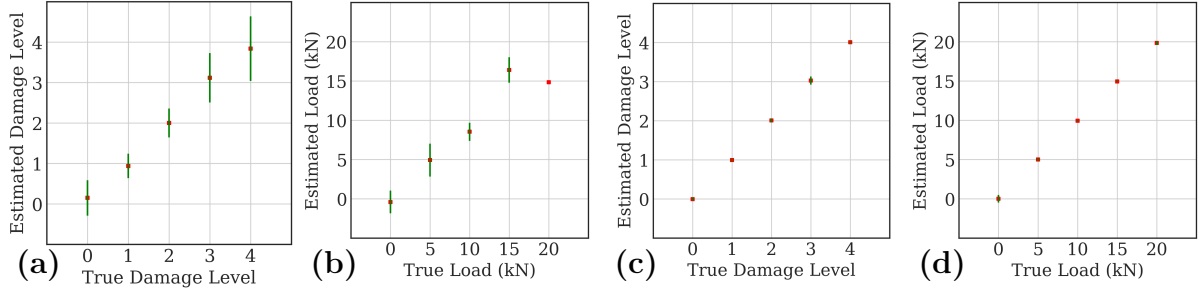

Building on the insights gained from the latent variable and coordinate plots, the state classifier is constructed using these variables instead of raw signals to simplify the model. Figure \ref{fig:state1} presents error bar plots of the state prediction results, where the red dots represent the mean predictions across realizations, and the green bars indicate the 95$\%$ confidence intervals (CIs).

In panels (a)-(b), the predicted damage levels closely match the ground truth values. However, the predicted loads show some deviation, particularly in the range of 10–20 kN. At 10 kN, the true value lies outside the 95$\%$ CI, and the largest error occurs at 20 kN, which can be attributed to the limited data available for this state. Panel (c)-(d) show that the use of CAE leads to more accurate state predictions with reduced uncertainties. While both methods yield reliable state estimates, CAE demonstrates superior accuracy in this test case.

\subsubsection{Signal Reconstruction}

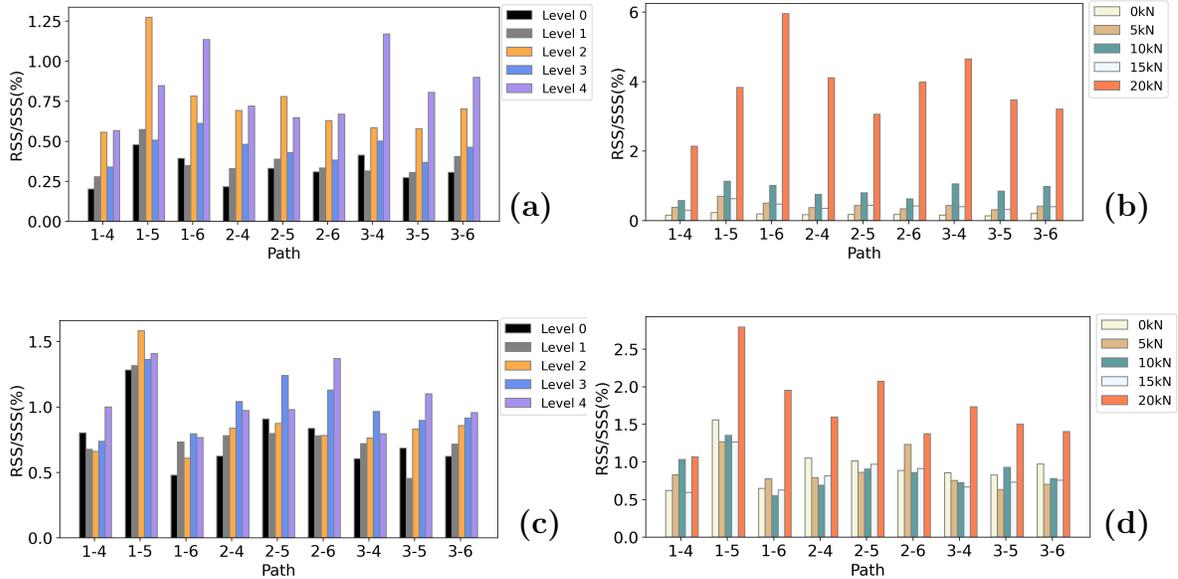
\begin{figure}[t!]
    \centering
    \begin{picture}(200,250)
    \put(-115,130){ \includegraphics[width=0.48\columnwidth]{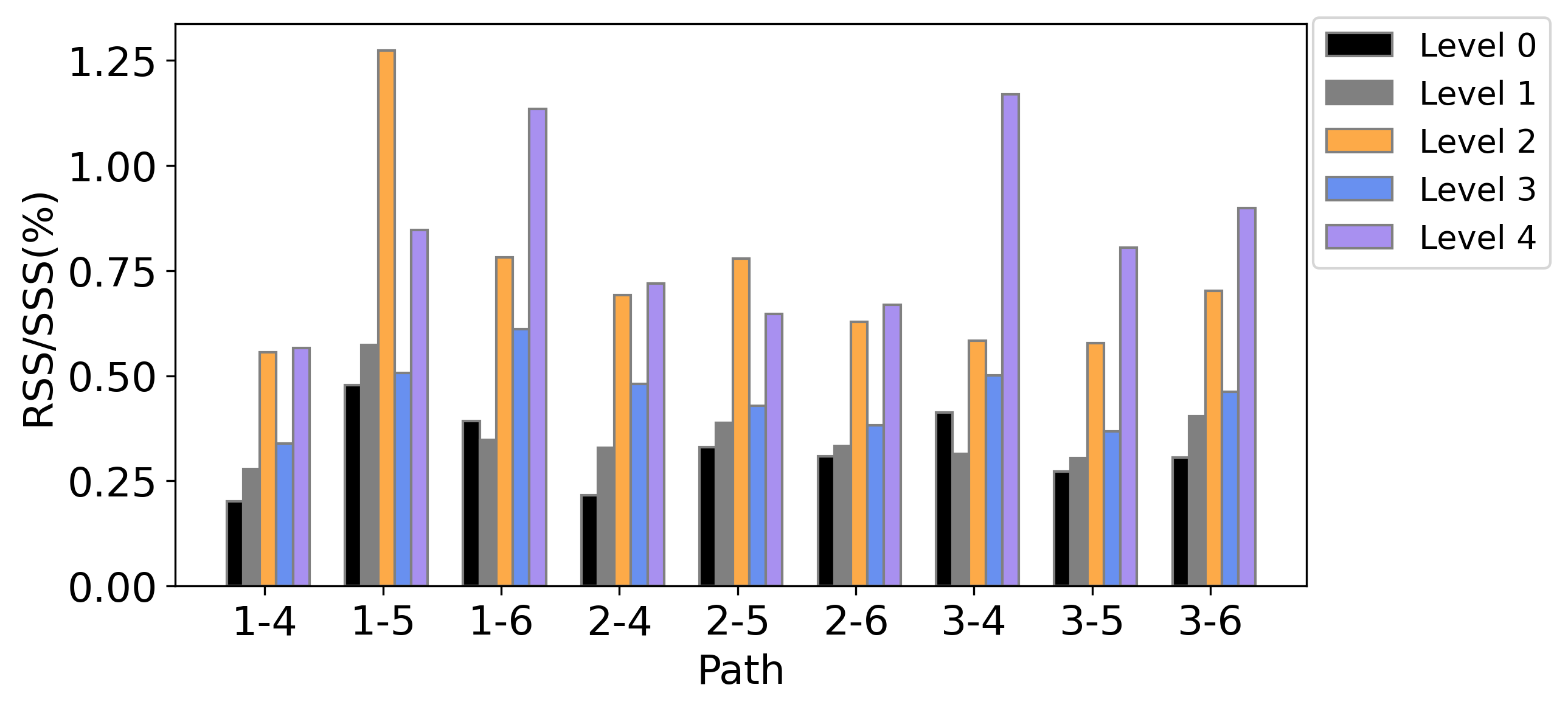}}
    \put(116,130){\includegraphics[width=0.46\columnwidth]{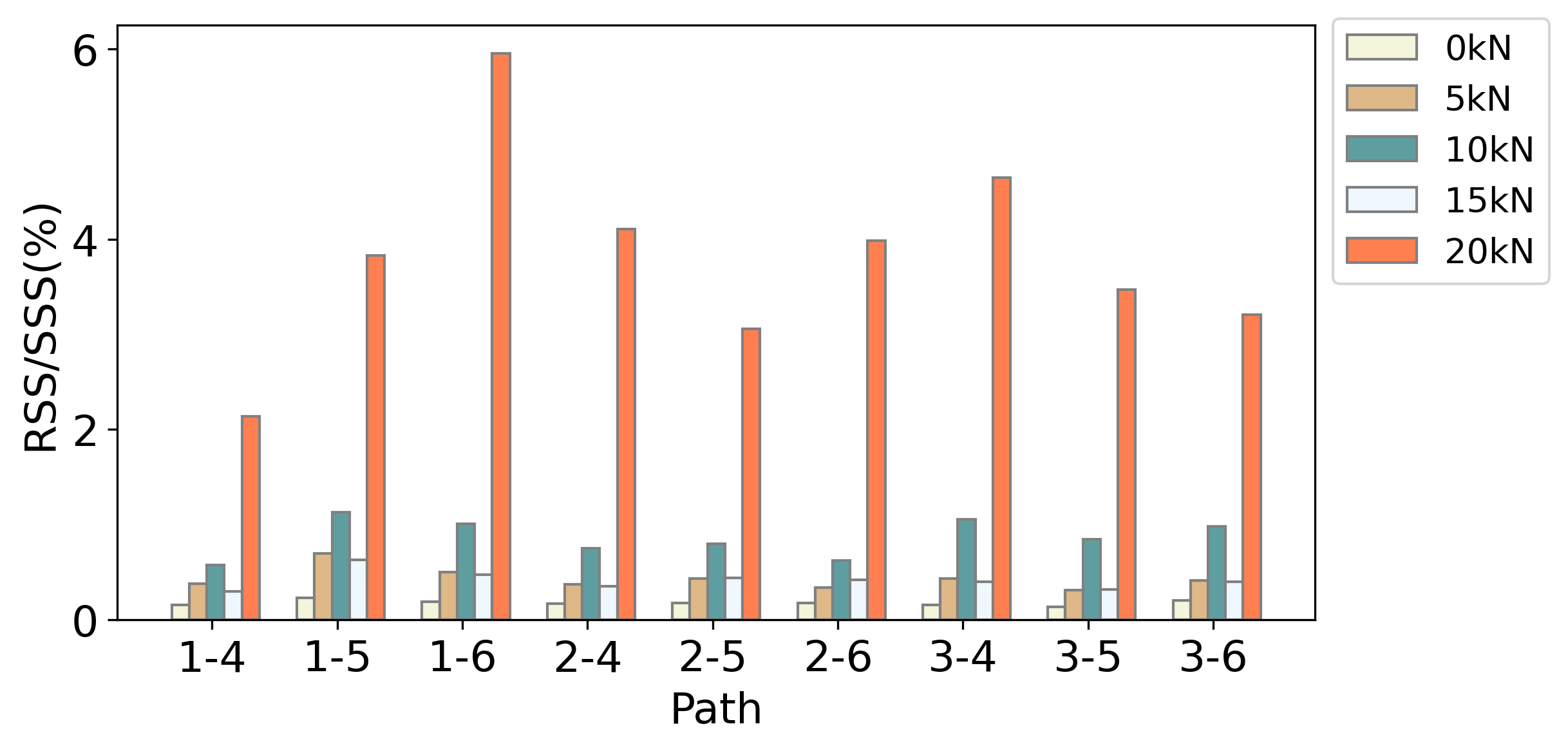}}
    \put(-115,10){ \includegraphics[width=0.48\columnwidth]{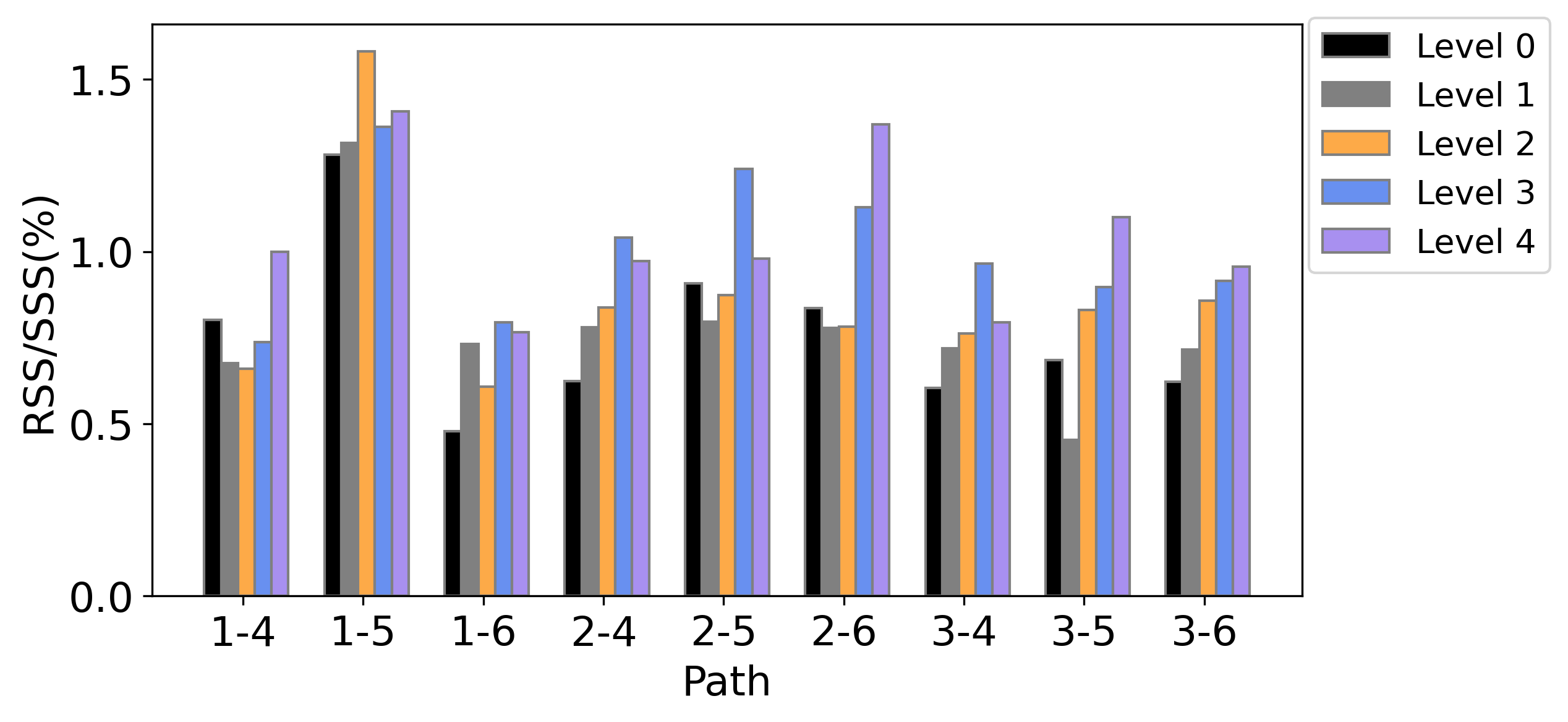}}
    \put(110,10){\includegraphics[width=0.48\columnwidth]{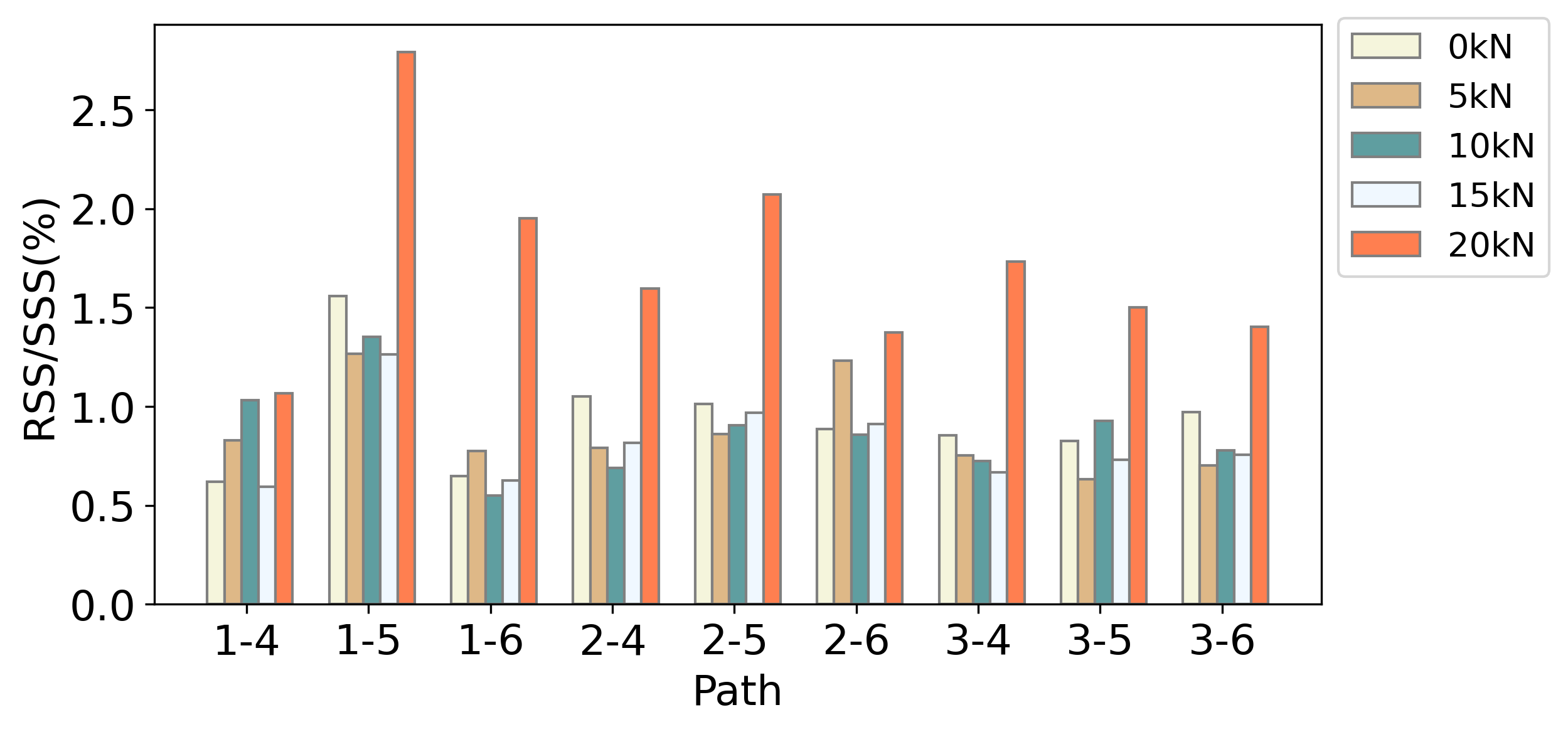}}

    \put(80,150){\large \textbf{(a)}}
    \put(305,150){\large \textbf{(b)}}
    \put(80,30){ \large \textbf{(c)}}
    \put(305,30){\large \textbf{(d)}}  
    \end{picture}
    % \vspace{-0.5cm}
    
    \caption{Signal reconstruction results from test case I. (a)-(b): results using PCE and Laplacian pyramids; (c)-(d): results using decoder.} 
\label{fig:recon1} %\vspace{-12pt}
\end{figure}

Figure \ref{fig:recon1} illustrates the overall signal reconstruction error for each state and each path. At first glance, the reconstruction errors are minimal, demonstrating the strong performance of both models. Panels (a) and (b) show results from diffusion maps, where the RSS/SSS values across all paths remain below 1.25$\%$ for each damage level. However, the error increases significantly at 20 kN, reaching up to 6$\%$, likely due to the limited data available for this state.

Panels (c) and (d) display results from CAE. The maximum reconstruction error for damage levels exceeds 1.5$\%$, while for loads, it surpasses 2.5$\%$. A comparison between the two methods reveals that, although CAE achieves better state classification due to its well-defined latent space, diffusion maps slightly outperform CAE in signal reconstruction accuracy.

\subsection{Test Case II}

In this case, the data set includes two different state factors including nine angles of attack and seven wind speeds. Unlike the static conditions in the first experiment, the data in this case were collected in a wind tunnel, where continuous vibration induced by airflow introduced significantly more noise and brought more challenges in identifying the state.

\subsubsection{Latent Space Representation}

\begin{figure}[t!]
    \centering
    \begin{picture}(200,240)
    \put(-140,120){ \includegraphics[width=0.34\columnwidth]{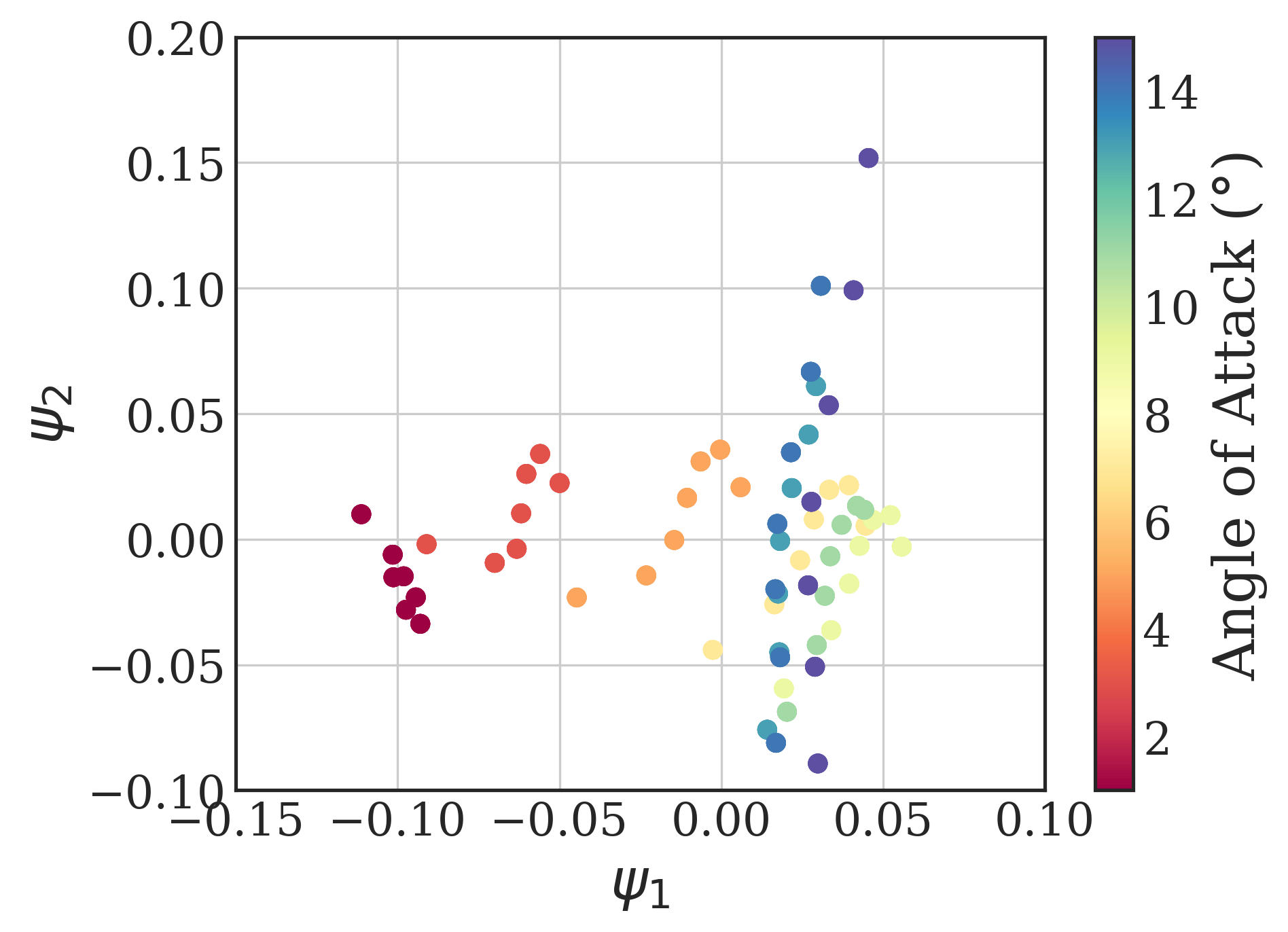}}
    \put(23,120){\includegraphics[width=0.32\columnwidth]{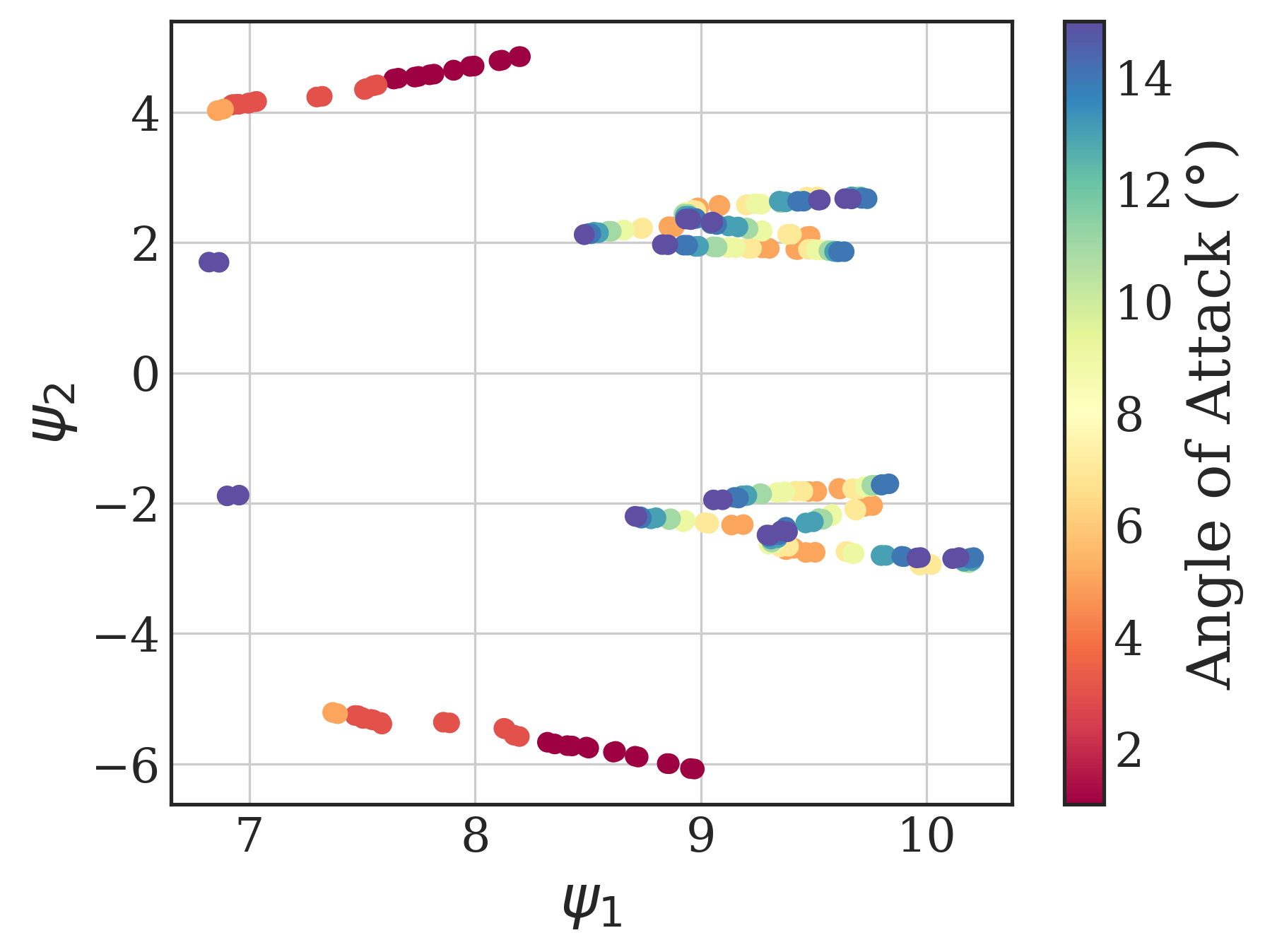}}
    \put(172,120){\includegraphics[width=0.33\columnwidth]{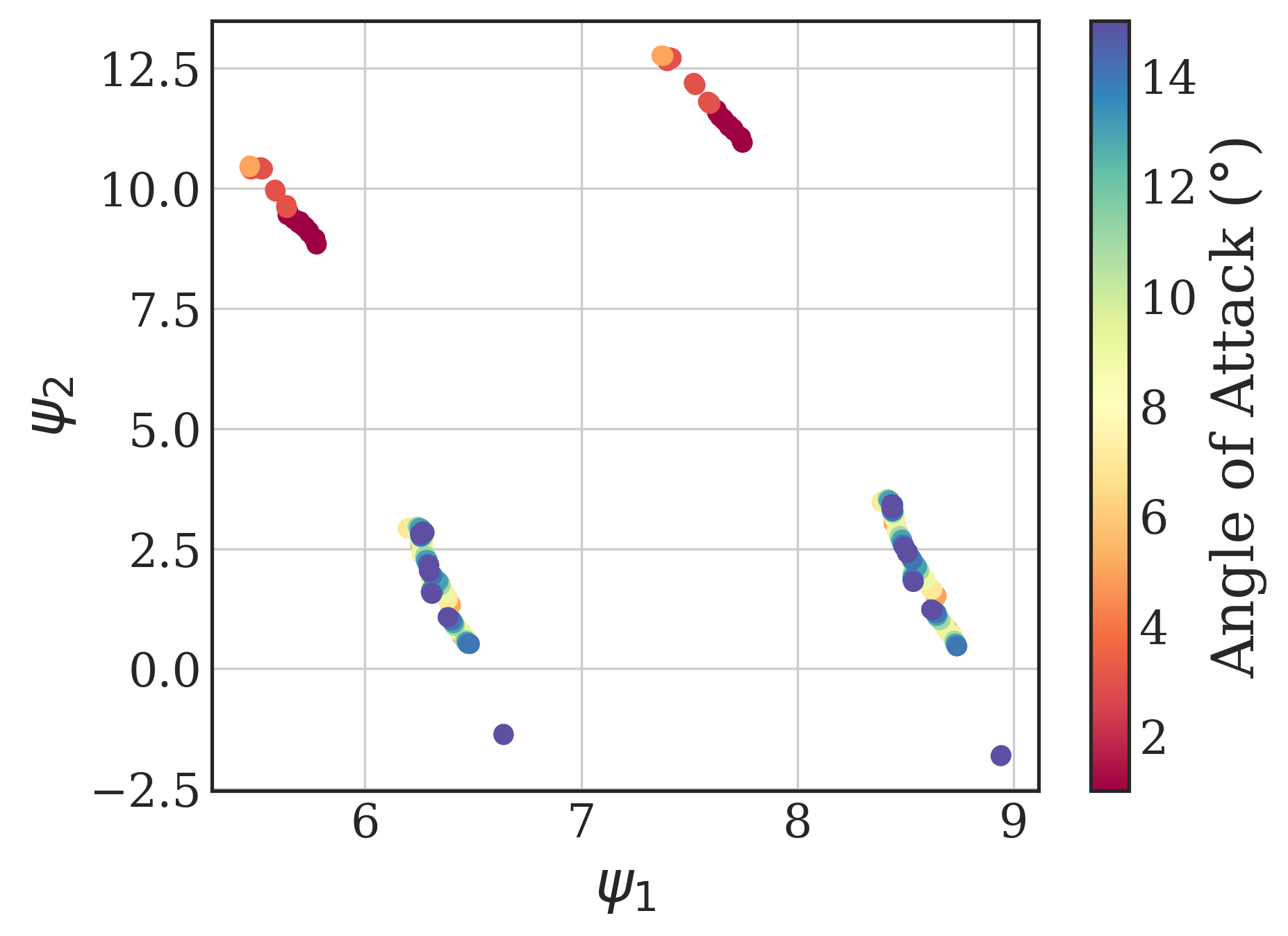}}
    \put(-140,0){ \includegraphics[width=0.34\columnwidth]{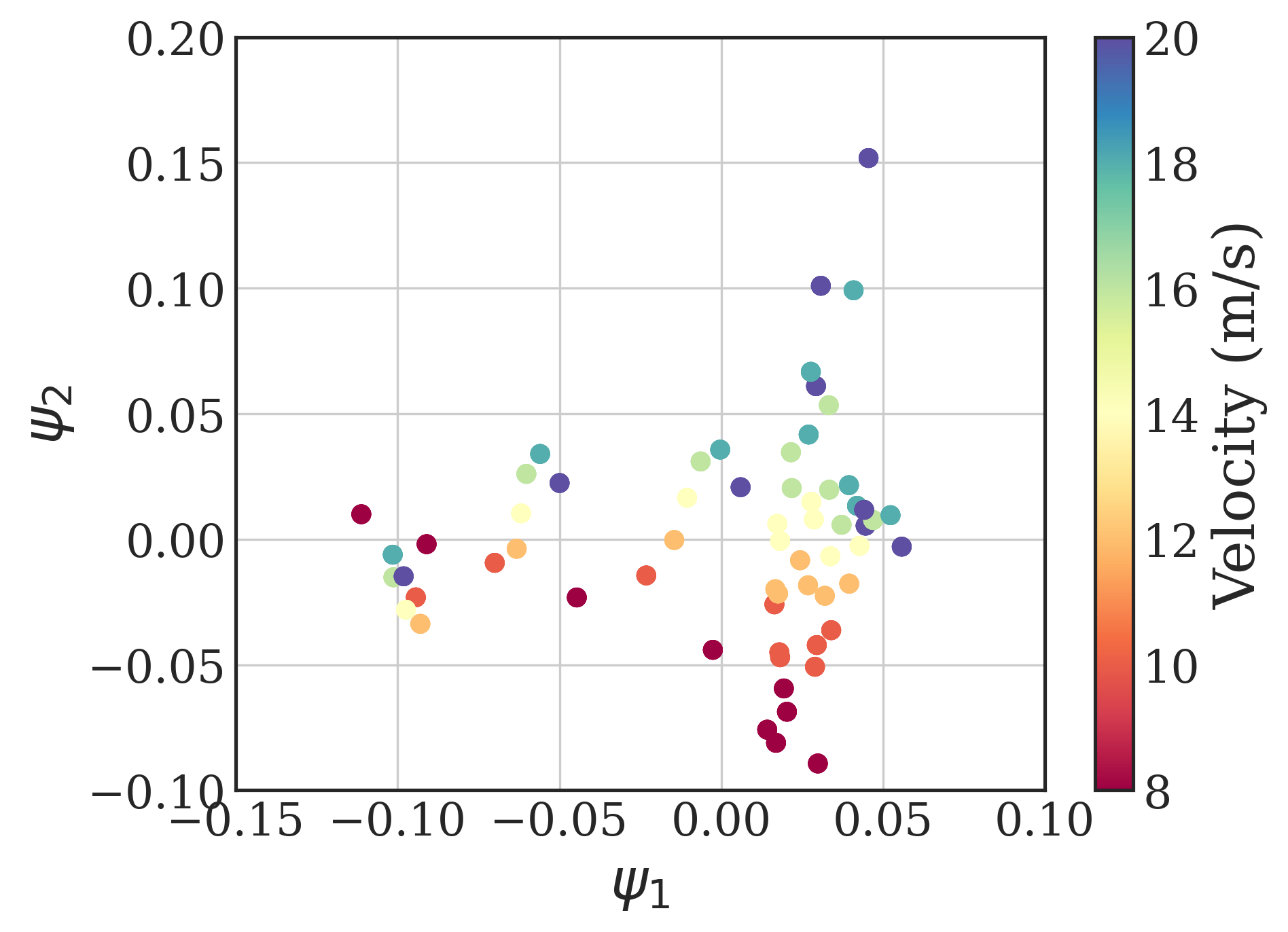}}
    \put(23,0){ \includegraphics[width=0.32\columnwidth]{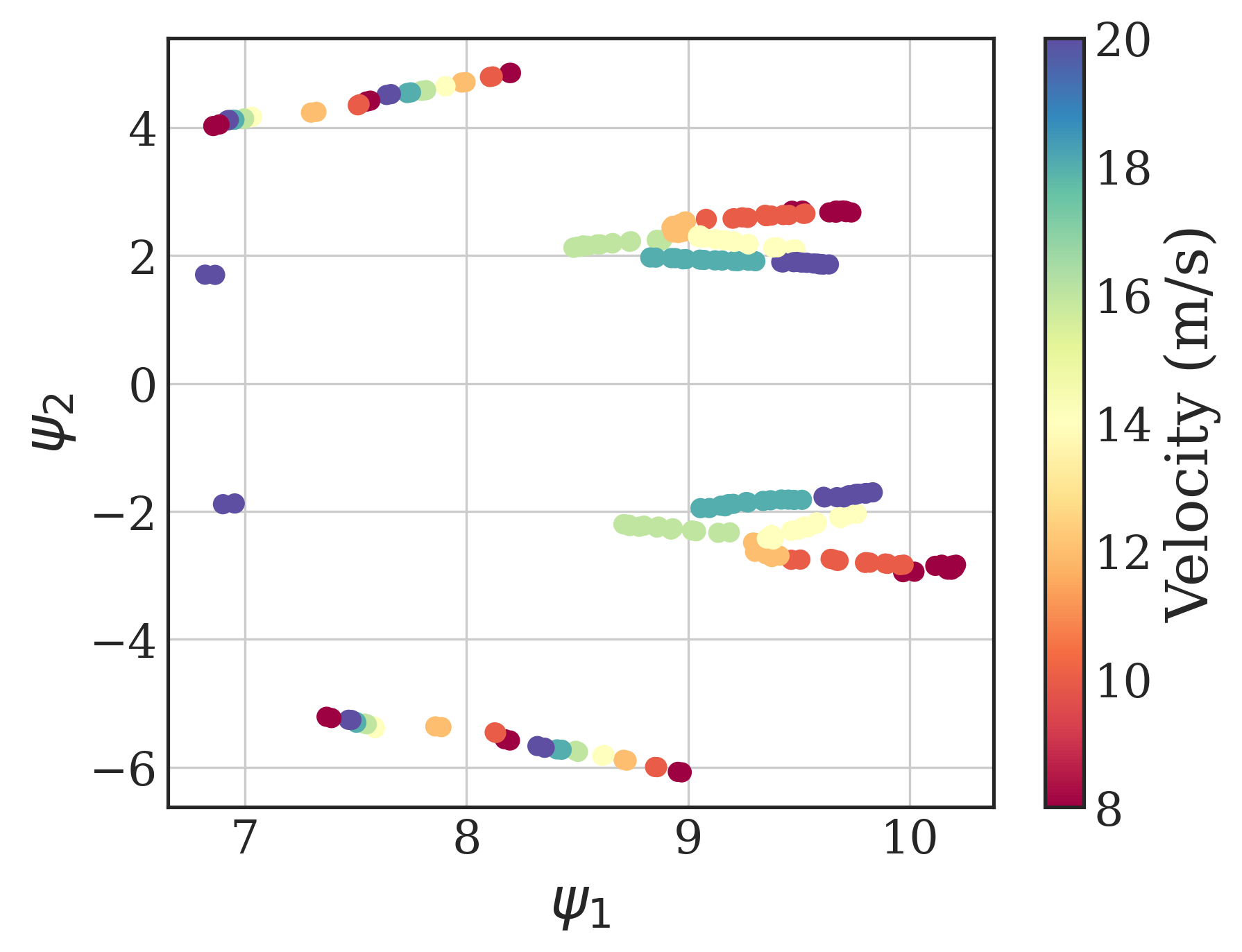}}
    \put(176,0){\includegraphics[width=0.33\columnwidth]{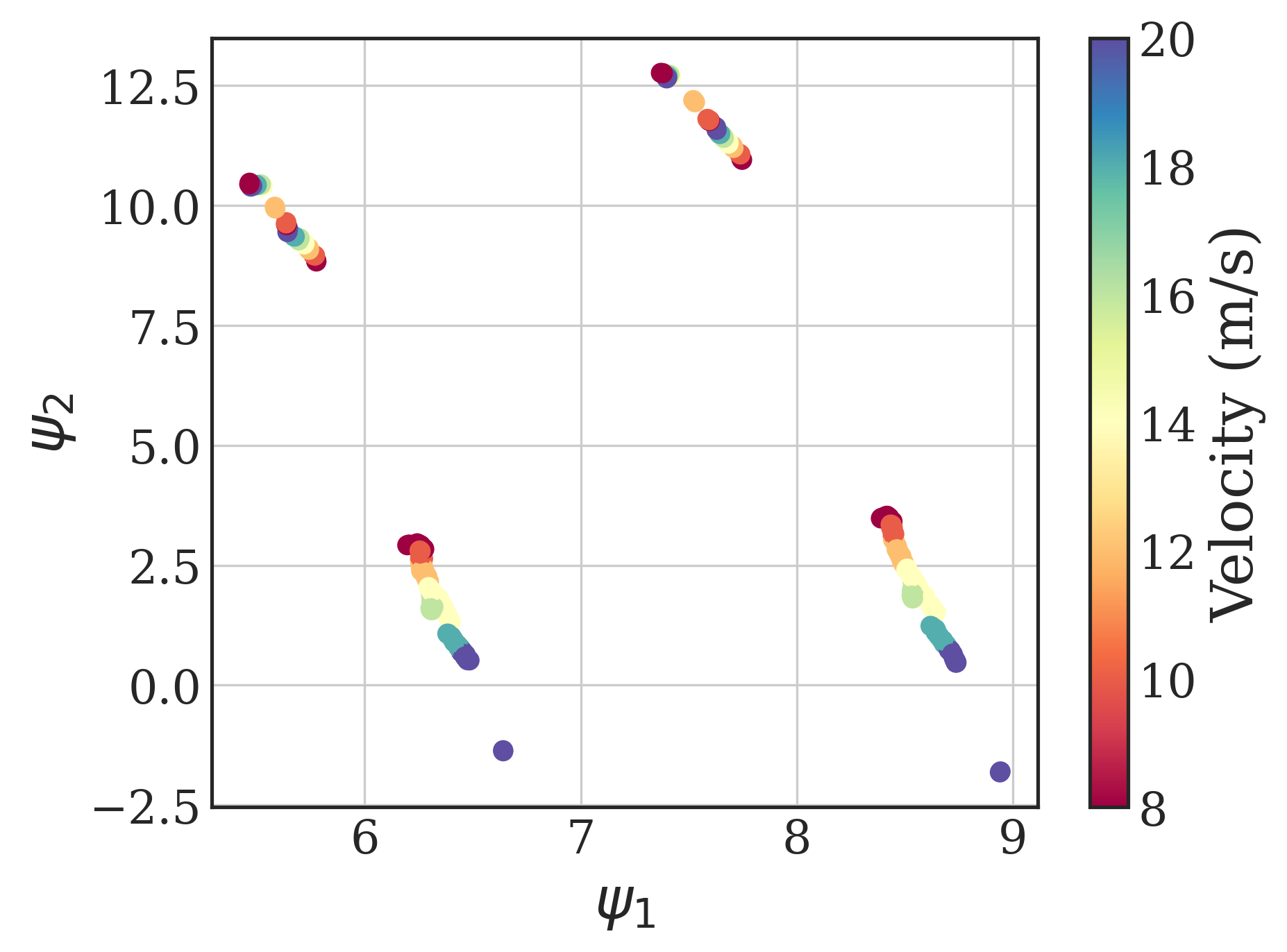}}
    
    %\put(40,-240){\includegraphics[width=0.7\columnwidth]{Figure/Candidacy_signal.png}}
    \put(10,130)
    {\large \textbf{(a)}}
    \put(12,10)
    {\large \textbf{(b)}}
    \put(160,130)
    {\large \textbf{(c)}}
    \put(162,10)
    {\large \textbf{(d)}}
    \put(314,130)
    {\large \textbf{(e)}}
    \put(316,10)
    {\large \textbf{(f)}}
    %\put(30,-40){\large \textbf{(c)}}    
    \end{picture}
    % \vspace{-0.5cm}
    
    \caption{Latent representation examples of various models from test case II for all testing data. (a)-(b): Diffusion map; (c)-(d): CAE; (e)-(f): VAE. It can be observed that the latent space representations from various states may overlap using diffusion map while the trend using CAE and VAE is more apparent.} 
\label{fig:case2lat1} %\vspace{-12pt}
\end{figure}
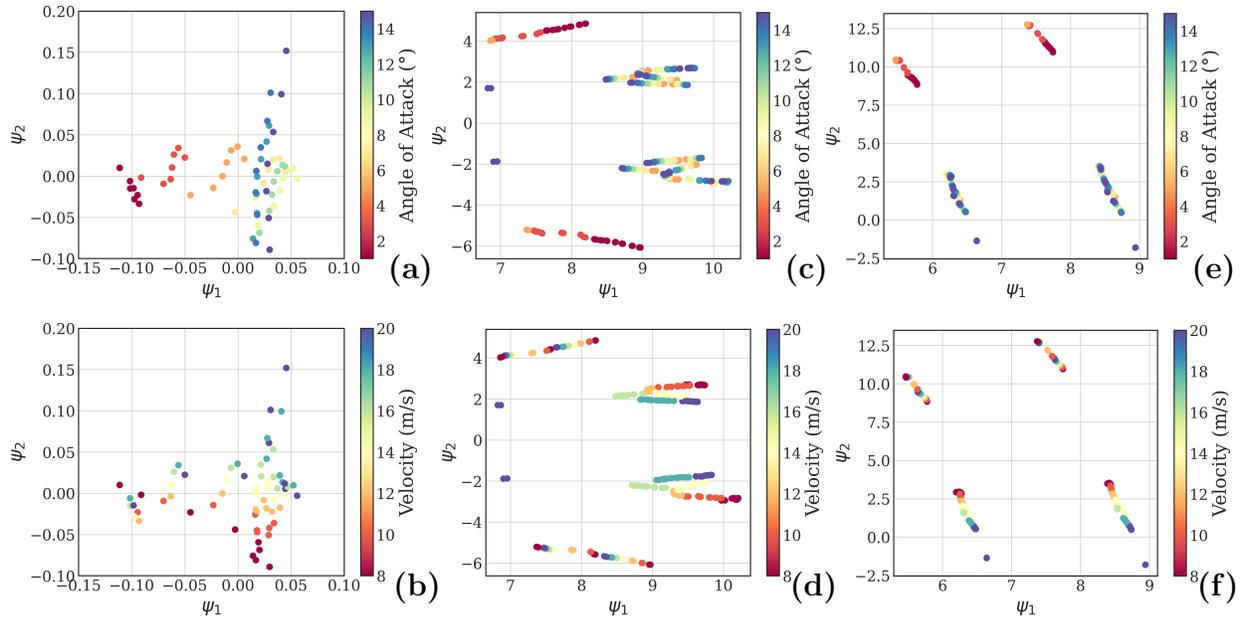 

%%%%%%%% new version starts%%%%%%%%%%%%%%
Three distinct data compression techniques (DMaps, CAE, and VAE) were implemented in the proposed framework, each generating a latent space representation from a given signal. 
Figure~\ref{fig:case2lat1} presents sample representations of the testing data from the three models, where each panel plots a pair of latent variables ($z_2$ vs.\ $z_1$) for various states indicated by different colors. All three methods in this example employ a two-dimensional latent space.

Panels~(a)--(b) show the diffusion map outputs for the full range of AoA and airspeeds, respectively. As either state factor changes, the colored markers form distinct clusters and exhibit a noticeable trajectory, indicating that the latent space is sensitive to the underlying states. This property allows building class estimation tools directly from the low-dimensional representations, thereby avoiding reliance on the original high-dimensional data. 
However, when certain states overlap---for instance, AoA = 12\textdegree{} and 14\textdegree{}---the resulting confusion may degrade prediction accuracy. Adding more latent dimensions could potentially help separate these overlapping states.

Panels~(c)--(d) depict results from the CAE, which display a more pronounced trajectory and better separation among different states than the DMap results. Finally, panels~(e)--(f) illustrate the VAE outputs. Although these compressed representations show a clear trend with respect to airspeed changes, they appear less sensitive to variations in AoA compared to the CAE shown in panel~(c).

\subsubsection{State Estimation}

\begin{figure}[t!]
    \centering
    \begin{picture}(200,330)
    % \put(-125,10){ \includegraphics[width=0.24\columnwidth]{Figures/case2/dmap/state/err_bar_aoa_.png}}
    % \put(-10,10){\includegraphics[width=0.24\columnwidth]{Figures/case2/dmap/state/err_bar_v_.png}}
    \put(-60,165){ \includegraphics[width=0.74\columnwidth]{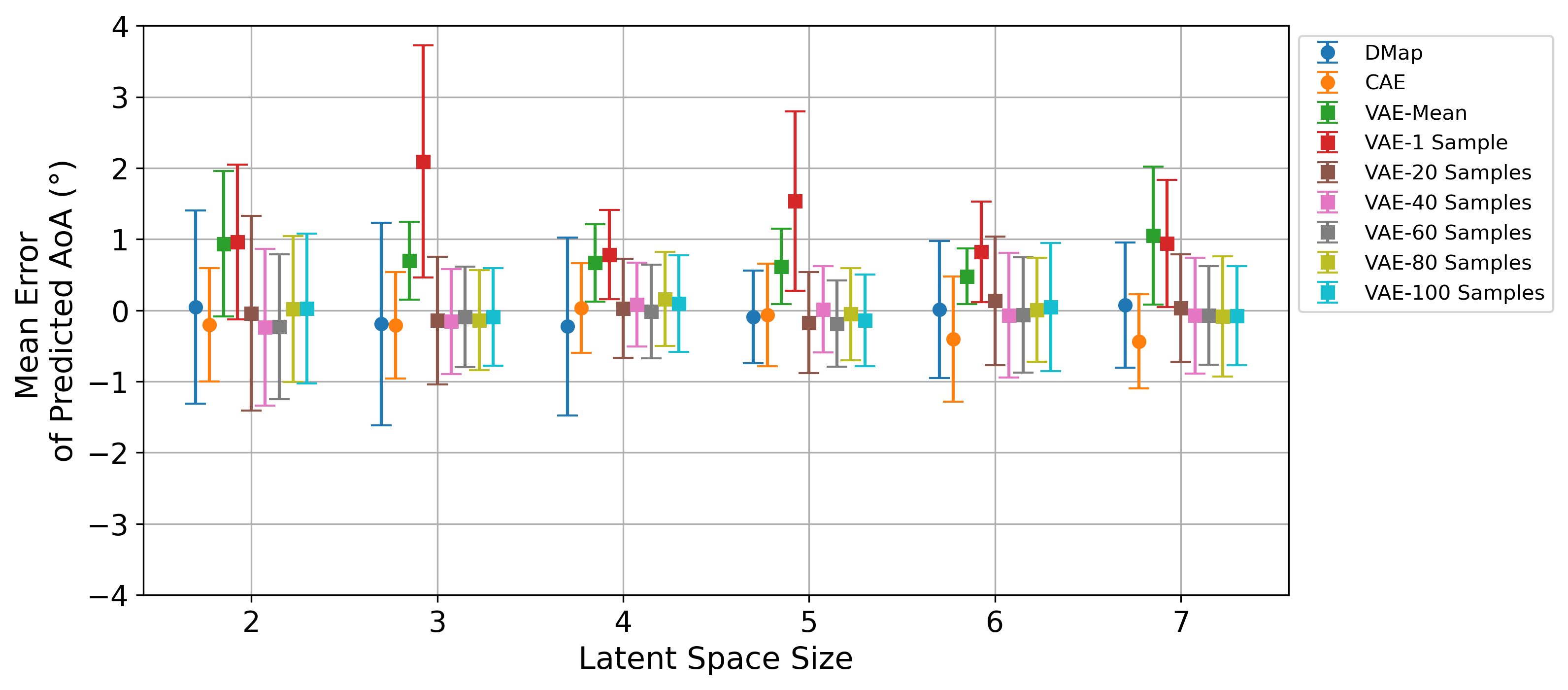}}
    \put(-60,10){\includegraphics[width=0.74\columnwidth]{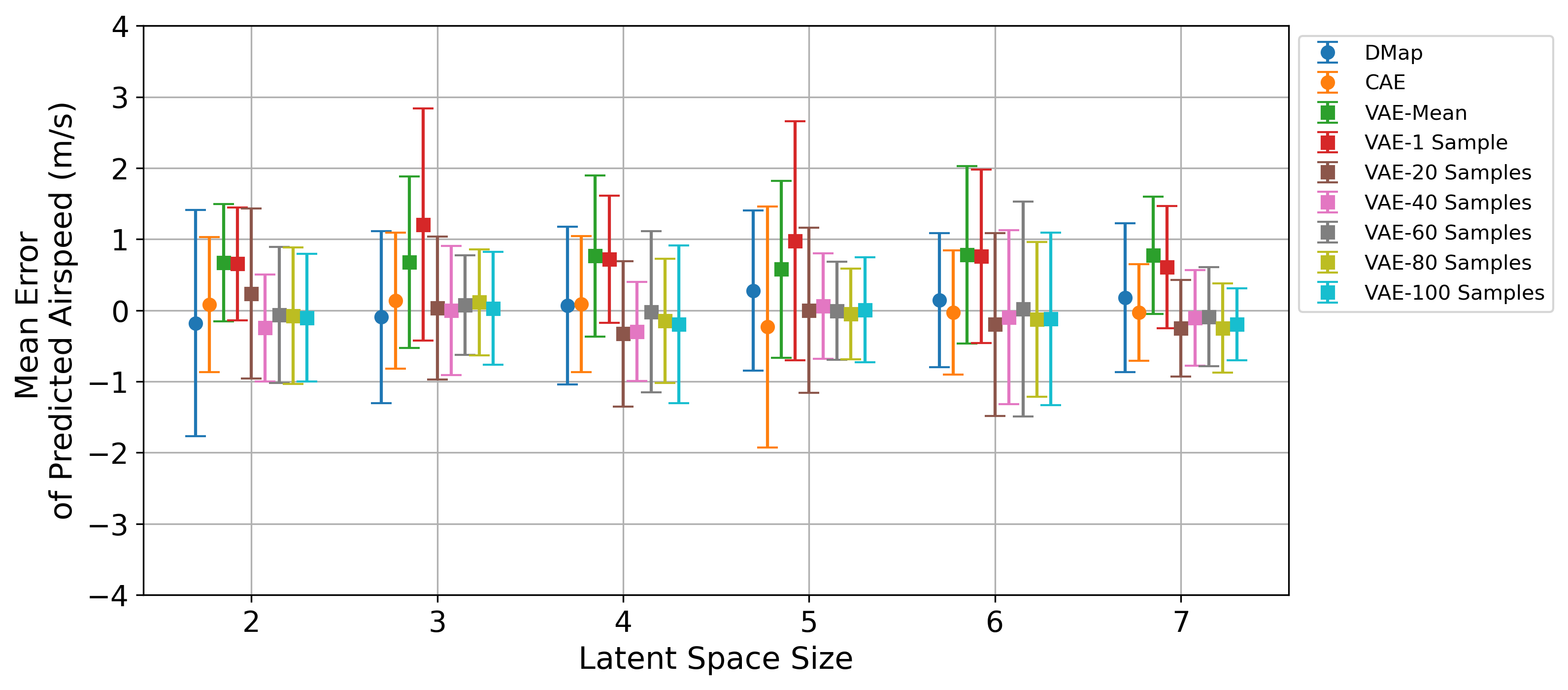}}

    \put(240,180){ \large \textbf{(a)}}
    \put(240,25){\large \textbf{(b)}}  
    % \put(110,15){\large \textbf{(c)}}  
    % \put(225,15){\large \textbf{(d)}}  
    \end{picture}
    % \vspace{-0.5cm}
    
    \caption{Error bar plots of overall state prediction results from test case II. (a): AoA prediction error from various models; (b): airspeed prediction error from various models. Among the three models examined, CAE and VAE exhibit lower prediction errors and reduced uncertainty compared to DMap.} 
    % One interesting observation here is, though does not have an obvious impact on the results of AoA prediction, adding more sampled data from VAE can improve airspeed prediction and surpass CAE.} 
\label{fig:case2all} %\vspace{-12pt}
\end{figure}

%%%%%%%%%%% predictive error for lat 5, sample 20
\begin{figure}[t!]
    \centering
    \begin{picture}(200,320)
% \put(-130,0)
    \put(-50,160){ \includegraphics[width=0.66\columnwidth]{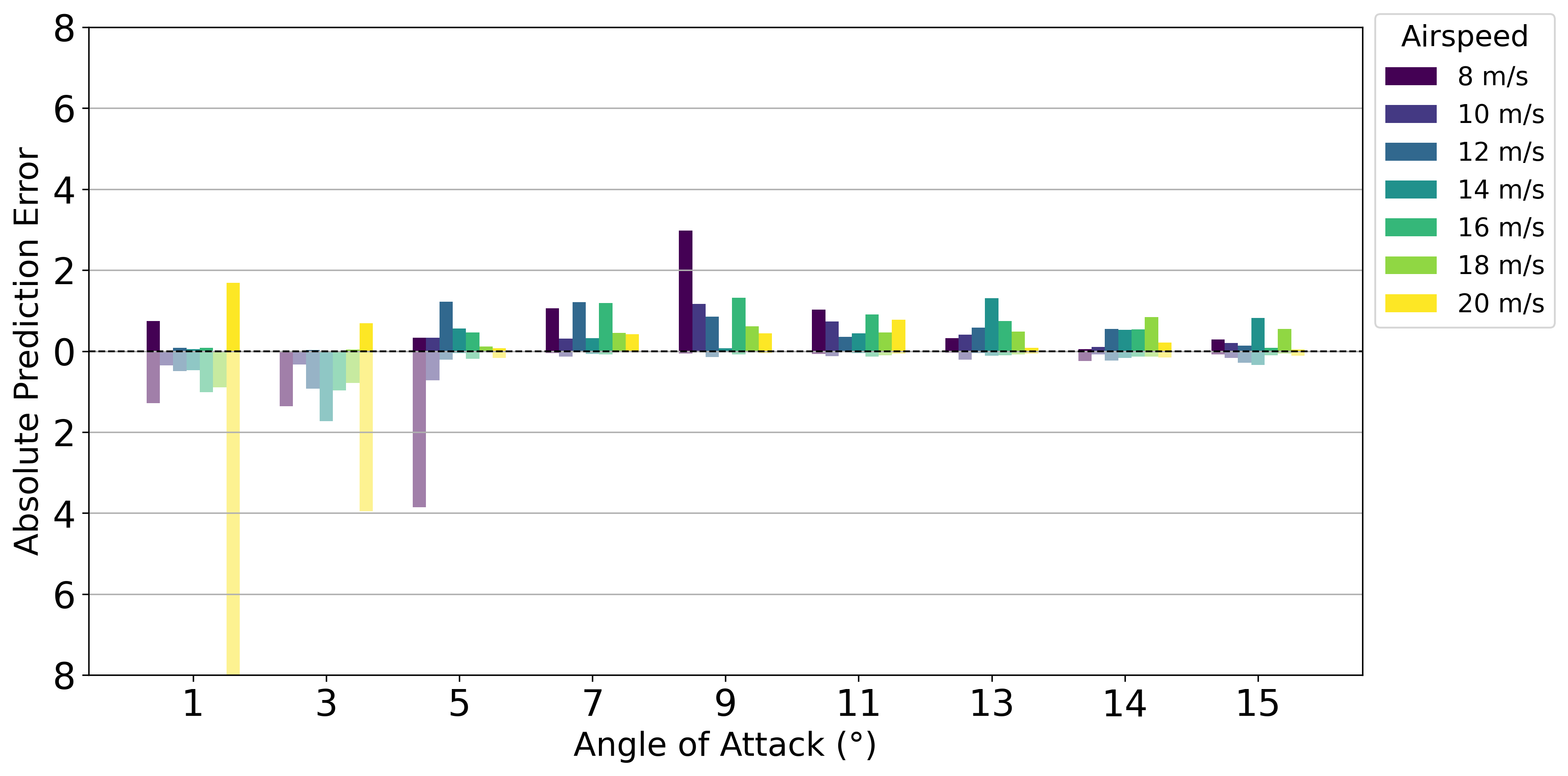}}
    \put(-50,0){\includegraphics[width=0.66\columnwidth]{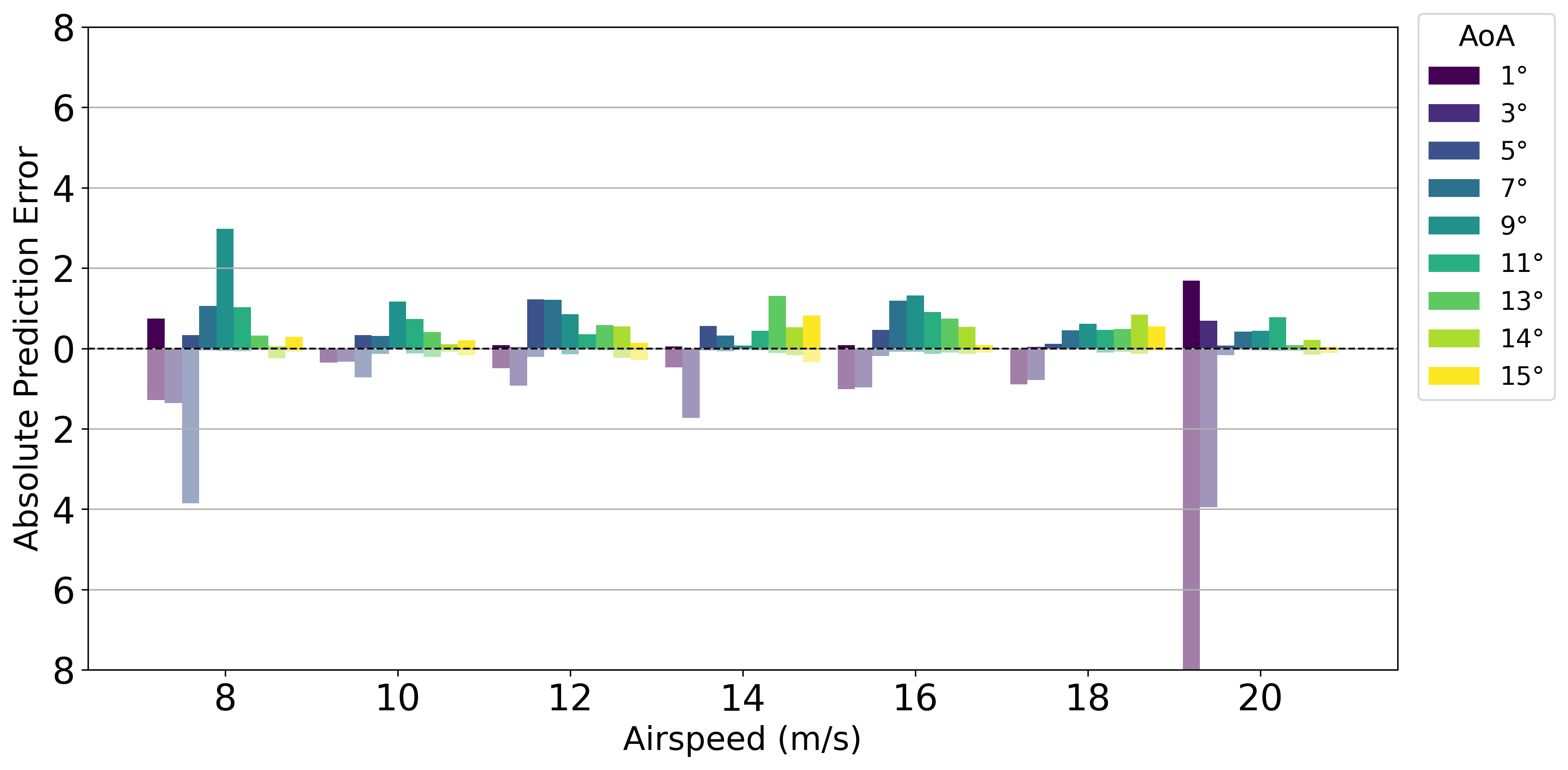}}

    % \put(-140,5){ \large \textbf{(a)}}
    % \put(100,5){\large \textbf{(b)}}
    \put(240,165){ \large \textbf{(a)}}
    \put(240,15){\large \textbf{(b)}}
    % \put(110,15){\large \textbf{(c)}}  
    % \put(225,15){\large \textbf{(d)}}  
    \end{picture}
    % \vspace{-0.5cm}
    
    \caption{Prediction error by AoA and airspeed, respectively, from VAE with 20 additional sampled data at each state when latent space size is 5. (a): prediction error by AoA; (b): prediction error by airspeed. A large error at AoA=1 (°), airspeed=20 (m/s) has been observed and further investigated.} 
\label{fig:case2lat5bar} %\vspace{-12pt}
\end{figure}
Based on the latent space data, separate FFNNs were trained for each model to assess their state prediction performance. Beyond simply comparing the three models, the unique characteristic of the VAE---its ability to generate a latent space distribution characterized by a mean and variance---was leveraged to sample additional data and expand the training set. 

Figure \ref{fig:case2all} summarizes the results. Panel (a) displays the average AoA prediction error for each model under varying latent space sizes. The DMap approach generally yields the highest error and widest uncertainty intervals, which aligns with its relatively small set of model parameters. By contrast, the CAE demonstrates a substantial drop in prediction error, indicating superior state estimation performance. Although the VAE is also a machine-learning-based approach, its prediction accuracy with limited sampled data is slightly lower than that of the CAE. This discrepancy arises because the VAE’s training objective includes both signal reconstruction and latent space distribution regularization, whereas the CAE focuses 
solely on reconstruction. Adding synthetic samples drawn from the VAE's latent distribution can improve AoA prediction accuracy compared to using only a single sampled instance. Panel (b) shows similar results for airspeed prediction; however, when the number of sampled data exceeds 40, the improvement in prediction accuracy becomes less significant.

%%%%%%%%%%%%%%%% explain bar plot of prediction
Figure~\ref{fig:case2lat5bar} illustrates the state prediction results when the latent space has five dimensions, covering both AoA and airspeed. Bars above the x-axis represent AoA prediction errors, while bars below the x-axis correspond to airspeed prediction errors. In panel (a), each bar is associated with a specific airspeed; in panel (b), each bar corresponds to one AoA. Generally, the AoA estimation error is higher than that of airspeed at most states. However, a substantial airspeed prediction error appears at AoA = 1\textdegree\ and airspeed = 
20~m/s. Further investigation was conducted to identify potential causes of this anomaly.

% %%%%%%%%%%%%%%%%%%%%%%%%%%
% % look into lat space to see why this happen for lat 5 no sample, all states
% %%%%%%%%%%%%%%%%%%%%%%%%%%%%

% \begin{figure}[t!]
%     \centering
%     \begin{picture}(200,280)
%     \put(-140,140){ \includegraphics[width=0.4\columnwidth]{figures/test2/state/vae_lat5/zarr_diffAoA_testz_1z_2_no_sample.png}}
%     \put(100,140){\includegraphics[width=0.4\columnwidth]{figures/test2/state/vae_lat5/zarr_diffv_testz_1z_2_no_sample.png}}
%     \put(-140,0){ \includegraphics[width=0.4\columnwidth]{figures/test2/state/vae_lat5/zarr_diffAoA_testz_2z_5_no_sample.png}}
%     \put(106,0){\includegraphics[width=0.39\columnwidth]{figures/test2/state/vae_lat5/zarr_diffv_testz_2z_5_no_sample.png}}
    
%     %\put(40,-240){\includegraphics[width=0.7\columnwidth]{Figure/Candidacy_signal.png}}
%     \put(50,156)
%     {\large \textbf{(a)}}
%     \put(290,156)
%     {\large \textbf{(b)}}
%     \put(50,16)
%     {\large \textbf{(c)}}
%     \put(290,16)
%     {\large \textbf{(d)}}
%     %\put(30,-40){\large \textbf{(c)}}    
%     \end{picture}
%     % \vspace{-0.5cm}
    
%     \caption{Latent representation examples from test case I. (a)-(b):  Coordinates obtained from DMaps; (c)-(d): Latent variables from encoder.} 
% \label{fig:lat1} %\vspace{-12pt}
% \end{figure} 

%%%%%%%%%%%%%%%%%%%%%%%%%%
% look into lat space to see why this happen for lat 5 aoa=1 v=20 & aoa=3 v=8; phi1 vs phi2
%%%%%%%%%%%%%%%%%%%%%%%%%%%%

\begin{figure}[t!]
    \centering
    \begin{picture}(200,240)
    \put(-140,120){ \includegraphics[width=0.34\columnwidth]{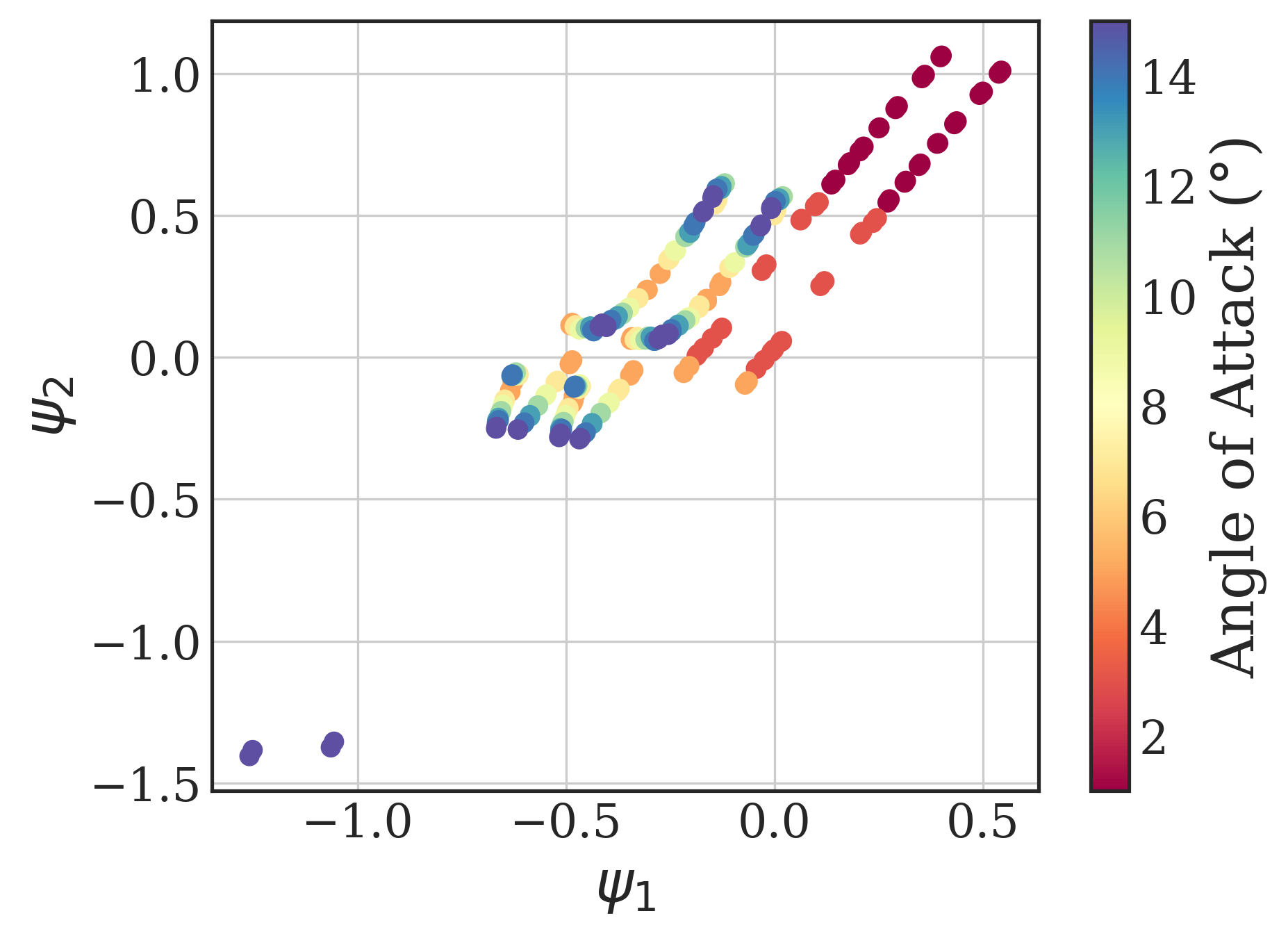}}
    \put(20,120){\includegraphics[width=0.34\columnwidth]{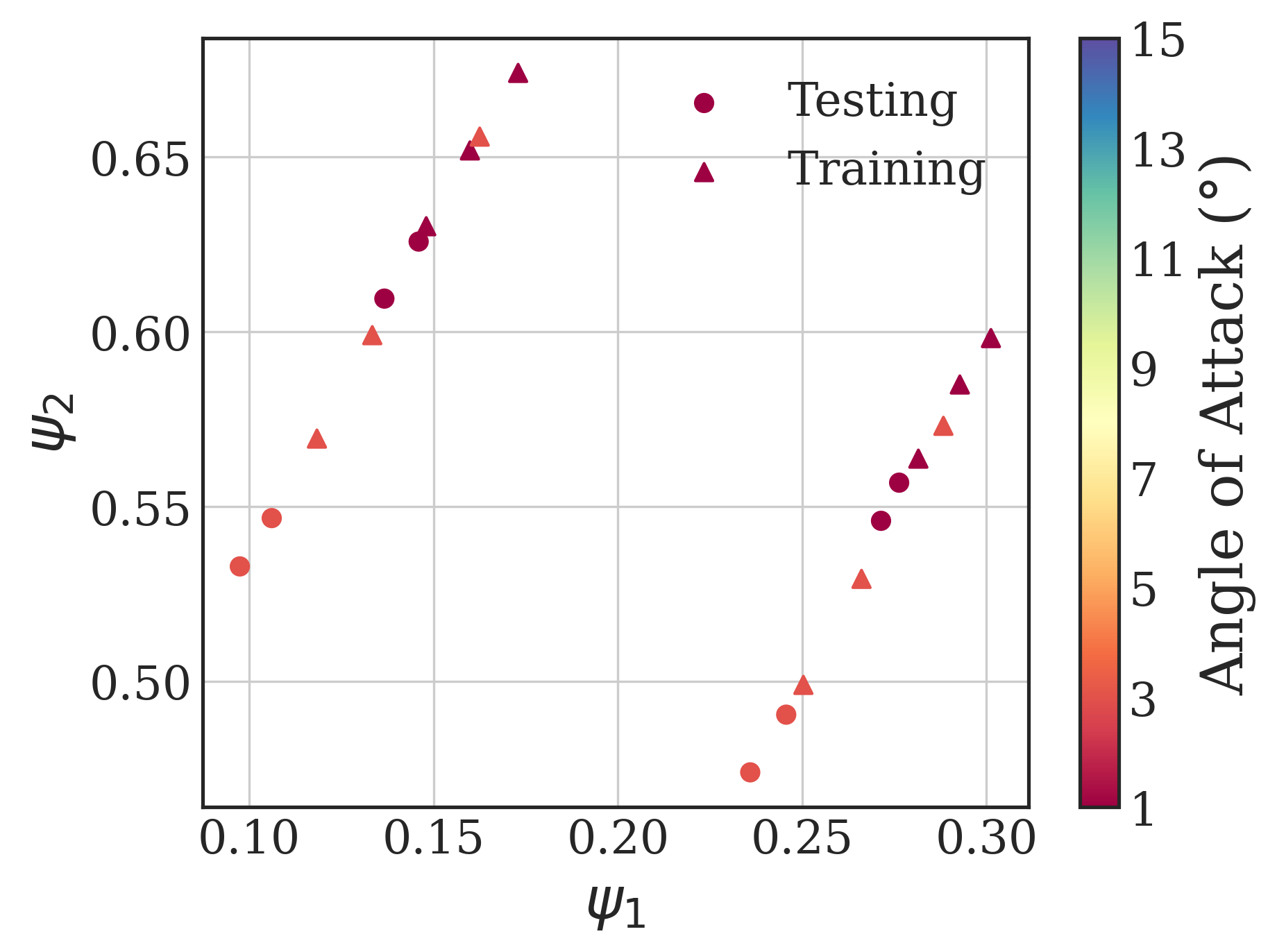}}
    \put(176,120){\includegraphics[width=0.34\columnwidth]{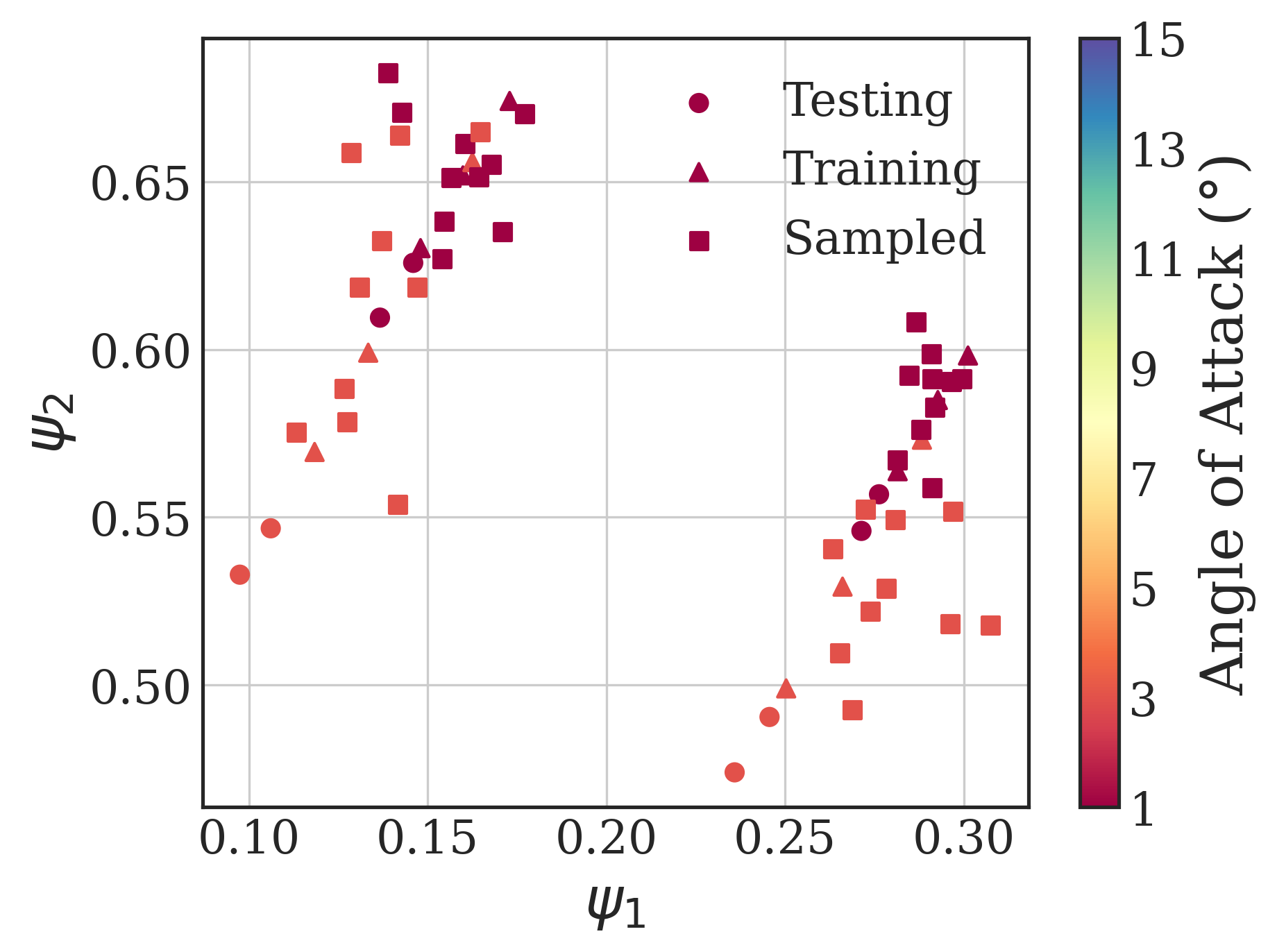}}
    \put(-140,0){ \includegraphics[width=0.34\columnwidth]{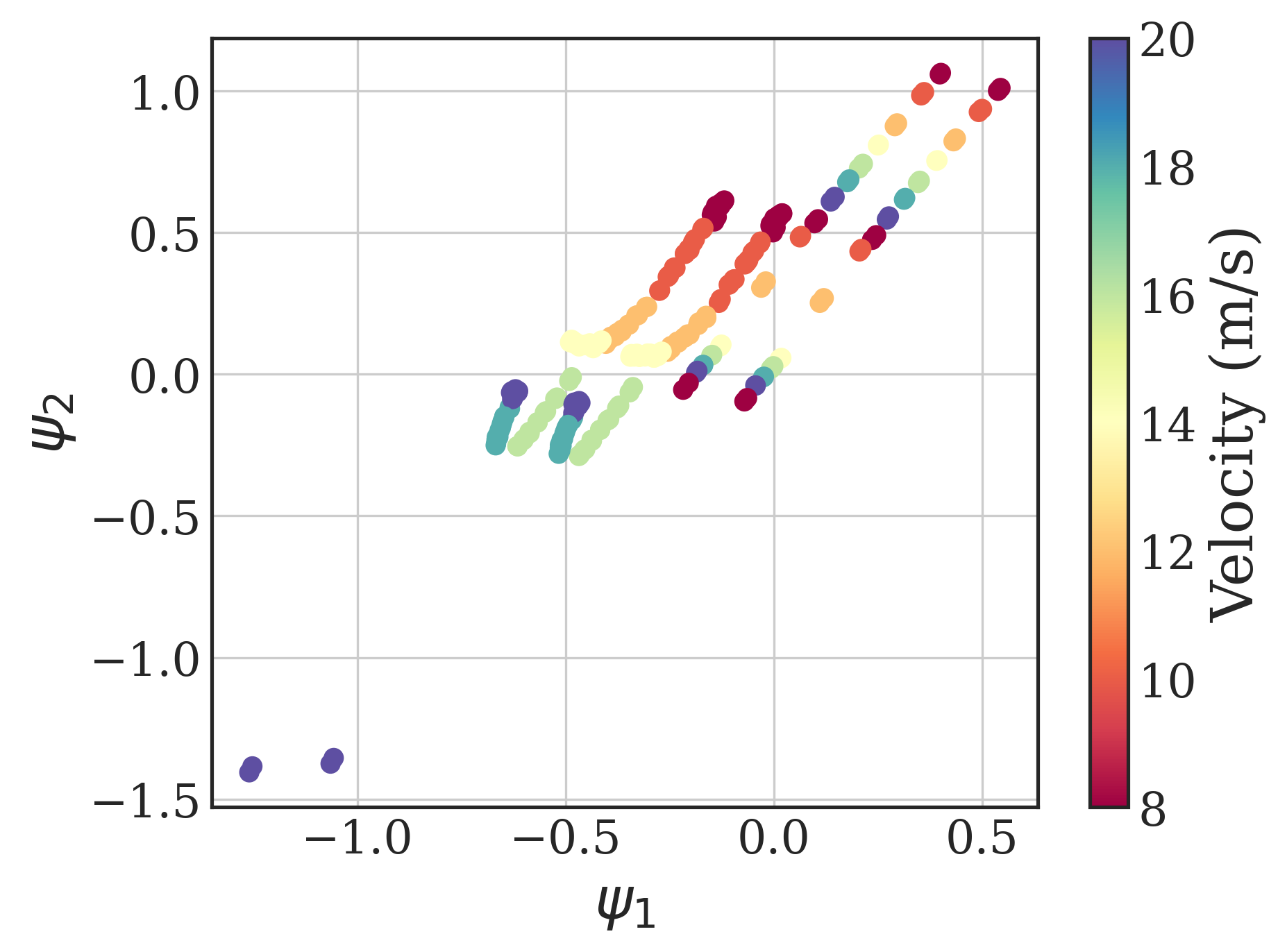}}
    \put(16,0){ \includegraphics[width=0.34\columnwidth]{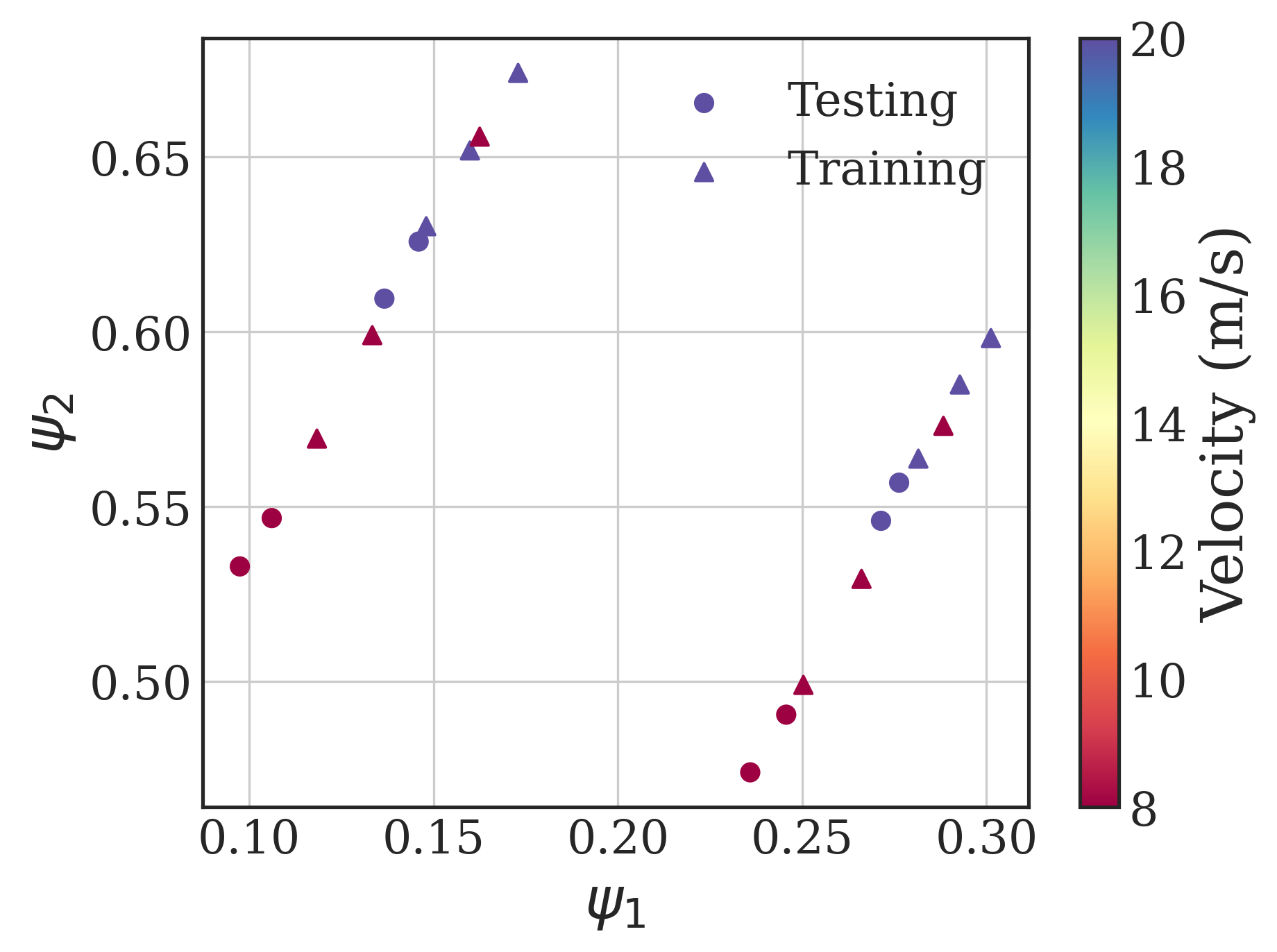}}
    \put(176,0){\includegraphics[width=0.34\columnwidth]{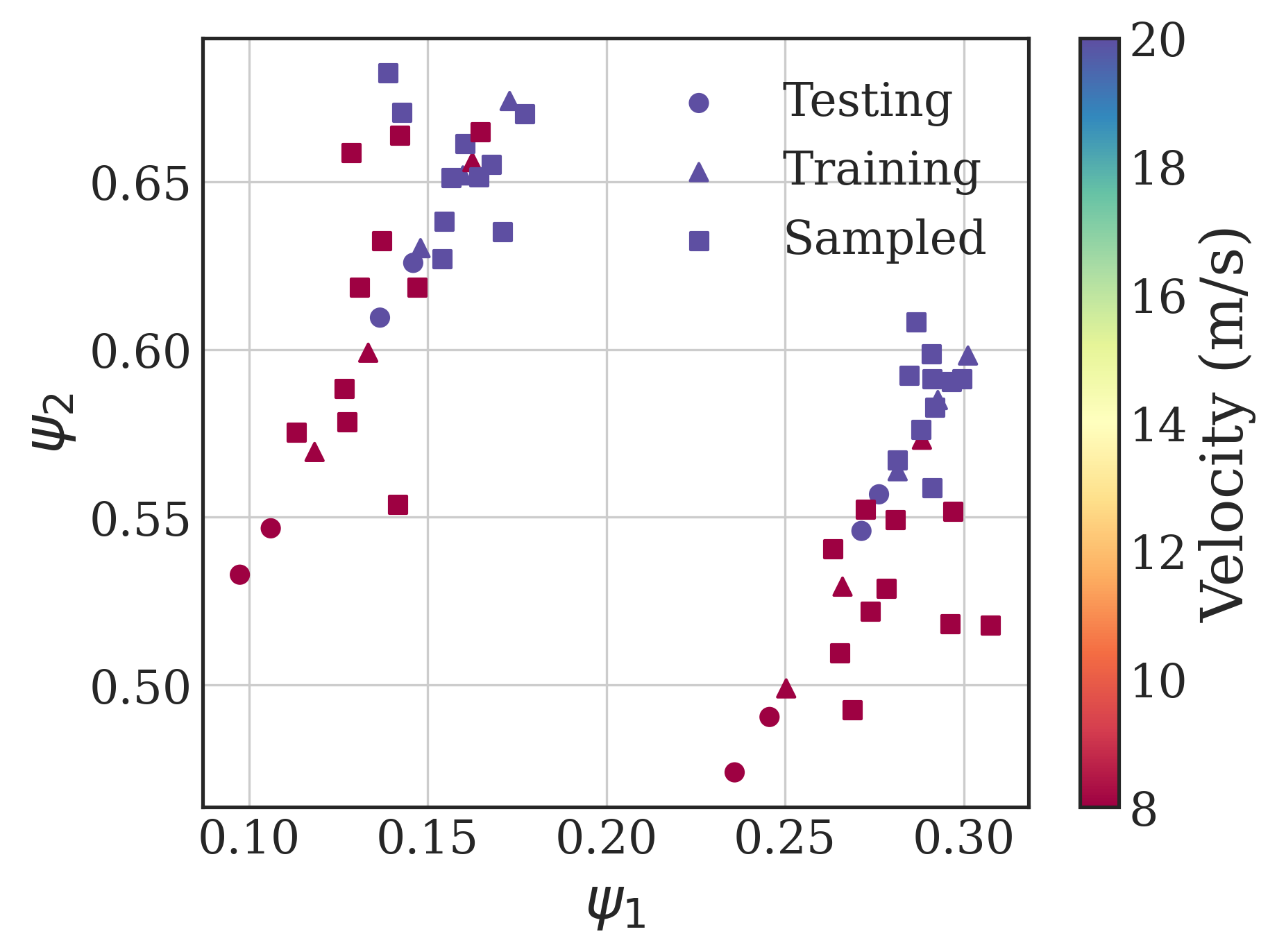}}
    
    %\put(40,-240){\includegraphics[width=0.7\columnwidth]{Figure/Candidacy_signal.png}}
    \put(10,130)
    {\large \textbf{(a)}}
    \put(10,10)
    {\large \textbf{(b)}}
    \put(170,130)
    {\large \textbf{(c)}}
    \put(172,10)
    {\large \textbf{(d)}}
    \put(326,130)
    {\large \textbf{(e)}}
    \put(328,10)
    {\large \textbf{(f)}}
    %\put(30,-40){\large \textbf{(c)}}    
    \end{picture}
    % \vspace{-0.5cm}
    
    \caption{Latent representation examples of latent space $z_2$ vs $z_1$ from VAE when latent space size is 5. (a)-(b): mean values under all states; (c)-(d): mean values under AoA=1 (°),v=20 (m/s) and AoA=3 (°),v=8 (m/s); (e)-(f): 20 sampled data under AoA=1 (°),v=20 (m/s) and AoA=3 (°),v=8 (m/s). It can be observed that the latent variables from the two states overlap, leading to a large prediction error.} 
\label{fig:case2lat5detail1} %\vspace{-12pt}
\end{figure} 

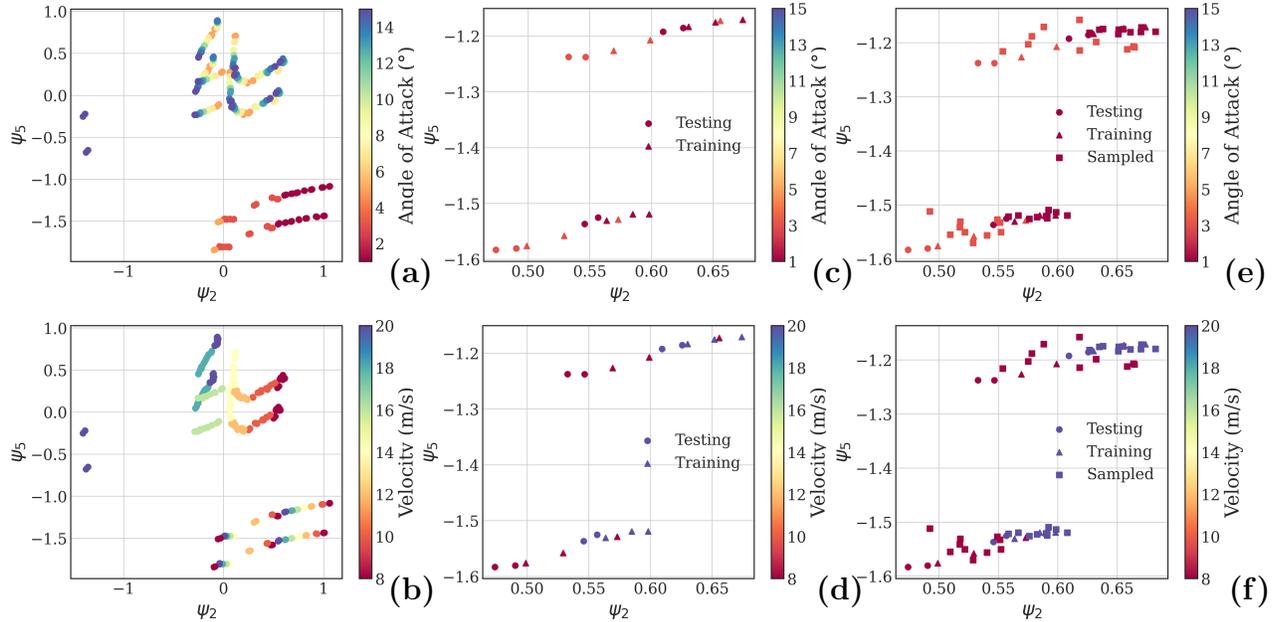
\begin{figure}[H]
    \centering
    \begin{picture}(200,240)
    \put(-140,120){ \includegraphics[width=0.34\columnwidth]{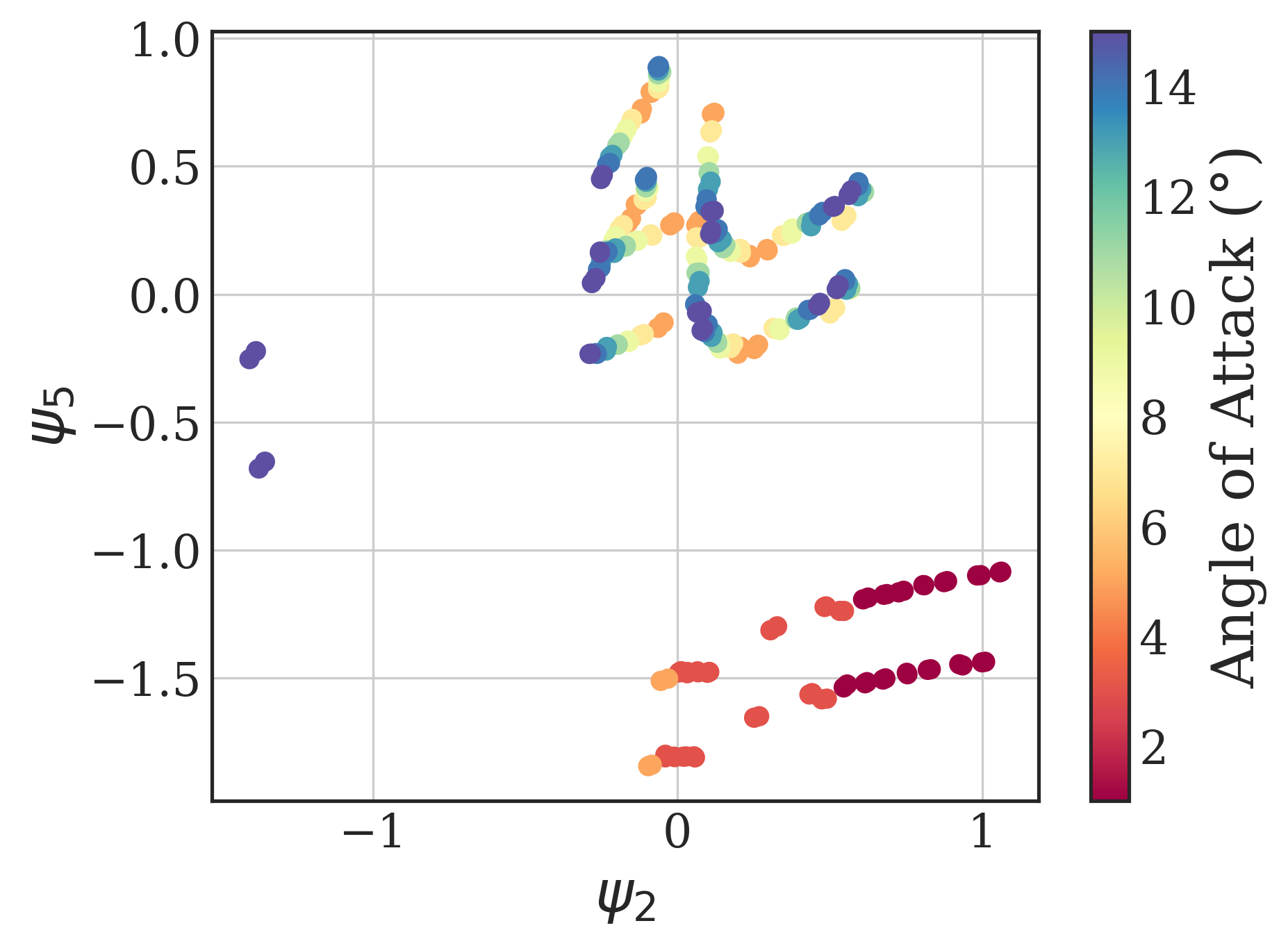}}
    \put(20,120){\includegraphics[width=0.34\columnwidth]{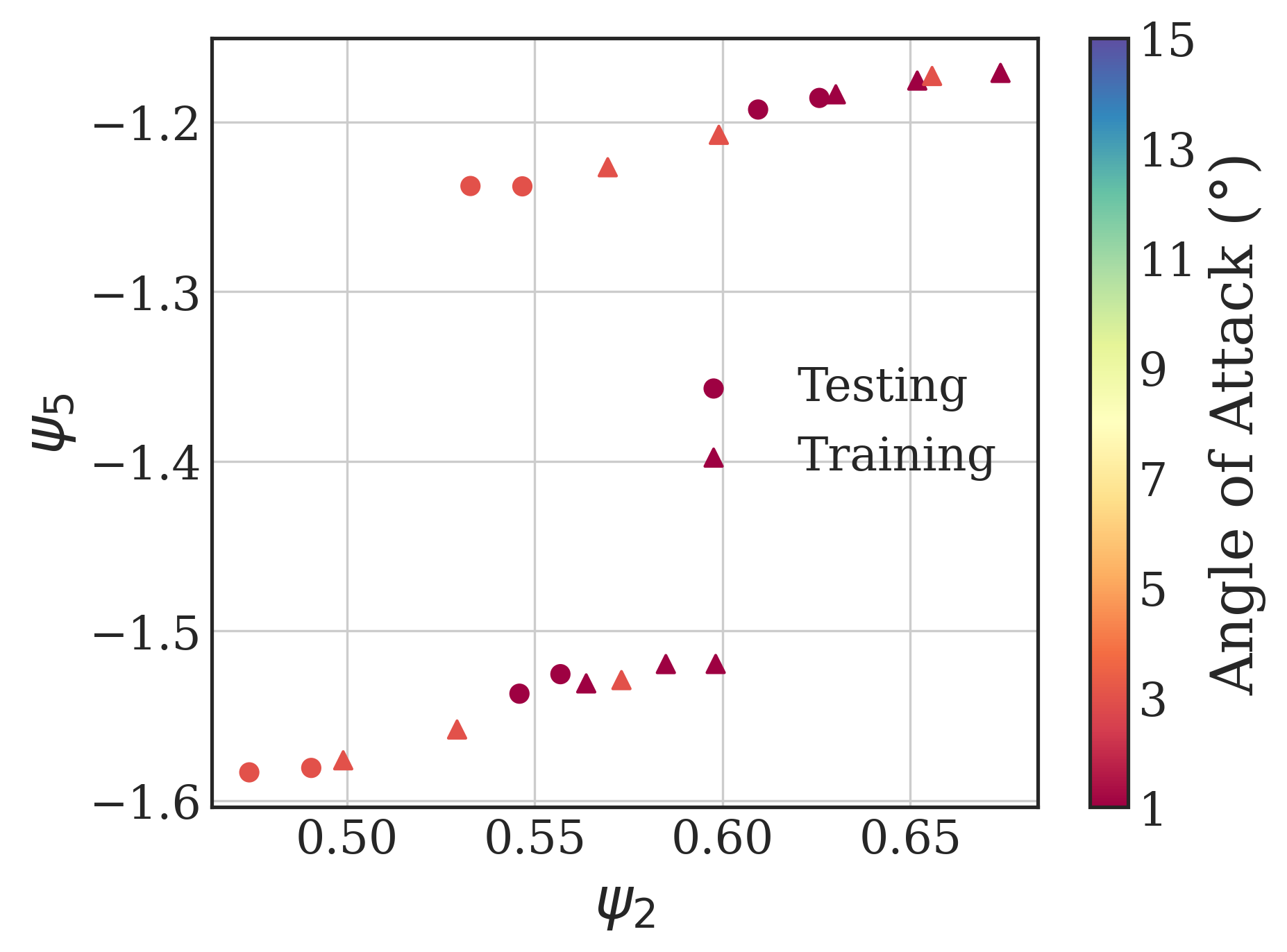}}
    \put(176,120){\includegraphics[width=0.34\columnwidth]{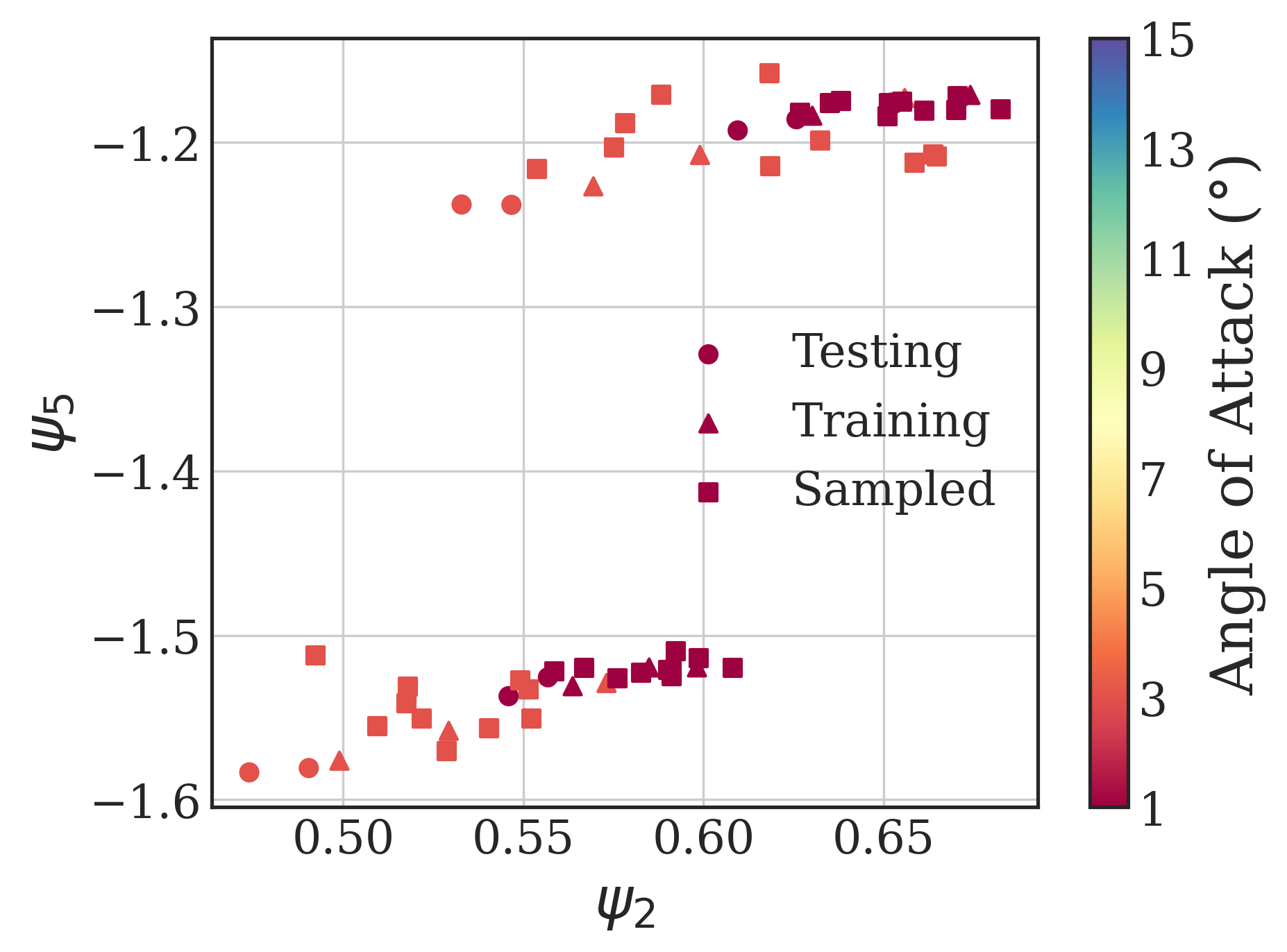}}
    \put(-140,0){ \includegraphics[width=0.34\columnwidth]{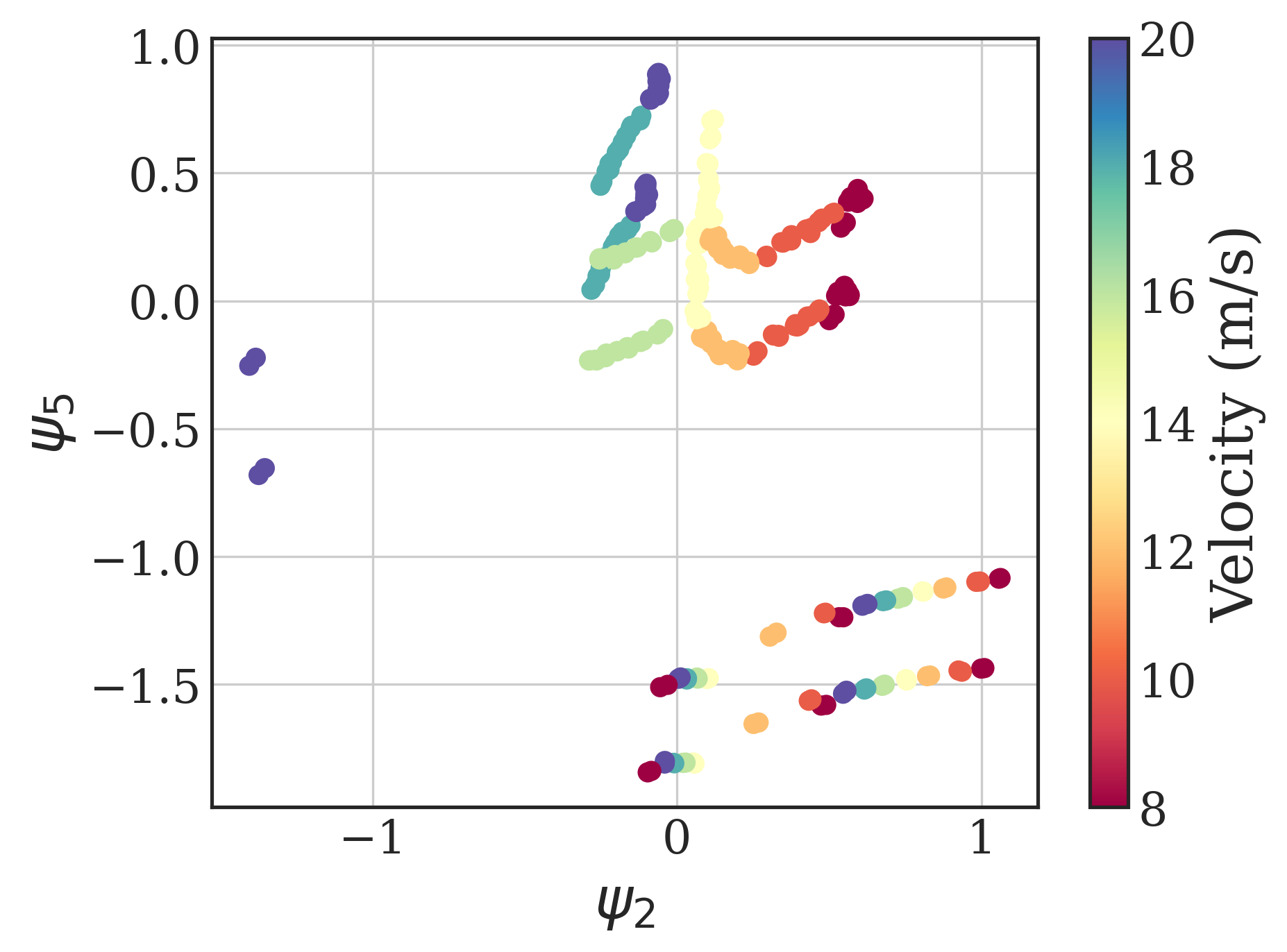}}
    \put(16,0){ \includegraphics[width=0.34\columnwidth]{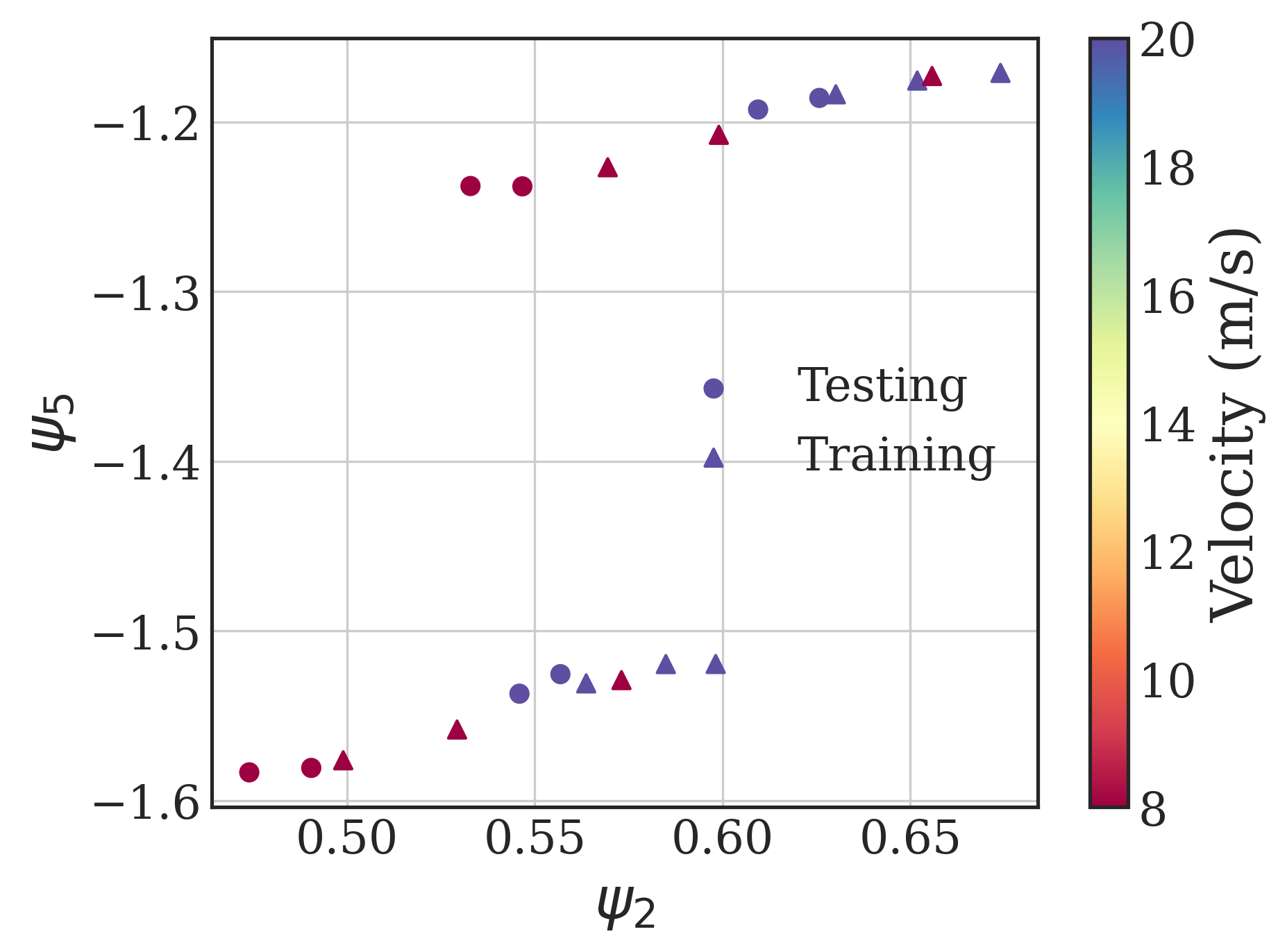}}
    \put(176,0){\includegraphics[width=0.34\columnwidth]{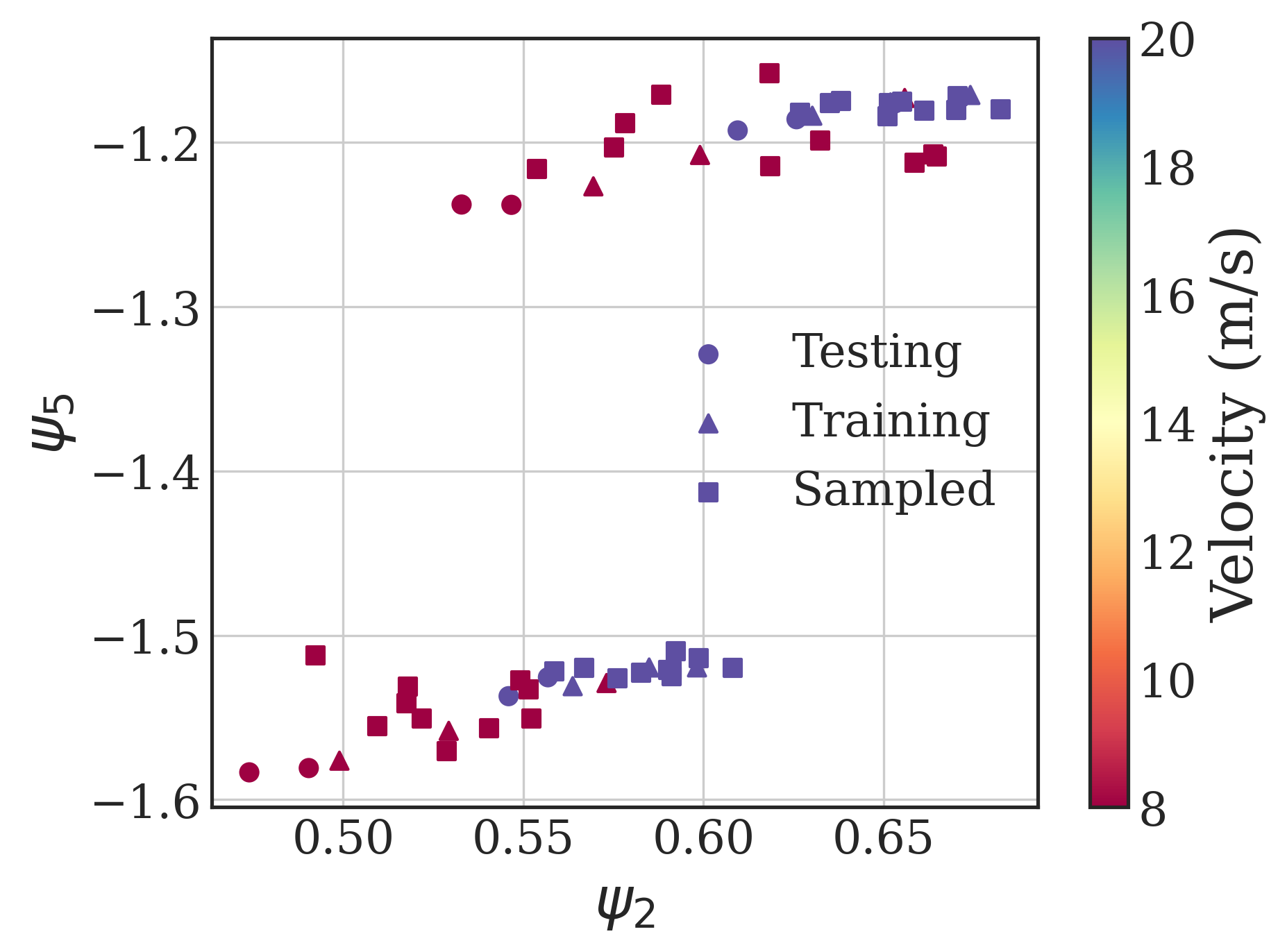}}
    
    %\put(40,-240){\includegraphics[width=0.7\columnwidth]{Figure/Candidacy_signal.png}}
    \put(10,130)
    {\large \textbf{(a)}}
    \put(10,10)
    {\large \textbf{(b)}}
    \put(170,130)
    {\large \textbf{(c)}}
    \put(172,10)
    {\large \textbf{(d)}}
    \put(326,130)
    {\large \textbf{(e)}}
    \put(328,10)
    {\large \textbf{(f)}}
    %\put(30,-40){\large \textbf{(c)}}    
    \end{picture}
    % \vspace{-0.5cm}
    
    \caption{Latent representation examples of latent space $z_5$ vs $z_2$ from VAE when latent space size is 5. (a)-(b): mean values under all states; (c)-(d): mean values under AoA=1 (°),v=20 (m/s) and AoA=3 (°),v=8 (m/s); (e)-(f): 20 sampled data under AoA=1 (°),v=20 (m/s) and AoA=3 (°),v=8 (m/s). It can be observed that the latent variables from the two states overlap, leading to a large prediction error.} 
\label{fig:case2lat5detail2} %\vspace{-12pt}
\end{figure} 
Figure \ref{fig:case2lat5detail1} presents examples of latent variable pairs at a five-dimensional latent space, illustrating why the airspeed prediction error is substantial in 
Figure \ref{fig:case2lat5bar}. Panels (a)--(b) depict the latent space coordinates under varying AoA and airspeeds, respectively. In panel~(b), one can observe a clear trajectory as velocity increases; specifically, the blue markers (corresponding to $v=20$~m/s) generally cluster away from the warm-colored markers that represent lower velocities. However, closer inspection reveals overlaps; for instance, panels (c)--(d) focus on just two states, (\textit{i.e.}, AoA=1\textdegree, $v=20$~m/s and AoA=3\textdegree, $v=8$~m/s), where the latent space points overlap. This overlap explains the considerable discrepancy between the predicted and actual airspeeds. Panels (e)--(f) display latent space mappings by adding twenty data sampled from the VAE’s latent distribution. Although the overlap persists, the prediction accuracy can increase from a probabilistic perspective.

%%%%%%%%%%%%%%%%%%%%%%%%%%
% look into lat space to see why this happen for lat 5 aoa=1 v=20 & aoa=3 v=8; phi1 vs phi2
%%%%%%%%%%%%%%%%%%%%%%%%%%%%

Figure \ref{fig:case2lat5detail2} exhibits the similar plots yet with a different latent variable pairs, i.e., $z_5$ vs. $z_2$, where similar conclusion can be drawn.

%%%%%%%%%%% predictive error for lat 5, sample 20, use z3 & z5 from lat 5

\begin{figure}[t!]
    \centering
    \begin{picture}(200,320)

    \put(-50,160){ \includegraphics[width=0.66\columnwidth]{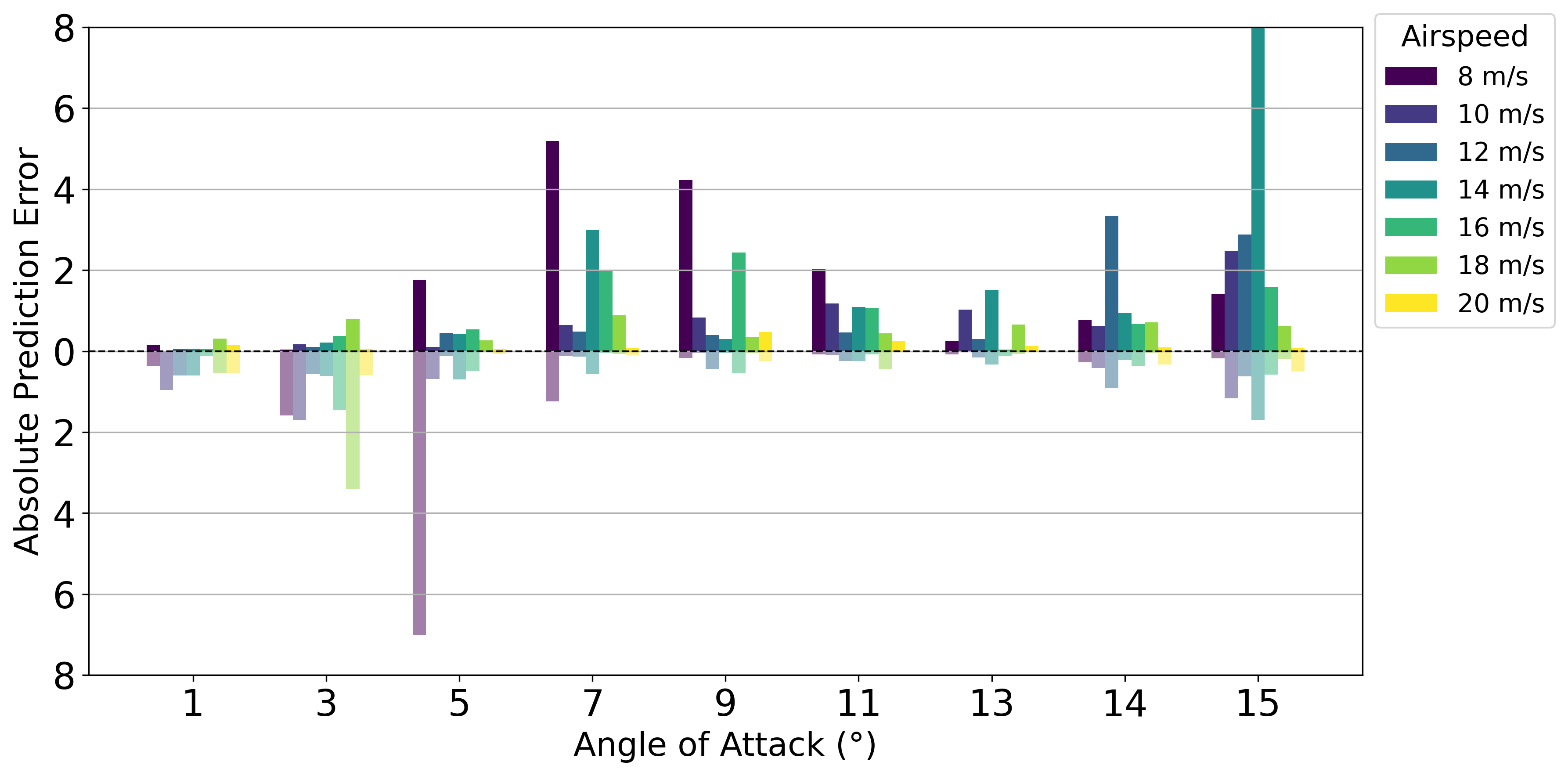}}
    \put(-50,0){\includegraphics[width=0.66\columnwidth]{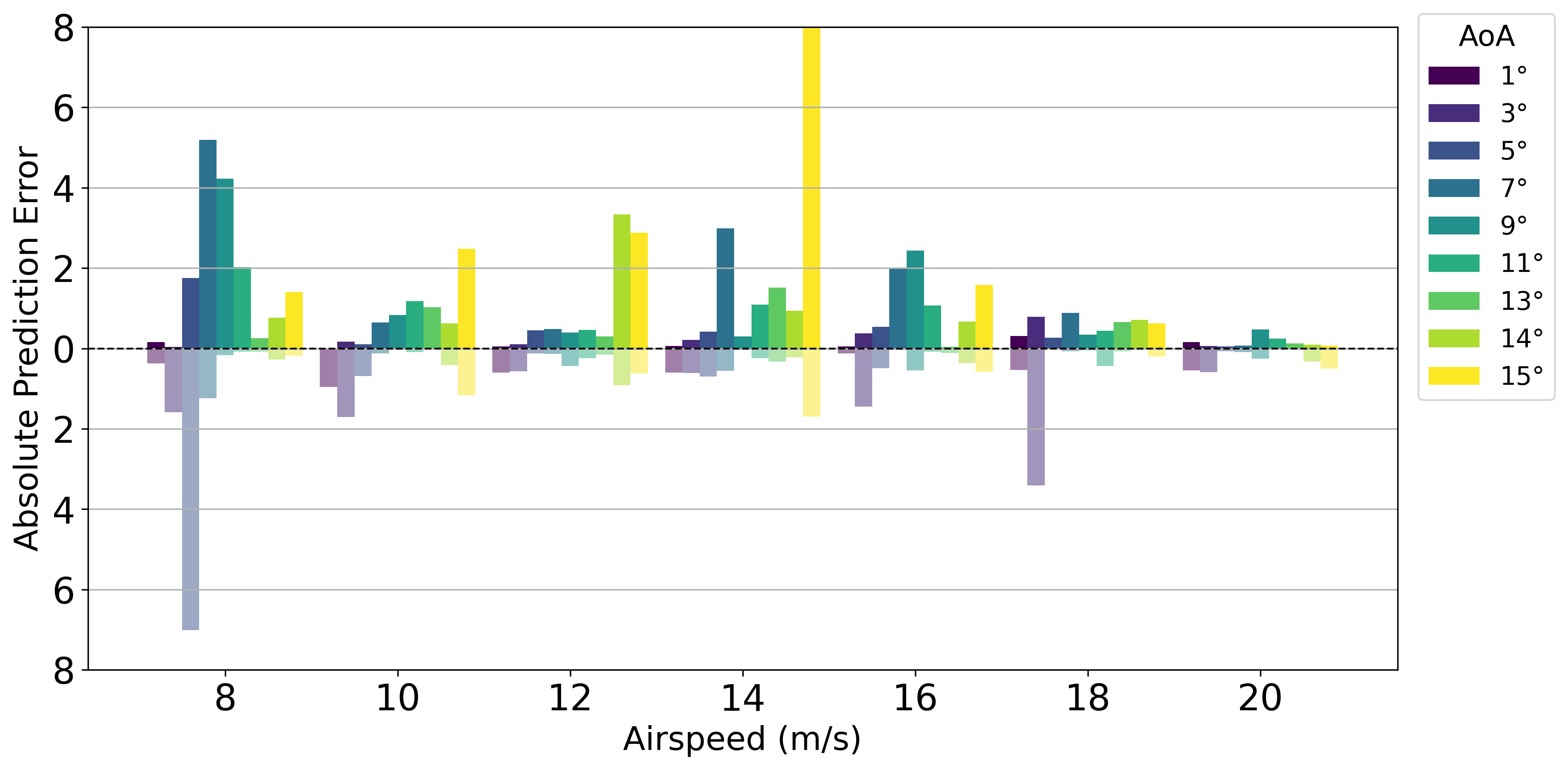}}

    \put(240,165){ \large \textbf{(a)}}
    \put(240,15){\large \textbf{(b)}}  
    % \put(110,15){\large \textbf{(c)}}  
    % \put(225,15){\large \textbf{(d)}}  
    \end{picture}
    % \vspace{-0.5cm}
    
    \caption{Prediction error by AoA and airspeed, respectively, from VAE with 20 additional sampled data at each state using only $z_3$ and $z_5$ when latent space size is 5. (a): prediction error by AoA; (b) prediction error by airspeed. The error at AoA=1 (°), airspeed=20 (m/s) decreases yet goes higher in other states.} 
\label{fig:case2lat5bar2} %\vspace{-12pt}
\end{figure}

Upon examining the five-dimensional latent space, the authors noted that plotting $z_5$ against $z_3$ offers better separation for the previously overlapping states. This observation suggests that using only these two variables, rather than the full five-dimensional latent space, may mitigate the prediction error at those particular states. Figure~\ref{fig:case2lat5bar2} presents the 
resulting prediction errors when only these two variables are employed for the state estimation 
FFNN. Although the airspeed prediction error at AoA = 1\textdegree{} and $v = 20$~m/s is indeed 
reduced (as indicated by the small yellow bar), the prediction error at other states increases. 
Therefore, truncating the latent space appears to be inadvisable.

%%%%%%%%%%%%%%%%%%%%%%%%%%%%

\subsubsection{Signal Reconstruction}

%%%%%%%%%%% signal reconstruction result for lat 5

\begin{figure}[t!]
    \centering
    \begin{picture}(200,230)
    \put(-122,150){ \includegraphics[width=0.48\columnwidth]{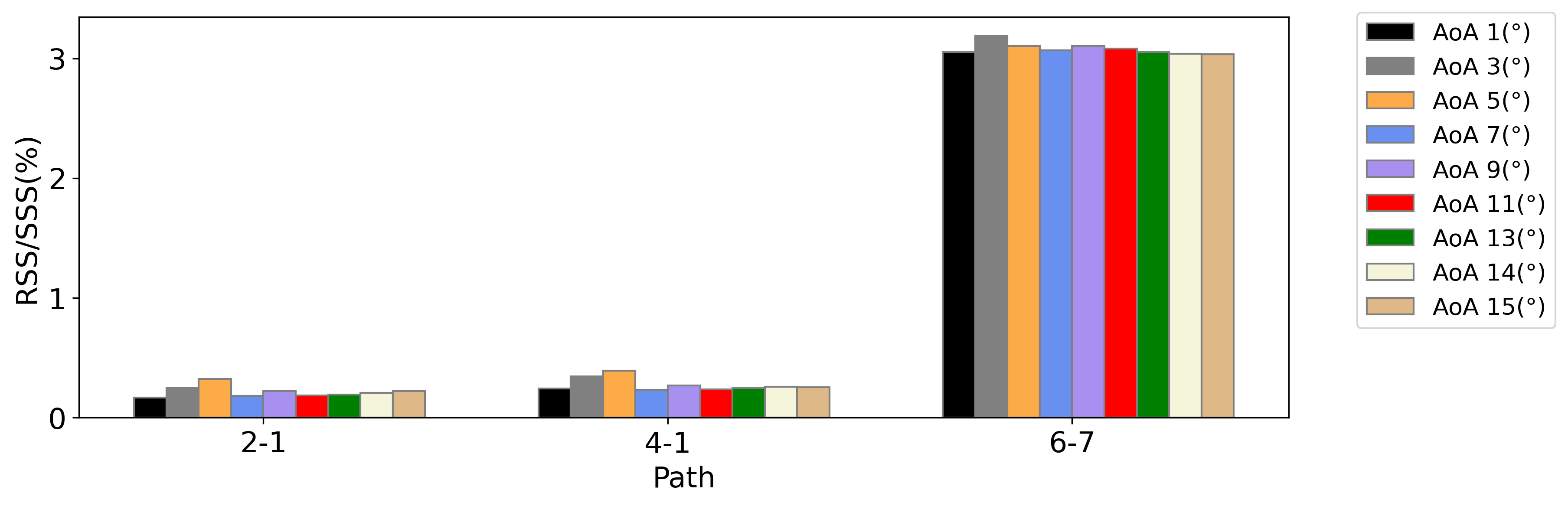}}
    \put(124,150){\includegraphics[width=0.48\columnwidth]{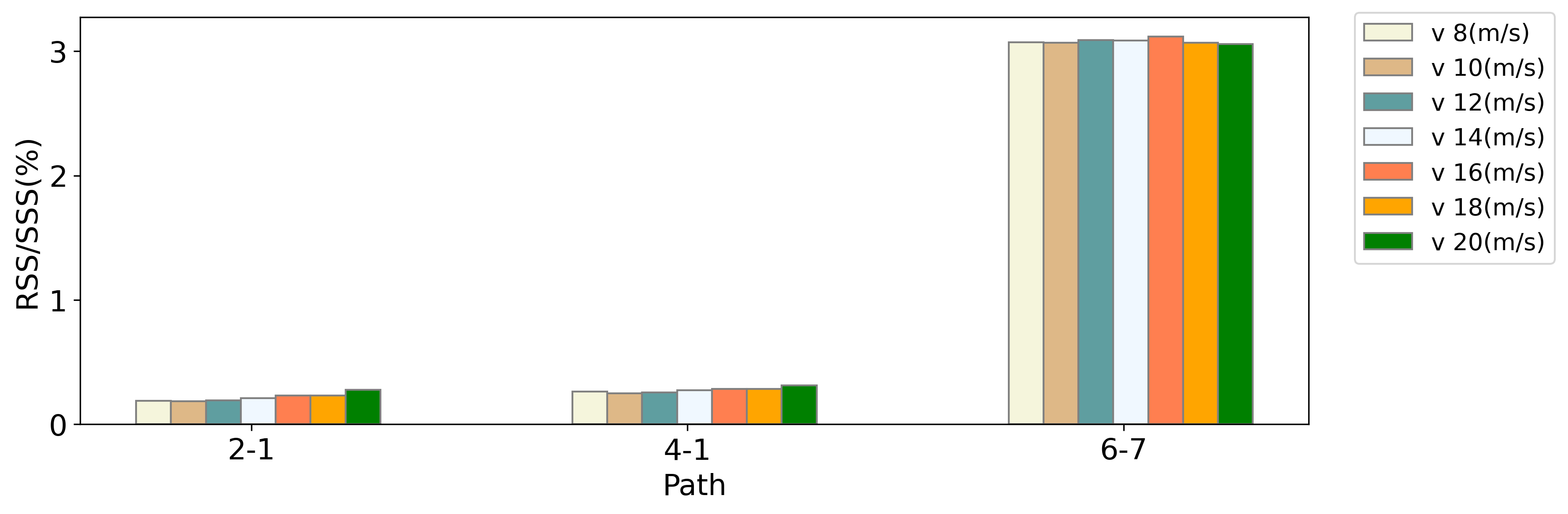}}
    \put(-130,75){ \includegraphics[width=0.5\columnwidth]{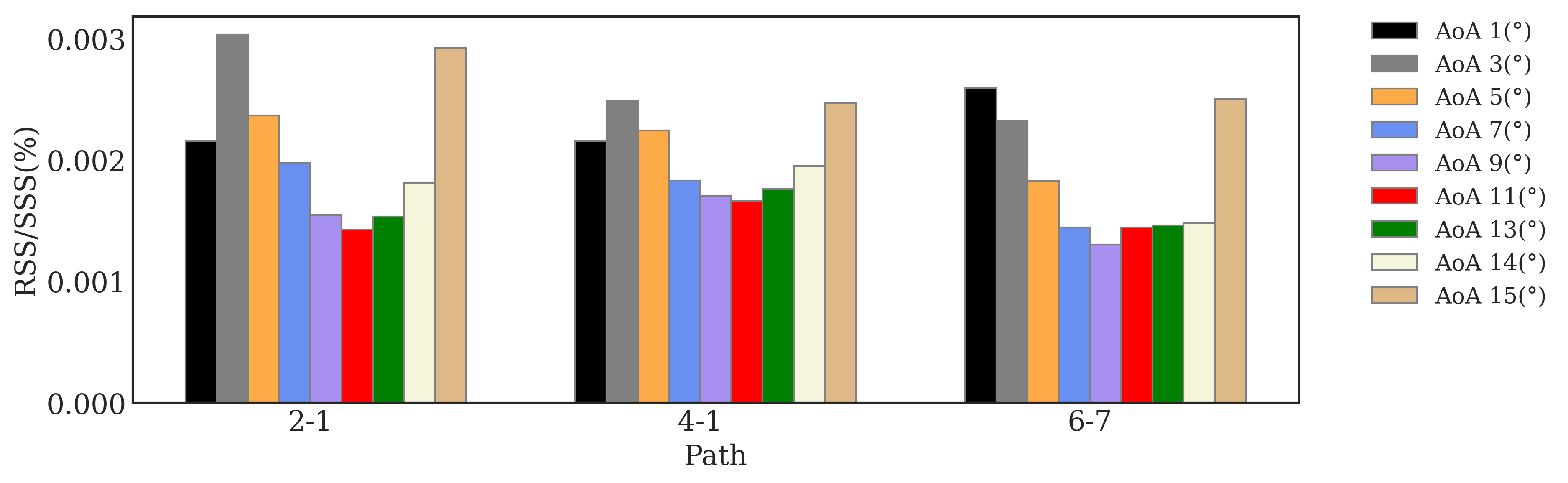}}
    \put(115,75){\includegraphics[width=0.5\columnwidth]{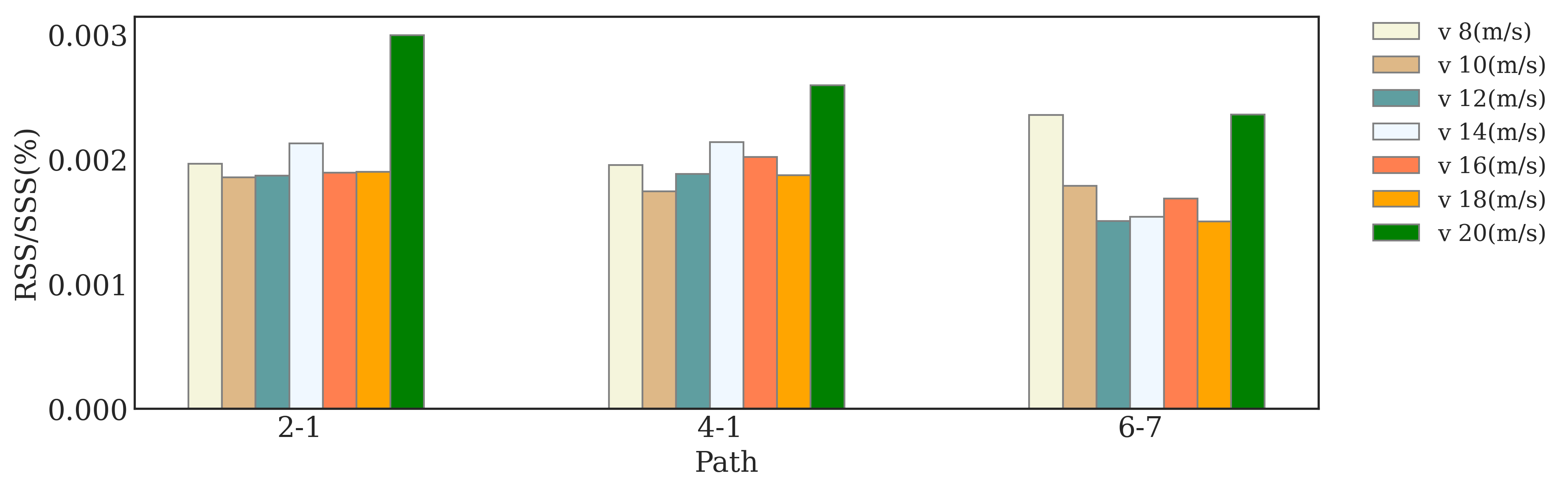}}
    \put(-130,0){ \includegraphics[width=0.5\columnwidth]{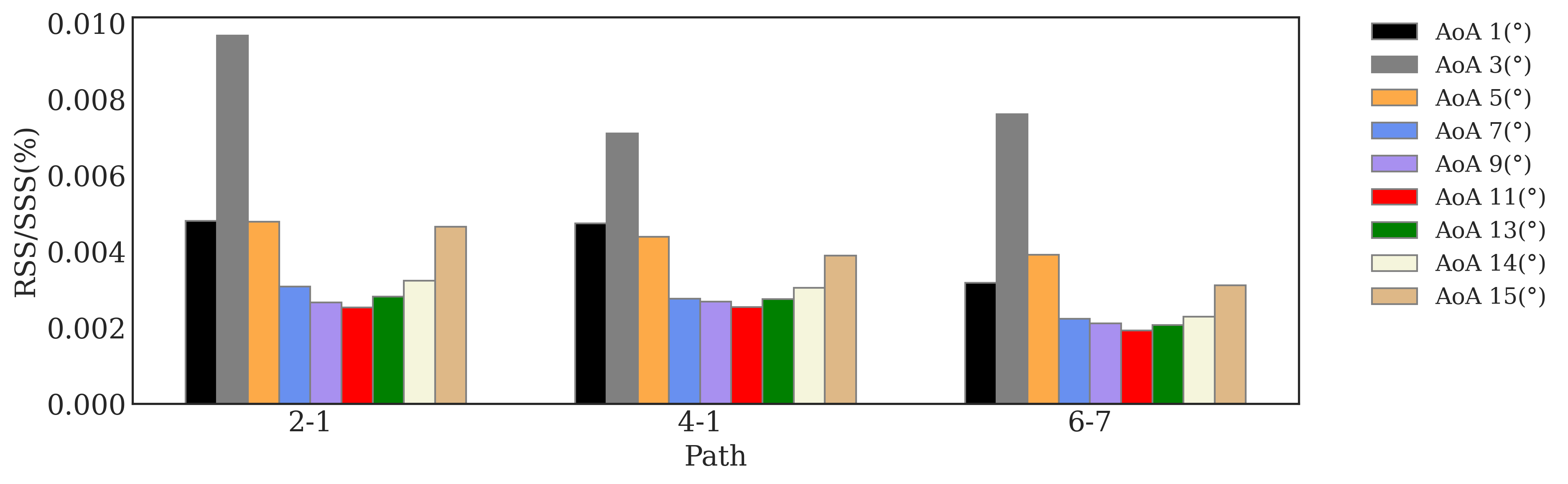}}
    \put(115,0){\includegraphics[width=0.5\columnwidth]{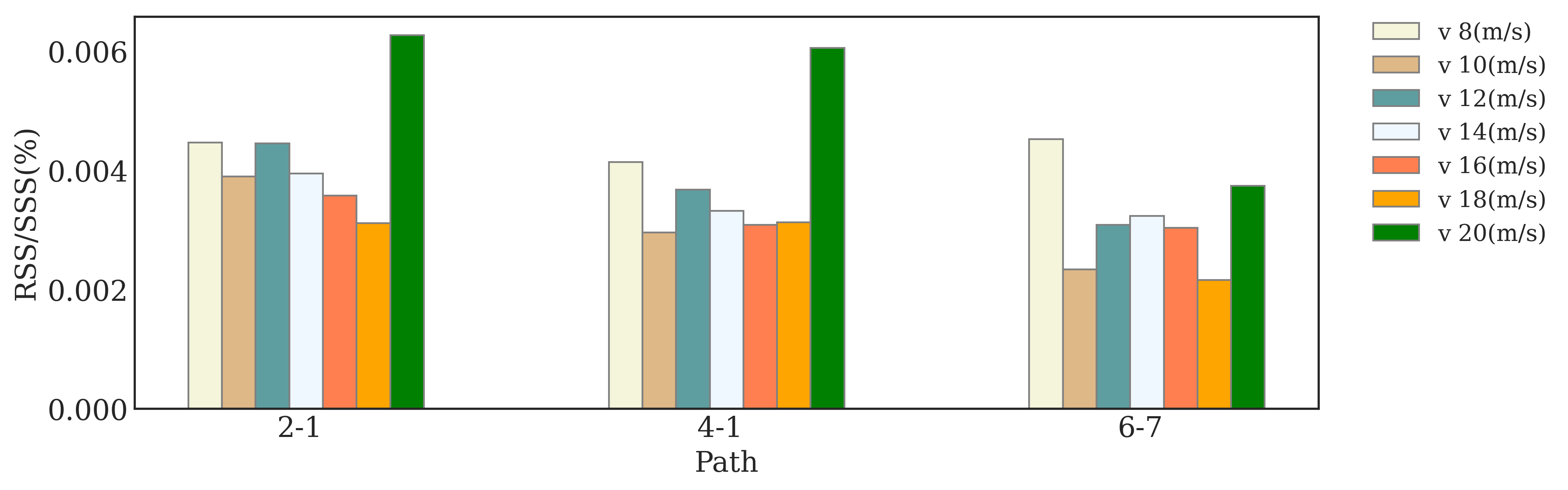}}
    
    \put(-140,150){ \large \textbf{(a)}}
    \put(100,150){\large \textbf{(b)}}  
    \put(-140,75){ \large \textbf{(c)}}
    \put(100,75){\large \textbf{(d)}}  
    \put(-140,5){ \large \textbf{(e)}}
    \put(100,5){\large \textbf{(f)}}  
    % \put(110,15){\large \textbf{(c)}}  
    % \put(225,15){\large \textbf{(d)}}  
    \end{picture}
    % \vspace{-0.5cm}
    
    \caption{Signal reconstruction error by angle of attack (AoA) and airspeed, respectively, for DMap, CAE, and VAE when the latent space has five dimensions. Panels (a)--(b) illustrate the DMap results, (c)--(d) show those for the CAE, and (e)--(f) present the VAE.}

\label{fig:case2signal} %\vspace{-12pt}
\end{figure}
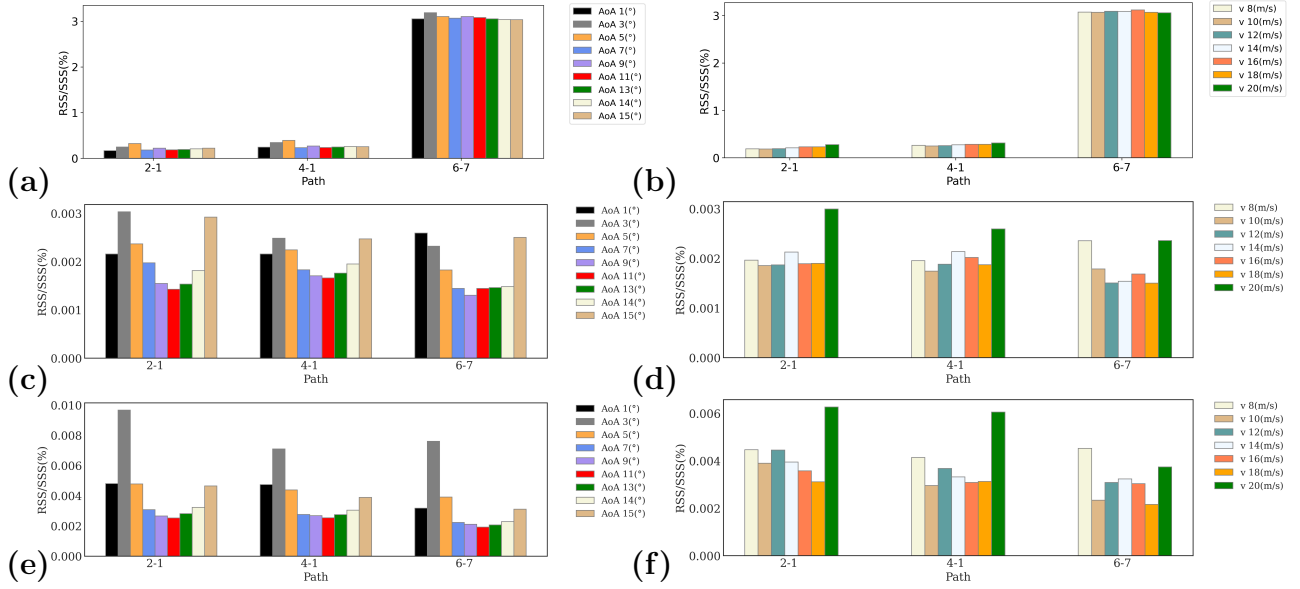

Figure \ref{fig:case2signal} presents the signal reconstruction results for the three data compression and expansion techniques. Panels~(a)--(b) display the average reconstruction error at different angles of attack (AoA) and airspeeds, respectively. Although the DMap approach maintains a consistently low error on sensor paths~2--1 and~4--1, it performs less reliably on path 6--7. By contrast, panels (c)--(d) indicate that the CAE achieves uniformly lower and more consistent errors across all tested states and sensor paths. Panels (e)--(f) illustrate the VAE’s results: although slightly less accurate than the CAE, the VAE nonetheless demonstrates sound overall reconstruction performance.

%-----------------------------------------------------------------------------------------------------
%                                        CONCLUDING REMARKS
%-----------------------------------------------------------------------------------------------------
\section{Concluding Remarks} \label{sec:conclusions}

A comprehensive approach for state prediction and signal reconstruction in structural health monitoring systems is put forward and thoroughly examined here. As multi-dimensional, nonlinear data become more common, the need for flexible data compression frameworks grows. Manifold learning techniques (e.g., diffusion maps) and deep learning models (e.g., autoencoders) reduce data to latent spaces without manually extracting features, thus preserving core information. In this latent form, the features are sensitive to both damage levels and environmental/operational factors, enabling simultaneous state and EOC estimation. Additionally, these latent representations retain enough key attributes to allow effective data recovery. To validate this framework, implementations of diffusion maps, CAE, and VAE were tested and compared for their dual functionality of state prediction and signal reconstruction. After successfully handling data under static conditions with good predictive accuracy and minimal reconstruction error, the models were further challenged by a noisy wind tunnel experiment. Overall, CAE yielded the smallest errors while diffusion maps demonstrated the weakest accuracy, reinforcing that deep learning–which generally has more tunable parameters–can learn subtleties less accessible to other methods. Although VAE introduced slightly higher errors than CAE, its performance could be further enhanced by leveraging the learned latent distribution for data augmentation.

%-----------------------------------------------------------------------------------------------------
%                                                    Acknowledgment
%-----------------------------------------------------------------------------------------------------
% \section*{Acknowledgment}

% This work is carried out at the Rensselaer Polytechnic Institute under the Vertical Lift Research Center of Excellence (VLRCOE) Program, grant number W911W61120012, with Dr. Mahendra Bhagwat and Dr. William Lewis as Technical Monitors. 

%-----------------------------------------------------------------------------------------------------
%                                                    References
%-----------------------------------------------------------------------------------------------------

\bibliographystyle{aiaa} 

\bibliography{references} % bibliography data 
%\bibliography{wing_references} % bibliography data 

%----------------------------------------------------------------------------------------------------
% Appendix
%----------------------------------------------------------------------------------------------------
% \appendix
% \appendixpage
% \addappheadtotoc

% \section{Additional results}

\end{document}